\documentclass[a4paper,reqno,9pt]{amsart}
\usepackage[margin=2.5cm,papersize={19.75cm,28cm}]{geometry}

\usepackage{amsmath}
  \usepackage{paralist}
  \usepackage{graphics} %% add this and next lines if pictures should be in esp format
  \usepackage{epsfig} %For pictures: screened artwork should be set up with an 85 or 100 line screen
\usepackage{graphicx}  \usepackage{epstopdf}%This is to transfer .eps figure to .pdf figure; please compile your paper using PDFLeTex or PDFTeXify.
 \usepackage[colorlinks=true]{hyperref}
\usepackage[all]{xy} %For commutative diagrams
\usepackage{amssymb} %For double arrows
\hypersetup{urlcolor=blue, citecolor=red}
\usepackage{multicol}
\usepackage[active]{srcltx}
\usepackage{amsthm}
\usepackage{amsmath}
\usepackage{wasysym}
\newcommand{\todo}[1]{\vspace{5 mm}\par \noindent
\marginpar{\textsc{ToDo}}
\framebox{\begin{minipage}[c]{0.89\columnwidth}
\tt #1 \end{minipage}}\vspace{5 mm}\par}

\newtheorem{theorem}{Theorem}[section]

\newtheorem{proposition}{Proposition}

\theoremstyle{definition}
\newtheorem{definition}[theorem]{Definition}
\newtheorem{remark}{Remark}

\newtheorem{example}{Example}

\newcommand{\llbracket}{\lbrack\! \lbrack}
\newcommand{\rrbracket}{\rbrack\! \rbrack}
\newcommand{\br}[2]{\lbrack\!\lbrack#1,#2\rbrack\!\rbrack}
\newcommand{\ab}{\lbrack\!\lbrack \;, \; \rbrack\!\rbrack}
\def\lcf{\lbrack\! \lbrack}
\def\rcf{\rbrack\! \rbrack}
\newcommand{\lp}{\left(}
\newcommand{\rp}{\right)}

\newcommand{\der}{\partial}

\newcommand{\R}{\mathbb{R}}      %Numeros reales
\newcommand{\Flder}{\rightarrow}

\usepackage{amssymb}

\newcommand{\proa}{A^*G \mbox{$\;$}_{\tau^*} \kern-3pt\times_\alpha
G \mbox{$\;$}_\beta \kern-3pt\times_{\tau^*} A^*G}

\def\inn{\mathop{i}\nolimits}
\newcommand{\we}{\wedge}

\newcommand{\al}{\mathfrak{g}}

\newcommand{\de}{\mbox{d}}

\newcommand{\te}{\mathcal{T}}

\title[Second-order constrained variational problems on Lie algebroids]{Second-order constrained variational problems on Lie algebroids: Applications to Optimal Control}

\author[Leonardo Colombo ]{}


 \email{ljcolomb@umich.edu}
\begin{document}
\maketitle

% Enter the first author's name and address:
\centerline{\scshape Leonardo Colombo}
\medskip
{\footnotesize
% please put the address of the first author
 \centerline{Department of Mathematics, University of Michigan}
 %\centerline{University of Michigan}
   \centerline{530 Church Street, 3828 East Hall}
   \centerline{ Ann Arbor, Michigan, 48109, USA}
} % Do not forget to end the {\footnotesize by the sign }

%\medskip

%\centerline{\scshape David Mart\'in de Diego}
%\medskip
%{\footnotesize
% % please put the address of the second  and third author
% \centerline{ Instituto de Ciencias Matem\'aticas (CSIC-UAM-UCM-UC3M)}
%   \centerline{Calle Nicol\'as Cabrera 15, Campus UAM, Cantoblanco}
%   \centerline{Madrid, 28049, Spain}
%}

\bigskip

% The name of the associate editor will be entered by an editorial staff
% "Communicated by the associate editor name" is not needed for special issue.
 %\centerline{(Communicated by the associate editor name)}

%The abstract of your paper
\begin{abstract}
The aim of this work is to study, from an intrinsic and geometric point of view, second-order constrained variational problems on Lie algebroids, that is, optimization problems defined by a cost function which depends on higher-order derivatives of admissible curves on a Lie algebroid. Extending the classical Skinner and Rusk formalism for the mechanics in the context of Lie algebroids, for second-order constrained mechanical systems, we derive the corresponding dynamical equations. We find a symplectic Lie subalgebroid where, under some mild regularity conditions, the second-order constrained variational problem, seen as a presymplectic Hamiltonian system, has a unique solution. We study the relationship of this formalism with the second-order constrained Euler-Poincar\'e and Lagrange-Poincar\'e equations, among others. Our study is applied to the optimal control of mechanical systems.\end{abstract}
\tableofcontents
%The title of your section 1
\section{Introduction}

%This work deals with mechanical control systems, giving emphasis to
%a particular class of mechanical control systems: underactuated
%mechanical systems. Underactuated mechanical systems are
%characterized by the fact that they have more degrees of freedom
%than actuators. The class of underactuated mechanical systems are
%abundant in real life for different reasons; for instance, as a
%result of design choices motivated by the search of less cost
%engineering devices or as a result of a failure regime in fully
%actuated mechanical systems. Underactuated systems include
%spacecrafts, underwater vehicles, mobile robots, helicopters,
%wheeled vehicles and underactuated manipulators.

%To analyze geometrically these control systems we will need an
%unifying concept: 
Lie algebroids have
deserved a lot of interest in recent years. Since a Lie algebroid is
a concept which unifies tangent bundles and Lie algebras, one can
suspect their relation with mechanics. More precisely, a Lie
algebroid over a manifold $Q$ is a vector bundle $\tau_E:E\to Q$
over $Q$ with a Lie algebra structure over the space
$\Gamma(\tau_E)$ of sections of $E$ and an application $\rho:E\to
TQ$ called anchor map satisfying some compatibility conditions (see
\cite{Mack}). Examples of Lie algebroids are the tangent bundle
over a manifold $Q$ where the Lie bracket is the usual Lie bracket
of vector fields and the anchor map is the identity function;
a real finite dimensional Lie algebras as vector bundles over a
point, where the anchor map is the null application; action Lie
algebroids of the type $pr_1:M\times\al\to M$ where $\al$ is a Lie
algebra acting infinitesimally over the manifold $M$ with a Lie
bracket over the space of sections induced by the Lie algebra
structure and whose anchor map is the action of $\al$ over $M$; and, the Lie-Atiyah algebroid $\tau_{TQ/G}:TQ/G\to
\widehat{M}=Q/G$ associated with the $G$-principal bundle $p:Q\to
\widehat{M}$ where the anchor map is induced by the tangent
application of $p$, $Tp:TQ\to T\widehat{M}$
\cite{LMM,Mack,Eduardoalg,We}.

In \cite{We} Alan Weinstein developed a generalized theory of
Lagrangian mechanics on Lie algebroids and he obtained the equations of
motion using the linear Poisson structure on the dual of the Lie
algebroid and the Legendre transformation associated with a regular
Lagrangian $L:E\to\R$. In \cite{We} also he asked about whether
it is possible to develop a formalism similar on Lie algebroids to
Klein's formalism \cite{Klein} in Lagrangian mechanics. This task
was obtained by Eduardo Mart\'inez in \cite{Eduardoalg}( see also \cite{Eduardo1}). The main notion is that of
prolongation of a Lie algebroid over a mapping introduced by Higgins
and Mackenzie in \cite{Mack}. A more general situation, the
prolongation of an anchored bundle $\tau_E:E\to Q$ was also
considered by Popescu in \cite{Popescu2,Popescu}.

The importance of Lie algebroids in mathematics is beyond doubt and
in the last years Lie algebroids has been a lot of applications in
theoretical physics and other related sciences. More concretely in
Classical Mechanics, Classical Field Theory and their applications. One of the main
characteristic concerning  that Lie algebroids are interesting in Classical Mechanics
lie in the fact that there are many different situations that can be
understand in a general framework using the theory of Lie algebroids
as systems with symmetries, systems over semidirect products,
Hamiltonian and Lagrangian systems, systems with constraints
(nonholonomic and vakonomic) and Classical Fields theory \cite{portugeses,Pepin2007,PepinEduardo,Bruce3,PepinMiguel,CoLeMaMa,CoMa,januzkatrina,MdLMdDJCM,JCsolo,Tom}. 

In \cite{LMM} M. de Le\'on, J.C Marrero and E. Mart\'inez have
developed a Hamiltonian description for the mechanics on Lie
algebroids and they have shown that the dynamics is obtained solving
an equation in the same way than in Classical Mechanics (see also
\cite{Eduardo1} and \cite{We}). Moreover, they shown that the
Legendre transformation $leg_{L}:E\to E^{*}$ associated to the
Lagrangian $L:E\to\R$ induces a Lie algebroid morphism and when the
Lagrangian is regular both formalisms are equivalent.
% Also they have
%extended the Tulczyjew's contruction \cite{Tulczy1,Tulczy2}
%to the framework of Lie algebroids and they have been introduced the
%notion of Lagrangian Lie subalgebroid of a symplectic Lie algebroid that we use in this work.
%Then they have shown that Euler-Lagrange equations and Hamilton
%equations over a Lie algebroids are just the local equations defined
%by certain Lagrangian submanifolds of a symplectic Lie algebroid
%associated to $E$. As a consequence they have deduced the
%Lagrange-Poincar\'e and Hamilton-Poincar\'e equations associated to
%a $G$-invariant Lagrangian and Hamiltonian, respectively.

Marrero and collaborators also have analyzed the case of
non-holonomic mechanics on Lie algebroids \cite{CoLeMaMa}. In
other direction, in \cite{IMMS} D. Iglesias, J.C. Marrero, D.
Mart\'in de Diego and D. Sosa have studied singular Lagrangian
systems and vakonomic mechanics from the point of view of Lie
algebroids obtained through the application of a constrained
variational principle. They have developed a constraint algorithm
for presymplectic Lie algebroids generalizing the well know
constraint algorithm of Gotay, Nester and Hinds \cite{GoNe, GoNesHinds78}
and they also have established the Skinner and Rusk formalism on Lie
algebroids. Some of the results given are as an extension of this framework for constrained second-order systems.

Our framework is based in the Skinner-Rusk formalism which
combines simultaneously some features of the Lagrangian and Hamiltonian classical
formalisms. The idea of this formulation was to obtain a common framework for both regular and
singular dynamics, obtaining simultaneously the Hamiltonian and Lagrangian formulations of
the dynamics. Over the years, however, Skinner and Rusk's framework was extended in many
directions: It was
originally developed for first-order autonomous mechanical systems \cite{Skinner-Rusk}, and later generalized to
non-autonomous dynamical systems \cite{art:Barbero_Echeverria_Martin_Munoz_Roman08, JorgeSKRK, Perecampos}, control systems \cite{Maria} and, more recently to classical field theories
\cite{Cedric, Narciso}.

Briefly,  in this formulation, one  starts with a differentiable manifold $Q$ as the
configuration space, and the Whitney sum $TQ\oplus T^*Q$ as the
evolution space (with canonical projections $\pi_1: TQ \oplus
T^*Q\longrightarrow TQ$ and $\pi_2: TQ \oplus T^*Q\longrightarrow
T^*Q$).  Define on $TQ\oplus T^*Q$ the presymplectic 2-form
$\Omega=\pi_2^*\omega_Q$, where $\omega_Q$ is the canonical
symplectic form on $T^*Q$, and observe that the rank of this
presymplectic form is everywhere equal to $2n$. If the dynamical
system under consideration admits a Lagrangian description, with
Lagrangian $L\in C^{\infty}(TQ)$, then one can obtain a
(presymplectic)-Hamiltonian representation on $TQ\oplus T^*Q$
given by the presymplectic 2-form $\Omega$ and the Hamiltonian
function $ H=\langle \pi_1,\pi_2\rangle -\pi_1^* L \; , $ where
$\langle \cdot , \cdot \rangle$ denotes the natural pairing
between vectors and covectors on $Q$.  In this Hamiltonian system
the dynamics is given by vector fields $X$, which are solutions to
the Hamiltonian equation $\inn_{X}\Omega=dH$. If $L$ is regular, then
there exists a unique vector field $X$ solution to the previous
equation, which is tangent to the graph of the Legendre map. In the singular case, it is necessary
to develop a constraint algorithm in order to find a submanifold
(if it exists) where there exists a well-defined dynamical vector field. 

Recently, higher-order variational problems have been studied for
their important applications in aeronautics, robotics,
computer-aided design, air traffic control, trajectory planning, and in general, problems of interpolation
and approximation of curves on Riemannian manifolds \cite{BlochCrouch2, splines1, HuBl, splines2, MR2580471, MaBl, DavidM, splines3}.
There are variational principles which involves higher-order
derivatives by Gay Balmaz et.al., \cite{GHMRV10,GHMRV12,GHR12}, (see also \cite{LR1}) since from it one can obtain the equations of motion for Lagrangians where the configuration space is a
higher-order tangent bundle. More recently, there have been an interest in study of the geometrical structures associated with higher order variational problems with the aim of a deepest understanding of those geometric sructures \cite{CoMdD,CoPr,PrietoRoman-Roy1,Eduardoho,pichon1,pichon2,pichon3} as well the relation of higher-order mechanics and graded bundles, \cite{Bruce1,Bruce2,Bruce3}.

In this work, we study a geometric framework, based on the Skinner and Rusk formalism, for constrained  second-order variational problems determined by a Lagrangian function, playing the role of cost function in an optimal control problem, which depends on derivatives of admissible curves on a Lie algebroid.  The strategy is to apply the geometric procedure described above in combination with an extension of the constraint algorithm developed by Gotay, Nester and Hinds \cite{GoNe, GoNesHinds78} in the setting of Lie algebroids \cite{IMMS}. Our work permits to obtain constrained second-order Euler-Lagrange equations, Euler-Poincar\'e, Lagrange-Poincar\'e equations in an unified framework and understand the geometric structures subjacent in second-order variational problems. We show how this study can be applied to the problem of finding  necessary conditions for optimality in optimal control problems
of mechanical system with symmetries, where trajectories are "parameterized" by the admissible controls and the necessary conditions for extremals in the optimal control problem are expressed using a "pseudo-Hamiltonian formulation" based on the Pontryagin maximun principle.

The paper is organized as follows. In Section \ref{LALG} we introduce some known notions concerning Lie algebroids that are necessary for further developments in this work. In section \ref{seccion3} we will use the
notion of Lie algebroid and prolongation of a Lie algebroid
described in \ref{LALG} to derive the Euler-Lagrange equations and
Hamilton equations on Lie algebroids. Next, after introduce the constraint algorithm for presymplectic Lie algebroids and study vakonomic mechanics on Lie algebroids, we study the geometric formalism for second-order constrained variational problems using and adaptation of the classical Skinner-Rusk formalism for the second-order constrained systems on Lie algebroids.  In section \ref{section4} we study optimal control problems of mechanical systems defined on Lie algebroids. Optimality conditions for the optimal control of the Elroy's Beanie are derived.
Several examples show how to apply the techniques along all the work.

\section{Lie algebroids and admissible elements}\label{LALG}

In this section, we introduce some known notions and develop new concepts concerning Lie
algebroids that are necessary for further developments in this work.  We
illustrate the theory with several examples. We refer the reader to
\cite{AC-AW:99,Mack} for more details about Lie algebroids
and their role in differential geometry.

\subsection{Lie algebroids, Lie subalgebroids and Cartan calculus on Lie algebroids}\label{LA}

\begin{definition}\label{alas}
Let $E$ be a vector bundle of rank $n$ over a manifold $M$ of
dimension $m$. A \emph{Lie algebroid structure} on the
vector bundle $\tau_{E}: E \to M$ is a $\R$-linear bracket $\lcf
\cdot, \cdot \rcf: \Gamma(\tau_{E}) \times \Gamma(\tau_{E}) \to
\Gamma(\tau_{E})$ on the space $\Gamma(\tau_{E})$, the
$C^{\infty}(M)$-module of sections of $E$, and a vector bundle
morphism $\rho: E \to TM$, the \emph{anchor map}, such that:
\begin{enumerate}
\item The bracket $\lcf \cdot, \cdot \rcf$ satisfies the Jacobi
identity, that is, $$\lcf X, \lcf Y, Z \rcf \rcf+\lcf Z, \lcf X, Y \rcf \rcf+\lcf Y, \lcf Z, X \rcf \rcf=0\quad \forall X,Y,Z\in\Gamma(\tau_{E}).$$
\item
If we also denote by $\rho: \Gamma(\tau_{E}) \to {\frak X}(M)$
the homomorphism of $C^{\infty}(M)$-modules induced by the
anchor map then
\begin{equation}\label{eqq}
\lcf X, fY \rcf = f \lcf X, Y \rcf + \rho(X)(f) Y, \;
\; \mbox{ for } X, Y \in \Gamma(\tau_{E}) \mbox{ and } f \in
C^{\infty}(M).
\end{equation}
\end{enumerate}
\end{definition}

We will said that the triple $(E, \lcf \cdot, \cdot \rcf, \rho)$ is a
\emph{Lie algebroid} over $M$.  In this context, sections of $\tau_E,$ play the role of vector
fields on $M$, and the sections of the dual bundle
$\tau_{E^{*}}:E^{*}\Flder M$ of 1-forms on $M$.

%\begin{remark}
%If $(A,\ab,\rho)$ is a Lie algebroid over $M,$ then the anchor map $\rho_\Gamma(\tau_{A})\Flder\mathfrak{X}(M)$ is a homomorphism between the Lie algebras $(\Gamma(\tau_{A}),\ab)$ and $(\mathfrak{X}(M),[\cdot,\cdot]).$
%\end{remark}

We may consider two type of distinguished functions: given $f\in
C^{\infty}(M)$ one may define a function $\tilde{f}$ on $E$ by
$\tilde{f}=f\circ\tau_E$, the \textit{basic functions}. And, given a
section $\theta$ of the dual bundle $\tau_{E^{*}}:E^{*}\Flder M$,
may be regarded as a \textit{lineal function} $\hat{\theta}$ on $E$
as $\hat{\theta}(e)=\langle\theta(\tau_{E}(e)),e\rangle$ for all
$e\in E$. In this sense, $\Gamma(\tau_E)$ is locally generated by
the differential of basic and linear functions.

If $X,Y,Z\in\Gamma(\tau_E)$ and $f\in C^{\infty}(M),$ then using the
Jacobi identity we obtain that \begin{equation}\label{eqq1}
{\br{\br{X}{Y}}{fZ}}=f{\br{X}{\br{Y}{Z}}}+[\rho(X),\rho(Y)](f)Z.
\end{equation}
Also, from \eqref{eqq} it follows that \begin{equation}\label{eqq2}
{\br{\br{X}{Y}}{fZ}}=f{\br{\br{X}{Y}}{Z}}+\rho{\br{X}{Y}}(f)Z.
\end{equation}
 Then, using \eqref{eqq1} and \eqref{eqq2} and the fact that
${\br{\cdot}{\cdot}}$ is a Lie bracket we conclude that
$$\rho{\br{X}{Y}}=[\rho(X),\rho(Y)],$$ that is, $\rho:\Gamma(\tau_E)\Flder\mathfrak{X}(M)$ is a homomorphism
between the Lie algebras $(\Gamma(\tau_E),{\br{\cdot}{\cdot}})$ and
$(\mathfrak{X},[\cdot,\cdot]).$

The algebra $\bigoplus_{k}\Gamma(\Lambda^{k} E^{*})$ of
multisections of $\tau_{E^{*}}$ plays the role of the algebra of the
differential forms and it is possible to define a
\textit{differential operator} as follow:

\begin{definition}
If $(E,{\br{\cdot}{\cdot}},\rho)$ is a Lie algebroid over $M$, one
can be define the \textit{differential of E},
$d^{E}:\Gamma(\bigwedge^{k}\tau_{E^{*}})\Flder\Gamma(\bigwedge^{k+1}\tau_{E^{*}}),$
as follows; \begin{eqnarray*}
d^{E}\mu(X_0,\ldots,X_k)&=&\sum_{i=0}^{k}(-1)^{i}\rho(X_i)(\mu(X_0,\ldots,\widehat{X}_i,\ldots,X_k))\\
&+&\sum_{i<j}(-1)^{i+j}\mu({\br{X}{Y}},X_0,\ldots,\widehat{X}_i,\ldots,\widehat{X}_j,\ldots,
X_k),
\end{eqnarray*} for $\mu\in\Gamma(\bigwedge^{k}\tau_{E^{*}})$ and
$X_0,\ldots,X_k\in\Gamma(\tau_E).$
\end{definition}

From the properties of Lie algebroids it follows that $d^{E}$ is a
cohomology operator, that is, $(d^{E})^2=0$ and
$d^{E}(\alpha\wedge\beta)=d^{E}\alpha\wedge\beta+(-1)^{k}\alpha\wedge
d^{E}\beta,$ for $\alpha\in\Gamma(\Lambda^{k}E^{*})$ and
$\beta\in\Gamma(\Lambda^{r}E^{*})$ (see \cite{Mack} for more details).

Conversely it is possible to recover the Lie algebroid structure of
$E$ from the existence of an exterior differential on
$\Gamma(\Lambda^{\bullet}\tau_{E^{*}})$. If $f:M\Flder\R$ is a real
smooth function, one can define the anchor map and the Lie bracket
as follows:

\begin{enumerate}
\item $d^{E}f(X)=\rho(X)f,$ for $X\in\Gamma(\tau_E),$
\item $i_{{\br{X}{Y}}}\theta=\rho(X)\theta(Y)-\rho(Y)\theta(X)-d^{E}\theta(X,Y)$ for all $X,Y\in\Gamma(\tau_{E})$ and $\theta\in\Gamma(\tau_{E^{*}}).$
\end{enumerate}

Moreover, from the last equality, the section
$\theta\in\Gamma(\tau_{E^{*}})$ is a \textit{1-cocycle} if and only
if $d^{E}\theta=0,$ or, equivalently,
$$\theta{\br{X}{Y}}=\rho(X)(\theta(Y))-\rho(Y)(\theta(X)),$$ for all
$X,Y\in\Gamma(\tau_{E})$.

We may also define the \textit{Lie derivative} with respect to a
section $X\in\Gamma(\tau_E)$ as the operator
$\mathcal{L}_{X}^{E}:\Gamma(\bigwedge^{k}\tau_{E^{*}})\Flder\Gamma(\bigwedge^{k}\tau_{E^{*}})$
given by $$\mathcal{L}^{E}_X\theta=i_{X}\circ d^{E}\theta+d^{E}\circ
i_X\theta,$$ for $\theta\in\Gamma(\Lambda^{k}\tau_{E^{*}}).$ One
also has the usual identities \begin{enumerate}
\item $d^{E}\circ\mathcal{L}^{E}_X=\mathcal{L}^{E}_X\circ d^{E},$
\item $\mathcal{L}^{E}_Xi_{Y}-i_{X}\mathcal{L}^{E}_Y=i_{{\br{X}{Y}}},$
\item $\mathcal{L}^{E}_X\mathcal{L}^{E}_Y-\mathcal{L}^{E}_Y\mathcal{L}^{E}_X=\mathcal{L}^{E}_{{\br{X}{Y}}}.$
\end{enumerate}

We take local coordinates $(x^i)$ on $M$ with $i=1,\ldots,m$ and a
local basis $\{e_A\}$ of sections of the vector bundle
$\tau_{E}:E\to M$ with $A=1,\ldots,n,$ then we have the
corresponding local coordinates on an open subset $\tau_{E}^{-1}(U)$
of $E,$ $(x^i,y^A)$ ($U$ is an open subset of $Q$), where $y^A(e)$
is the $A$-th coordinate of $e\in E$ in the given basis i.e., every
$e\in E$ is expressed as
$e=y^{1}e_{1}(\tau_{E}(e))+\ldots+y^{n}e_{n}(\tau_{E}(e))$.

Such coordinates determine the local functions $\rho_A^i$,
$\mathcal{C}_{AB}^{C}$ on $M$ which contain the local information of
the Lie algebroid structure, and accordingly they are called {\it
structure functions of the Lie algebroid.} These are given by
\begin{equation}\label{estruct0}
\rho(e_A)=\rho_A^i\frac{\partial }{\partial
x^i}\;\;\;\mbox{ and }\;\;\; \lcf
e_A,e_B\rcf=\mathcal{C}_{AB}^C e_C.
\end{equation}

These functions should satisfy the relations
\begin{equation}\label{estruc1}
\rho_A^j\frac{\partial \rho_B^i}{\partial x^j}
-\rho_B^j\frac{\partial \rho_A^i}{\partial x^j}=
\rho_C^i\mathcal{C}_{AB}^C \end{equation}
and \begin{equation}\label{estruc2}
\sum_{cyclic(A,B,C)}\left[\rho_{A}^i\frac{\partial
\mathcal{C}_{BC}^D}{\partial x^i} + \mathcal{C}_{AF}^D
\mathcal{C}_{BC}^F\right]=0,
\end{equation} which are usually called {\it the structure equations.}

If $f\in C^\infty(M)$, 
\begin{equation}\label{diff0}
d^E f=\frac{\partial f}{\partial x^i}\rho_A^i e^A,
\end{equation}
where $\{e^A\}$ is the dual basis of $\{e_A\}$. If $\theta\in \Gamma(\tau_{E^*})$ and $\theta=\theta_C e^C$ it
follows that
\begin{equation}\label{diff1}
d^E \theta=\left(\frac{\partial \theta_C}{\partial
x^i}\rho^i_B-\frac{1}{2}\theta_A
\mathcal{C}^A_{BC}\right)e^{B}\wedge e^C.
\end{equation}
In particular, \[d^E x^i=\rho_A^ie^A,\;\;\; d^E
e^A=-\frac{1}{2}C_{BC}^{A} e^B\wedge e^C.\]

\subsubsection{Examples of Lie algebroids}

\begin{example}
 Given a \textit{finite dimensional real Lie algebra} $\mathfrak{g}$ and 
$M=\{m\}$ be a unique point, we consider the vector bundle
$\tau_{\mathfrak{g}}:\mathfrak{g}\Flder M.$ The sections of this
bundle can be identified with the elements of $\mathfrak{g}$ and
therefore we can consider as the Lie bracket the structure of the
Lie algebra induced by $\al$, and denoted by $[\cdot,\cdot]_{\mathfrak{g}}$. Since
$TM=\{0\}$ one may consider the anchor map $\rho\equiv 0$. The triple
$(\mathfrak{g},[\cdot,\cdot]_{\mathfrak{g}},0)$ is a Lie algebroid
over a point.
\end{example}
\begin{example}
Consider a \textit{tangent bundle} of a manifold $M.$ The sections
of the bundle $\tau_{TM}:TM\Flder M$ are the set of vector
fields on $M$. The anchor map $\rho:TM\Flder TM$ is the identity
function and the Lie bracket defined on $\Gamma(\tau_{TM})$ is
induced by the Lie bracket of vector fields on $M.$
\end{example}
\begin{example}
Let $\phi:M\times G\Flder M$ be an action of $G$ on the manifold $M$ where
$G$  is a Lie group. The induced anti-homomorphism between the
Lie algebras $\mathfrak{g}$ and $\mathfrak{X}(M)$ by the action is determined by 
$\Phi:\mathfrak{g}\Flder\mathfrak{X}(M)$,
$\xi\mapsto\xi_M$, where $\xi_{M}$ is the infinitesimal generator of the
action for $\xi\in\mathfrak{g}.$

The vector bundle $\tau_{M\times\mathfrak{g}}:M\times\mathfrak{g}\Flder
M$ is a Lie algebroid over $M$. The anchor map $\rho:M\times\mathfrak{g}\Flder TM$, is defined by $\rho(m,\xi)=-\xi_{M}(m)$ and
the Lie bracket of sections is given by the Lie algebra structure on
$\Gamma(\tau_{M\times\mathfrak{g}})$ as
$${\br{\hat{\xi}}{\hat{\eta}}}_{M\times\al}(m)=(m,[\xi,\eta])=\widehat{[\xi,\eta]}$$
for $m\in M$, where $\hat{\xi}(m)=(m,\xi)$, $\hat{\eta}(m)=(m,\eta)$ for
$\xi,\eta\in\al$. The triple $(M\times\al, \rho, {\br{\cdot}{\cdot}}_{M\times\al})$ is called \textit{Action Lie algebroid}.
\end{example}
\begin{example}\label{Atiyah case}
Let $G$ be a Lie group and we assume that $G$ acts free and properly on $M$. We
denote by $\pi:M\Flder \widehat{M}=M/G$ the associated principal
bundle. The tangent lift of the action gives a free and proper
action of $G$ on $TM$ and $\widehat{TM}=TM/G$ is a quotient
manifold. The quotient vector bundle
$\tau_{\widehat{TM}}:\widehat{TM}\Flder \widehat{M}$ where
$\tau_{\widehat{TM}}([v_m])=\pi(m)$ is a Lie algebroid over
$\widehat{M}.$ The fiber of $\widehat{TM}$ over a point $\pi(m)\in \widehat{M}$ is isomorphic to $T_{m}M.$

The Lie bracket is defined on the space
$\Gamma(\tau_{\widehat{TM}})$ which is isomorphic to the Lie
subalgebra of $G$-invariant vector fields, that is,
$$\Gamma(\tau_{\widehat{TM}})=\{X\in\mathfrak{X}(M)\mid
X \hbox{ is $G$-invariant}\}.$$ Thus, the Lie bracket on
$\widehat{TM}$ is the bracket of $G$-invariant vector fields.
The anchor map $\rho:\widehat{TM}\Flder T\widehat{M}$ is given by
$\rho([v_m])=T_{m}\pi(v_m).$ Moreover, $\rho$ is a Lie algebra
homomorpishm satisfying the compatibility condition since the
$G$-invariant vector fields are $\pi$-projectable. This Lie
algebroid is called \textit{Lie-Atiyah algebroid} associated with
the principal bundle $\pi:M\Flder\widehat{M}.$

Let $\mathcal{A}:TM\to\al$ be a principal connection in the principal bundle $\pi:M\to\widehat{M}$ and $B:TM\oplus TM\to\al$ be the curvature of $\mathcal{A}.$ The connection determines an isomorphism $\alpha_{\mathcal{A}}$ between the vector bundles
$\widehat{TM}\to\widehat{M}$ and $T\widehat{M}\oplus\widetilde{\al}\to \widehat{M}$, where $\widetilde{\al}=(M\times\al)/G$ is the adjoint bundle associated with the principal bundle $\pi:M\to\widehat{M}$ (see \cite{CeMaRa} for example).

We choose a local trivialization of the principal bundle
$\pi:M\to\widehat{M}$ to be $U\times G,$ where $U$ is an open subset
of $\widehat{M}.$ Suppose that $e$ is the identity of $G$, $(x^{i})$
are local coordinates on $U$ and $\{\xi_{A}\}$ is a basis of $\al.$

Denote by $\{\overleftarrow{\xi_{A}}\}$ the corresponding
left-invariant vector field on $G$, that is,
$$\overleftarrow{\xi_{A}}(g)=(T_{e}L_{g})(\xi_{A})$$ for $g\in G$ where
$L_{g}:G\to G$ is the left-translation on $G$ by $g.$ If
$$\mathcal{A}\left(\frac{\partial}{\partial
x^{i}}\Big{|}_{(x,e)}\right)=\mathcal{A}_{i}^{A}(x)\xi_{A},\quad\mathcal{B}\left(\frac{\partial}{\partial
x^{i}}\Big{|}_{(x,e)},\frac{\partial}{\partial
x^{j}}\Big{|}_{(x,e)}\right)=\mathcal{B}_{ij}^{A}(x)\xi_{A},$$ for
$i,j\in\{1,\ldots,m\}$ and $x\in U,$ then the horizontal lift of the
vector field $\frac{\partial}{\partial x^{i}}$ is the vector field
on $\pi^{-1}(U)\simeq U\times G$ given by
$$\left(\frac{\partial}{\partial
x^{i}}\right)^{h}=\frac{\partial}{\partial
x^{i}}-\mathcal{A}_{i}^{A}\overleftarrow{\xi_{A}}.$$

Therefore, the vector fields on $U\times G$
$$e_{i}=\frac{\partial}{\partial
x^{i}}-\mathcal{A}_{i}^{A}\overleftarrow{\xi_{A}}\hbox{ and }
e_{B}=\overleftarrow{\xi_{B}}$$ are $G$-invariant under the action
of $G$ over $M$ and define a local basis
$\{\hat{e}_{i},\hat{e}_{B}\}$ on
$\Gamma(\widehat{TM})=\Gamma(\tau_{T\widehat{M}\oplus\tilde{\al}}).$
The corresponding local structure functions of
$\tau_{\widehat{TM}}:\widehat{TM}\to\widehat{M}$ are
\begin{eqnarray*}
\mathcal{C}_{ij}^{k}&=&\mathcal{C}_{iA}^{j}=-\mathcal{C}_{Ai}^{j}=\mathcal{C}_{AB}^{i}=0,\quad\mathcal{C}_{ij}^{A}=-\mathcal{B}_{ij}^{A},\quad\mathcal{C}_{iA}^{C}=-\mathcal{C}_{Ai}^{C}=c_{AB}^{C}\mathcal{A}_{i}^{B},\\
\mathcal{C}_{AB}^{C}&=&c_{AB}^{C},\quad\rho_{i}^{j}=\delta_{ij},\quad\rho_{i}^{A}=\rho_{A}^{i}=\rho_{A}^{B}=0,
\end{eqnarray*} being $\{c_{AB}^{C}\}$ the constant structures of $\al$ with respect to the basis $\{\xi_{A}\}$ (see \cite{LMM} for more details). That is,
$$\lcf\hat{e}_{i},\hat{x}_{j}\rcf_{\widehat{TM}}=-\mathcal{B}_{ij}^{C}\hat{e}_{C},\quad\lcf\hat{e}_{i},\hat{e}_{A}\rcf_{\widehat{TM}}=c_{AB}^{C}\mathcal{A}_{i}^{B}\hat{e}_{C},\quad \lcf\hat{e}_{A},\hat{e}_{B}\rcf_{\widehat{TM}}=c_{AB}^{C}\hat{e}_{C},$$
$$\rho_{\widehat{TM}}(\hat{e}_{i})=\frac{\partial}{\partial x^{i}},\quad\rho_{\widehat{TM}}(\hat{e}_{A})=0.$$ The basis $\{\hat{e}_{i},\hat{e}_{B}\}$ induce local coordinates $(x^{i},y^{i},\bar{y}^{B})$ on $\widehat{TM}=TM/G.$

\end{example}

Next, we introduce the notion of Lie subalgebroid associated with a Lie algebroid.

\begin{definition}
Let $(E,\lcf\cdot,\cdot \rcf_E,\rho_E)$ be a Lie algebroid over $M$
and $N$ is a submanifold of $M.$ A Lie subalgebroid of $E$ over $N$
is a vector subbundle $B$ of $E$ over $N$ $$
\xymatrix{
B\ar[d]_{\tau_{B}} \ar@{^{(}->}[rr]^{j} & \ &E \ar[d]^{\tau_{E}} \\
N\ar@{^{(}->}[rr]^{i} & \ & M
}$$ such that $\rho_{B}=\rho_{E}\mid_{B}:B\Flder TN$ is well define and;
given $X,Y\in\Gamma(B)$ and
$\widetilde{X},\widetilde{Y}\in\Gamma(E)$ arbitrary extensions of
$X,Y$ respectively, we have that
$(\lcf\widetilde{X},\widetilde{Y}\rcf_E)\mid_{N}\in\Gamma(B).$
\end{definition}

\subsubsection{Examples of Lie subalgebroids}

\begin{example}
Let $E$ be a Lie algebroid over $M.$ Given a submanifold $N$ of $M,$
if $B=E\mid_{N}\cap(\rho\mid_{N})^{-1}(TN)$ exists as a vector
bundle, it will be a Lie subalgebroid of $E$ over $N,$ and will be
called \textit{Lie algebroid restriction of E to N} (see
\cite{Mack}).
\end{example}

\begin{example}
Let $N$ be a submanifold of $M.$ Then, $TN$ is a Lie subalgebroid of $TM.$

Now, let $\mathcal{F}$ be a completely integrable distribution on a
manifold $M.$ $\mathcal{F}$ equipped with the bracket of vector
fields is a Lie algebroid over $M$ since
$\tau_{E}\mid_{\mathcal{F}}:\mathcal{F}\Flder M$ is a vector bundle
and if $\mathcal{F}$ is a foliation,
$(\Gamma(\mathcal{F}),[\cdot,\cdot])$ is a Lie algebra. The anchor
map is the inclusion $i_{\mathcal{F}}:\mathcal{F}\Flder TM$
($i_{\mathcal{F}}$ is a Lie algebroid monomorphism).

Moreover, if $N$ is a submanifold of $M$ and $\mathcal{F}_{N}$ is a
foliation on $N,$ then $\mathcal{F}_{N}$ is a Lie subalgebroid of
the Lie algebroid $\tau_{TM}:TM\Flder M$.

\end{example}

\begin{example}
Let $\mathfrak{g}$ be a Lie algebra and $\mathfrak{h}$ be a Lie
subalgebra. If we consider the Lie algebroid induced by
$\mathfrak{g}$ and $\mathfrak{h}$ over a point, then $\mathfrak{h}$
is a Lie subalgebroid of $\mathfrak{g}.$
\end{example}

\begin{example}
Let $M\times\mathfrak{g}\Flder M$ be an action Lie algebroid and
let $N$ be a submanifold of $M.$ Let $\mathfrak{h}$ be a Lie
subalgebra of $\mathfrak{g}$ such that the infinitesimal generators
of the elements of $\mathfrak{h}$ are tangent to $N;$ that is, the
application \begin{eqnarray*}&&\mathfrak{h}\Flder\mathfrak{X}(N)\\
&&\xi\mapsto \xi_{N} \end{eqnarray*} is well defined. Thus, the
action Lie algebroid $N\times\mathfrak{h}\Flder N$ is a Lie
subalgebroid of $M\times\mathfrak{g}\Flder M.$
\end{example}

\begin{example}
Suppose that the Lie group $G$ acts free and properly on $M$. Let
$\pi:M\Flder M/G=\widehat{M}$ be the associated $G-$principal
bundle. Let $N$ be a $G-$invariant submanifold of $M$ and
$\mathcal{F}_{N}$ be a $G-$invariant foliation over $N.$ We may
consider the vector bundle
$\widehat{\mathcal{F}_{N}}=\mathcal{F}_{N}/G\Flder N/G=\widehat{N}$
and endow it with a Lie algebroid structure. The sections of
$\widehat{\mathcal{F}_{N}}$ are $$\Gamma(\widehat{\mathcal{F}}_{N})=\{X\in\mathfrak{X}(N)\mid X\hbox{ is $G$-invariant and }X(q)\in\mathcal{F}_{N}(q), \forall q\in N\}.$$
The standard bracket of vector fields on $N$ induces a Lie algebra
structure on $\Gamma(\widehat{\mathcal{F}}_{N}).$ The anchor map is
the canonical inclusion of $\widehat{\mathcal{F}}_{N}$ on
$T\widehat{N}$ and $\widehat{\mathcal{F}}_{N}$ is a Lie subalgebroid
of $\widehat{TM}\Flder\widehat{M}.$
\end{example}

\subsection{$E$-tangent bundle to a Lie algebroid $E$}\label{prolongationliealgebroidsubsection}

We consider the prolongation over the canonical projection of the
Lie algebroid E over $M$, that is, 
$$\mathcal{T}^{\tau_E}E=\bigcup_{e\in E}\left(E_{\rho}\times_{T\tau_E} T_{e}E\right)=\bigcup_{e\in E}\{(e',v_e)\in E\times
T_{e}E\mid\rho(e')=(T_{e}\tau_E)(v_e)\},$$ where
$T\tau_E:TE\rightarrow TM$ is the tangent map to $\tau_E$.

In fact, $\mathcal{T}^{\tau_E}E$ is a Lie algebroid of rank $2n$
over $E$  where $\tau_{E}^{(1)}:\mathcal{T}^{\tau_E}E\Flder E$ is
the vector bundle projection,
$\tau_{E}^{(1)}(b,v_{e})=\tau_{TE}(v_{e})=e,$ and the anchor map is
$\rho_1:\mathcal{T}^{\tau_E}E\rightarrow TE$ is given by the projection
over the second factor. The bracket of sections of this new Lie
algebroid will be denoted by $\lcf\cdot,\cdot \rcf_{\tau_E^{(1)}}$
(See \cite{Eduardoalg} for more details).

If we denote by $(e,e',v_e)$ an element
$(e',v_e)\in\mathcal{T}^{\tau_{E}}E$ where $e\in E$ and where $v$ is
tangent; we rewrite the definition for the prolongation of the Lie
algebroid as the subset of $E\times E\times TE$ by $$\mathcal{T}^{\tau_E}E= \{(e,e',v_e)\in E\times E\times TE\mid \rho(e')=(T\tau_E)(v_e),
v_e\in T_{e}E
 \hbox{ and }\tau_E(e)=\tau_E(e')\}.$$ Thus, if $(e,e',v_e)\in \mathcal{T}^{\tau_E}E;$ then $\rho_1(e,e',v_{e})=(e,v_e)\in T_{e}E,$ and $\tau_E^{(1)}(e,e',v_e)=e\in E$.

Next, we introduce two canonical operations that we have on a Lie
algebroid $E$. The first one is obtained using the Lie algebroid
structure of $E$ and the second one is a consequence of $E$ being a
vector bundle. On one hand, if $f\in C^{\infty}(M)$ we will denote by $f^{c}$ the
\textit{complete lift} to $E$ of $f$ defined by
$f^{c}(e)=\rho(e)(f)$ for all $e\in E$. Let $X$ be a section of
$E$ then there exists a unique vector field $X^{c}$ on $E$, the
\textit{complete lift} of $X$, satisfying the two following
conditions:
\begin{enumerate}
\item $X^c$ is $\tau_E$-projectable on $\rho(X)$ and \item
$X^c(\hat{\alpha})=\widehat{\mathcal{L}^E_X\alpha}$,
\end{enumerate}
for every $\alpha\in \Gamma(\tau_{E^*})$ (see \cite{GU}). Here, if
$\beta\in\Gamma(\tau_{E^*})$ then $\hat\beta$ is the linear function
on $E$ defined by
\[
\hat\beta(e)=\langle\beta(\tau_E(e)),e\rangle,\;\;\; \mbox{ for all  $e\in E$.}
\]
 We may introduce {\it the
complete
 lift } $X^{\bf c}$ of a section $X\in \Gamma(\tau_E)$ as the sections of $\tau_{E}^{(1)}:\mathcal{T}^{\tau_{E}} E\to E$
 given by
 \begin{equation}\label{Defvc}
 X^{\bf c}(e)=(X(\tau_{E}(e)),X^{c}(e))
 \end{equation}
 for all $e\in E$ (see \cite{Eduardoalg}).

Given a section $X\in\Gamma(\tau_E)$ we define
the \textit{vertical lift} as the vector field
$X^{v}\in\mathfrak{X}(E)
$ given by
$$X^{v}(e)=X(\tau_{E}(e))_{e}^{v}, \hbox{ for }e\in E,$$ where
$_{e}^{v}:E_{q}\Flder T_{e}E_{q}$ for $q=\tau_{E}(e)$ is the
canonical isomorphism between the vector spaces $E_{q}$ and
$T_{e}E_{q}$.

Finally we may introduce the \textit{vertical lift} $X^{\bf v}$ of a
section $X\in\Gamma(\tau_E)$ as a section of $\tau_{E}^{(1)}$ given
by
$$X^{\bf v}(e)=(0,X^{v}(e))\hbox{ for }e\in E.$$
With these definitions we have the properties (see \cite{GU} and
\cite{Eduardoalg})
\begin{equation}\label{cvstru}
[X^c,Y^c]=\lcf X,Y\rcf^c,\;\;\; [X^c,Y^v]=\lcf X,Y\rcf^v, \;\;\;
[X^v,Y^v]=0\end{equation} for all $X,Y\in\Gamma(\tau_{E})$.

If $(x^{i})$ are local coordinates on an open subset $U$ of $M$ and
$\{e_{A}\}$ is a basis of sections of $\tau_{E}$ then we have
induced coordinates $(x^{i},y^{A})$ on $E$.
%and thus $\{e_{A}^{\bf
%v},e_{A}^{\bf c}\}$ is a local basis of sections of
%$\tau_{E}^{(1)}$. For $X\in\Gamma(\tau_{E})$ locally given by
%$X=X^{A}e_{A}$ we have
%$$X^{\bf v}=X^{A}e_{A}^{\bf v}\hbox{ and } X^{\bf
%c}=\rho_{B}^{i}\frac{\partial X^{A}}{\partial x^{i}}y^{B}e_{A}^{\bf
%v}+X^{A}e_{A}^{\bf c},$$ where $\rho_{A}^{i}$ are the components of
%the anchor map with respect to the basis $\{e_{A}\}.$
From the basis $\{e_{A}\}$ we may define a local basis
$\{e_{A}^{(1)},e_{A}^{(2)}\}$ of sections of $\tau_{E}^{(1)}$  given by

%where $e_{A}^{(1)}=e_{A}^{\bf
%c}+\mathcal{C}_{AB}^{C}y^{B}e_{C}^{\bf v}$ and
%$e_{A}^{(2)}=e_{A}^{\bf v}$

$$e_{A}^{(1)}(e)=\left(e,e_{A}(\tau_{A}(e)),\rho_{A}^{i}\frac{\partial}{\partial x^{i}}\Big{|}_{e}\right),\quad e_{A}^{(2)}(e)=\left(e,0,\frac{\partial}{\partial y^{A}}\Big{|}_{e}\right),$$
for $e\in(\tau_{E})^{-1}(U)$ with $U$ an open subset of $M$ (see
\cite{LMM} for more details).

From this basis we have that the structure of Lie algebroid is
determined by

\begin{align*}
\rho_{1}(e_{A}^{(1)}(e))&=\left(e,\rho_{A}^{i}\frac{\partial}{\partial x^{i}}\Big{|}_{e}\right),\qquad\rho_{1}(e_{A}^{(2)}(e))=\left(e,\frac{\partial}{\partial y^{A}}\Big{|}_{e}\right)\\[8pt]
\lcf e_{A}^{(1)},e_{B}^{(1)}\rcf_{\tau_{E}^{(1)}}=&{\mathcal
C}_{AB}^{C}e_{C}^{(1)},\\[8pt]
\lcf e_{A}^{(1)},e_{B}^{(2)}\rcf_{\tau_E^{(1)}}=&\lcf e_{A}^{(2)},
e_{B}^{(2)}\rcf_{\tau_E^{(1)}}=0,
\end{align*}for all $A$, $B$ and $C$; where ${\mathcal C}_{AB}^{C}$ are the
structure functions of $E$ determined by the Lie bracket
$\lcf\cdot,\cdot\rcf$ with respect to the basis $\{e_{A}\}$.

Using $\{e_{A}^{(1)},e_{A}^{(2)}\}$ one may introduce local
coordinates $(x^{i},y^{A};z^{A},v^{A})$ on $E$. If $V$ is a section
of $\tau_{E}^{(1)}$, locally it is determined by 
$$V(x,y)=(x^{i},y^{A},z^{A}(x,y),v^{A}(x,y));
$$ therefore the
expression of $V$ in terms of the basis
$\{e_{A}^{(1)},e_{A}^{(2)}\}$ is
$V=z^{A}e_{A}^{(1)}+v^{A}e_{A}^{(2)}$ and the vector field
$\rho_1(V)\in\mathfrak{X}(E)$ has the expression
$$\rho_{1}(V)=\rho_{A}^{i}z^{A}(x,y)\frac{\partial}{\partial
x^{i}}\Big{|}_{(x,y)}+v^{A}(x,y)\frac{\partial}{\partial
y^{A}}\Big{|}_{(x,y)}.$$
%Moreover, if $e\in A,$ then $$\rho_{1}(e_{\alpha}^{(1)})(e)=\left(e,\rho_{\alpha}^{i}\frac{\partial}{\partial x^{i}}\Big{|}_{e}\right);\qquad\rho_{1}(e_{\alpha}^{(2)})(e)=\left(e,\frac{\partial}{\partial y^{\alpha}}\Big{|}_{e}\right).$$

%\begin{remark}
%The notation $e_{\alpha}^{(1)}$ denotes complete lifts of sections and $e_{\alpha}^{(2)}$ denotes vertical lifts of sections.
%\end{remark}

Moreover, if $\{e_{(1)}^{A},e_{(2)}^{A}\}$ denotes the dual basis of
$\{e_{A}^{(1)},e_{A}^{(2)}\},$ 

\begin{align}
d^{\mathcal{T}^{\tau_E}E}F(x^{i},y^{A})&=\rho_{A}^{i}\frac{\partial F}{\partial x^{i}}e_{(1)}^{A}+\frac{\partial F}{\partial y^{A}}e_{(2)}^{A},\nonumber\\
d^{\mathcal{T}^{\tau_E}E}e_{(1)}^{C}&=-\frac{1}{2}\mathcal{C}_{AB}^{C}e_{(1)}^{A}\wedge e_{(1)}^{B},\quad d^{\mathcal{T}^{\tau_E}E}e_{(2)}^{C}=0.\nonumber
\end{align}

\begin{example}
In the case of $E=TM$ one may identify $\mathcal{T}^{\tau_E}E$ with
$TTM$ with the standard Lie algebroid structure over $TM$.
\end{example}

\begin{example}\label{ejemploprolongado1}
Let $\al$ be a real Lie algebra of finite dimension. $\al$ is a Lie algebroid over a single point $M=\{q\}.$ The anchor map of $\al$ is zero constant function, and from the anchor map we deduce that
$$\mathcal{T}^{\tau_{\al}}\al=\{(\xi_1,\xi_2,v_{\xi_1})\in\al\times T\al\}\simeq \al\times\al\times\al\simeq 3\al.$$ The vector bundle projection $\tau_{\al}^{(1)}:3\al\to\al$ is given by $\tau_{\al}^{(1)}(\xi_1,\xi_2,\xi_3)=\xi_1$ with anchor map $\rho_{1}(\xi_1,\xi_2,\xi_3)=(\xi_1,\xi_3)\simeq v_{\xi_1}\in T_{\xi_1}\al.$

Let $\{e_{A}\}$ be a basis of the Lie algebra $\al,$ this basis
induces local coordinates $y^{A}$ on $\al,$ that is,
$\xi=y^{A}e_{A}$. Also, this basis induces a basis of sections of
$\tau_{\al}^{(1)}$ as
$$e_{A}^{(1)}(\xi)=(\xi,e_{A},0),\quad e_{A}^{(2)}(\xi)=\left(\xi,0,\frac{\partial}{\partial y^{A}}\right).$$ Moreover $$\rho_{1}(e_{A}^{(1)})(\xi)=(\xi,0),\quad \rho_{1}(e_{A}^{(2)})(\xi)=\left(\xi,\frac{\partial}{\partial y^{A}}\right).$$

The basis $\{e_{A}^{(1)},e_{A}^{(2)}\}$ induces adapted coordinates
$(y^{A},z^{A},v^{A})$ in $3\al$ and therefore a
section $Y$ on $\Gamma(\tau_{\al}^{(1)})$ is written as
$Y(\xi)=z^{A}(\xi)e_{A}^{(1)}+v^{A}(\xi)e_{A}^{(2)}$. Thus, the
vector field $\rho_{1}(Y)\in\mathfrak{\al}$ has the expression
$\rho_{1}(Y)=v^{A}(\xi)\frac{\partial}{\partial
y^{A}}\Big{|}_{\xi}$. Finally, the Lie algebroid structure on
$\tau_{\al}^{(1)}$ is determined by the Lie bracket $\lcf
(\xi,\tilde{\xi}),(\eta,\tilde{\eta})\rcf=([\xi,\eta],0).$
\end{example}

\begin{example}\label{ejemploprolongado2}
We consider a Lie algebra $\al$ acting on a manifold $M,$ that is,
we have a Lie algebra homomorphism $\al\to\mathfrak{X}(M)$ mapping
every element $\xi$ of $\al$ to a vector field $\xi_{M}$ on $M.$
Then we can consider the action Lie algebroid $E=M\times\al$.
Identifying $TE=TM\times T\al=TM\times 2\al$, an element of
the prolongation Lie algebroid to $E$ over the bundle projection is
of the form $(a,b,v_{a})=((x,\xi),(x,\eta),(v_{x},\xi,\chi))$ where
$x\in M$, $v_{x}\in T_{x}M$ and $(\xi,\eta,\chi)\in 3\al$. The
condition $T\tau_{\al}(v)=\rho(b)$ implies that
$v_{x}=-\eta_{M}(x).$ Therefore we can identify the prolongation Lie
algebroid with $M\times\al\times\al\times\al$ with projection onto
the first two factors $(x,\xi)$ and anchor map
$\rho_{1}(x,\xi,\eta,\chi)=(-\eta_{M}(x),\xi,\chi)$. Given a base
$\{e_{A}\}$ of $\al$ the basis $\{e_{A}^{(1)},e_{A}^{(2)}\}$ of
sections of $\mathcal{T}^{\tau_{M\times\al}}(M\times\al)$ is given
by
$$e_{A}^{(1)}(x,\xi)=(x,\xi,e_{A},0),\quad
e_{A}^{(2)}(x,\xi)=(x,\xi,0,e_{A}).$$ Moreover,
$$\rho_{1}(e_{A}^{(1)}(x,\xi))=\left(x,-(e_{A})_{M}(x),\xi,0\right),\quad\rho_{2}(e_{A}^{(2)}(x,\xi))=(x,0,\xi,e_{A}).$$

Finally, the Lie bracket of two constant sections is given by
$\lcf (\xi,\tilde{\xi}),(\eta,\tilde{\eta})\rcf=([\xi,\eta],0).$

\end{example}

\begin{example}\label{ejemploprolongado3}
Let us describe the $E$-tangent bundle to $E$ in the case of $E$
being an Atiyah algebroid induced by a trivial principal $G-$bundle
$\pi:G\times M\to M.$ In such case, by left trivialization we get
the Atiyah algebroid, the vector bundle $\tau_{\al\times
TM}:\al\times TM\to TM.$ If $X\in\mathfrak{X}(M)$ and $\xi\in\al$
then we may consider the section $X^{\xi}:M\to\al\times TM$ of the
Atiyah algebroid by $$X^{\xi}(q)=(\xi,X(q))\hbox{ for }q\in M.$$
Moreover, in this sense $$\lcf X^{\xi},Y^{\xi}\rcf_{\al\times
TM}=([X,Y]_{TM},[\xi,\eta]_{\al}),\quad\rho(X^{\xi})=X.$$

On the other hand, if $(\xi,v_{q})\in\al\times T_{q}M$, then the
fiber of $\mathcal{T}^{\tau_{\al\times TM}}(\al\times TM)$ over
$(\xi,v_q)$ is
\begin{align}
\mathcal{T}^{\tau_{\al\times TM}}_{(\xi,v_q)}(\al\times TM)=\bigg\{((\eta,u_q),(\tilde{\eta},X_{v_{q}}))\in\al\times& T_{q}M\times\al\times T_{v_q}(TM)\nonumber\\
&\hbox{such that } u_{q}=T_{v_q}\tau_{\al\times TM}(X_{v_q})\bigg\}.\nonumber\end{align} This implies that
$\mathcal{T}^{\tau_{\al\times TM}}_{(\xi,v_q)}(\al\times TM)$ may be
identified with the space $2\al\times T_{v_q}(TM).$ Thus,
the Lie algebroid $\mathcal{T}^{\tau_{\al\times TM}}(\al\times TM)$
may be identified with the vector bundle
$\al\times 2\al\times TTM\to\al\times TM$ whose vector
bundle projection is $$(\xi,((\eta,\tilde{\eta}),X_{v_q}))\mapsto
(\xi,v_q)$$ for
$(\xi,((\eta,\tilde{\eta}),X_{v_q}))\in\al\times 2\al\times
TTM.$ Therefore, if $(\eta,\tilde{\eta})\in 2\al$ and
$X\in\mathfrak{X}(TM)$ then one may consider the section
$((\eta,\tilde{\eta}),X)$ given by
$$((\eta,\tilde{\eta}),X)(\xi,v_{q})=(\xi,((\eta,\tilde{\eta}),X(v_q)))\hbox{
for }(\xi,v_q)\in\al\times T_{q}M.$$ Moreover,
$$\lcf((\eta,\tilde{\eta}),X),((\xi,\tilde{\xi}),Y)\rcf_{\tau_{\al\times
TM}^{(1)}}=(([\eta,\xi]_{\al},0),[X,Y]_{TM}),$$ and the anchor map
$\rho_1:\al\times 2\al\times TTM\to\al\times\al\times TTM$
is defined as
$$\rho_{1}(\xi,((\eta,\tilde{\eta}),X))=((\xi,\tilde{\eta}),X).$$

\end{example}

\subsection{$E$-tangent bundle of the dual bundle of a Lie algebroid}

Let $(E,\ab,\rho)$ be a Lie algebroid of rank $n$ over a manifold of
dimension $m.$  Consider the projection of the dual $E^{*}$ of $E$
over ·$M$, $\tau_{E^{*}}:E^{*}\rightarrow M,$ and define the
prolongation $\mathcal{T}^{\tau_{E^{*}}}E$ of $E$ over
$\tau_{E^{*}}$; that is,
$$\mathcal{T}^{\tau_{E^{*}}}E=\bigcup_{\mu\in E^{*}}\{(e,v_{\mu})\in
E\times T_{\mu}E^{*}\mid \rho(e)=T\tau_{E^{*}}(v_{\mu})\}.$$
$\mathcal{T}^{\tau_{E^{*}}}E$ is a Lie algebroid over $E^{*}$ of
rank $2n$ with vector bundle projection
$\tau_{E^{*}}^{(1)}:\mathcal{T}^{\tau_{E^{*}}}E\Flder E^{*}$ given
by $\tau_{E^{*}}^{(1)}(e,v_{\mu})=\mu$, for $(e,v_{\mu})\in
\mathcal{T}^{\tau_{E^{*}}}E$.

As before, if we now denote by $(\mu,e,v_{\mu})$ an element
$(e,v_{\mu})\in\mathcal{T}^{\tau_{E^{*}}}E$ where $\mu\in E^{*}$, we
rewrite the definition of the prolongation Lie algebroid as the
subset of $E^{*}\times E\times TE^{*}$ by
$$\mathcal{T}^{\tau_{E^{*}}}E= \{(\mu,e,v_{\mu})\in E^{*}\times
E\times TE^{*}\mid \rho(e)=(T\tau_{E^{*}})(v_{\mu}), v_{\mu}\in
T_{\mu}E^{*} \hbox{ and }\tau_{E^{*}}(\mu)=\tau_{E}(e)\}.$$
%Also one could be
%identify (see \cite{LMM} and \cite{Mac}) this Lie algebroid by the
%pullback vector bundle $\rho^{*}(TA^{*})$ of
%$T\tau_{A^{*}}^{(0)}:TA^{*}\rightarrow TM$ over $\rho,$ that is,
%$$\rho^{*}(TA^{*})=\mathcal{T}^{\tau_{A^{*}}^{(0)}}A.$$

If $(x^i)$ are local coordinates on an open subset $U$ of $M$,
$\{e_{A}\}$ is a basis of sections of the vector bundle
$(\tau_E)^{-1}(U)\to U$ and $\{e^{A}\}$ is its dual basis, then
$\{\tilde{e}_{A}^{(1)},\tilde{e}_{A}^{(2)}\}$ is a basis of sections
of the vector bundle $\tau_{E^{*}}^{(1)}$, where $$
\tilde{
e}_{A}^{(1)}(\mu)=\left(\mu,e_{A}(\tau_{E^{*}}(\mu)),\rho_{A}^i\displaystyle\frac{\partial
}{\partial x^i}\Big{|}_{\mu}\right),\quad
(\tilde{e}^{A})^{(2)}(\mu)=\left(\mu,0,\displaystyle\frac{\partial }{\partial
p_{A}}\Big{|}_{\mu}\right),$$ for $\mu\in (\tau_{E^{*}})^{-1}(U).$ Here, $(x^i,p_{A})$ are the
local coordinates on $E^*$ induced by the local coordinates $(x^i)$
and the basis of sections of $E^{*},$ $\{e^{A}\}.$

Using the local basis $\{\tilde{e}_{A}^{(1)},(\tilde{
e}^{A})^{(2)}\}$, one may introduce, in a natural way, local
coordinates $(x^i,p_{A};z^{A},v_{A})$ on
$\mathcal{T}^{\tau_{E^{*}}}E.$ If $\omega^*$ is a point of
$\mathcal{T}^{\tau_{E^{*}}}E$ over $(x,p)\in E^{*}$, then
\[
\omega^*(x,p)=z^{A}\tilde{e}_{A}^{(1)}(x,p) + v_{A}(\tilde{e}^{A})^{(2)}(x,p).
\]

Denoting by $\rho_{\tau_{E^{*}}^{(1)}}$ the anchor map of the Lie
algebroid $\mathcal{T}^{\tau_{E^{*}}}E\to E^{*}$ locally given by
$$\rho_{\tau_{E^{*}}^{(1)}}(x^{i},p_A,z^{A},v_A)=(x^{i},p_{A},\rho_{A}^{i}z^{A},v^{A}),$$
we have that

$$\rho_{\tau_{E^{*}}^{(1)}}(\tilde{e}_{A}^{(1)})(\mu)=\left(\mu,\rho_{A}^i\displaystyle\frac{\partial
}{\partial x^i}\bigg{|}_{\mu}\right),\quad \rho_{\tau_{E^{*}}^{(1)}}((\tilde{e}^{A})^{(2)})(\mu)=\left(\mu,\displaystyle\frac{\partial
}{\partial p_{A}}\bigg{|}_{\mu}\right).$$ Therefore, we have that the corresponding vector field $\rho_{\tau_{E^{*}}^{(1)}}(V)$ for the section determined by $V=(x^{i},p_{A},z^{A}(x,p),v_{A}(x,p))$ is given by $$\rho_{\tau_{E^{*}}^{(1)}}(V)=\rho_{A}^{i}z^{A}\frac{\partial}{\partial x^{i}}\bigg{|}_{\mu}+v_{A}\frac{\partial}{\partial p_{A}}\bigg{|}_{e^{*}}.$$

Finally, the structure of the Lie algebroid
$(\mathcal{T}^{\tau_{E^{*}}}E,\lcf\cdot,\cdot\rcf_{\tau_{{E}^{*}}^{(1)}},\rho_{\tau_{E^{*}}^{(1)}}),$
is determined by the bracket of sections
$$\lcf\tilde{e}_{A}^{(1)},\tilde{e}_{B}^{(1)}\rcf_{\tau_{E^{*}}^{(1)}}={\mathcal
C}_{AB}^{C}\tilde{e}_{C}^{(1)},\quad
\lcf\tilde{e}_{A}^{(1)},(\tilde{e}^{B})^{(2)}\rcf_{\tau_{E^{*}}^{(1)}}=\lcf (\tilde{e}^{A})^{(2)}, (\tilde{e}^{B})^{(2)}\rcf_{\tau_{E^{*}}^{(1)}}=0,$$

\noindent for all $A,B$ and $C$. Thus, if we denote by
$\{\tilde{e}^{A}_{(1)},(\tilde{e}_{A})_{(2)}\}$ is the dual basis of
$\{\tilde{e}_{A}^{(1)},(\tilde{e}^{A})^{(2)}\}$, then
%\begin{equation}\label{dif*}
\begin{align}d^{\mathcal{T}^{\tau_{E^{*}}}E}f(x^{i},p_{A})&=
\rho_{A}^i\displaystyle{\frac{\partial f}{\partial x^i}\tilde{e}^{A}_{(1)}}+ \displaystyle{\frac{\partial f}{\partial p_{A}}(\tilde{e}_{A})_{(2)}},\nonumber\\
d^{\mathcal{T}^{\tau_{E^{*}}}E}\tilde{e}^{C}_{(1)}&=\displaystyle{-\frac{1}{2}
{\mathcal C}_{AB}^{C}\tilde{e}^{A}_{(1)}\wedge\tilde{e}^{B}_{(1)}},\qquad
 d^{\mathcal{T}^{\tau_{E^{*}}}E}(\tilde{e}_{C})_{(2)}=0,\nonumber
\end{align} for $f\in C^\infty(E^*).$ We refer to
\cite{LMM} for further details about the Lie algebroid structure of
the E-tangent bundle of the dual bundle of a Lie algebroid.

\begin{example}
In the case of $E=TM$ one may identify $\mathcal{T}^{\tau_{E^{*}}}E$
with $T(T^{*}M)$ with the standard Lie algebroid structure.
\end{example}

\begin{example}
Let $\al$ be a real Lie algebra of finite dimension. Then $\al$ is a
Lie algebroid over a single point. Using that the anchor map is zero
we have that $\mathcal{T}^{\tau_{\al^{*}}}\al$ may be identified
with the vector bundle
$pr_{1}:\al^{*}\times(\al\times\al^{*})\to\al^{*}.$ Under this
identification the anchor map is given by
$$\rho_{\tau_{\al^{*}}^{(1)}}:\al^{*}\times(\al\times\al^{*})\to
T\al^{*}\simeq\al^{*}\times\al^{*},\quad(\mu,(\xi,\alpha))\mapsto(\mu,\alpha)$$
and the Lie bracket of two constant sections
$(\xi,\alpha),(\eta,\beta)\in\al\times\al^{*}$ is the constant
section $([\xi,\eta],0).$
\end{example}

\begin{example}
Let $A=M\times\al\to M$ be an action Lie algebroid over $M$ and $(q,\mu)\in M\times\al^{*}.$ It follows that the prolongation may be identified with the trivial vector bundle $(M\times\al^{*})\times(\al\times\al^{*})\to M\times\al^{*}$ since

$$\mathcal{T}^{\tau_{(M\times\al^{*})}}M\times\al=\{((q,\xi),(X_q,\alpha))\in M\times\al\times T_{q}M\times\al^{*}\Big{|}-\xi_{M}(q)=X_{q}\}\simeq \al\times\al^{*}.$$ The anchor map $\rho_{\tau_{M\times\al^{*}}^{(1)}}:(M\times\al^{*})\times(\al\times\al^{*})\to TM\times(\al^{*}\times\al^{*})$ is given by $$\rho_{\tau_{M\times\al^{*}}^{(1)}}((q,\mu),(\xi,\alpha))=(-\xi_{M}(q),\mu,\alpha).$$ Moreover, the Lie bracket of two constant sections $(\xi,\alpha),(\eta,\beta)\in\al\times\al^{*}$ is just the constant section $([\xi,\eta],0).$
\end{example}

\begin{example}
Let us describe the $A$-tangent bundle to $A$ in the case of $A$ being an Atiyah algebroid induced by a trivial principal $G-$bundle
$\pi:G\times M\to M.$ In such case, by left trivialization we have that the Atiyah algebroid is the vector bundle $\tau_{\al\times TM}:\al\times TM\to TM.$
If $X\mathfrak{X}(M)$ and $\xi\in\al$ then we may consider the section $X^{\xi}:M\to\al\times TM$ of the Atiyah algebroid by $$X^{\xi}(q)=(\xi,X(q))\hbox{ for }q\in M.$$ Moreover, in this sence $$\lcf X^{\xi},Y^{\xi}\rcf_{\al\times TM}=([X,Y]_{TM},[\xi,\eta]_{\al}),\quad\rho(X^{\xi})=X.$$

If $(\mu,\alpha_{q})\in\al^{*}\times T_{q}^{*}M$ then the fiber of $\mathcal{T}^{\tau_{(\al\times TM)^{*}}}(\al\times TM)$ over $(\mu,\alpha_q)$ is
\begin{align}
\mathcal{T}^{\tau_{(\al\times TM)^{*}}}_{(\mu,\alpha_q)}(\al\times TM)=\bigg\{((\eta,u_q),(\beta,X_{\alpha_{q}}))\in&\al\times T_{q}M\times\al^{*}\times T_{v_q}(T^{*}M)\nonumber\\
&\hbox{ such that }u_{q}=T_{\alpha_q}\tau_{(\al\times TM)^{*}}(X_{\alpha_q})\bigg\}\nonumber.\end{align} This implies that
$\mathcal{T}^{\tau_{(\al\times TM)^{*}}}_{(\mu,\alpha_q)}(\al\times TM)$ may be identified with the vector space $(\al\times\al^{*})\times T_{\alpha_q}(T^{*}M).$ Thus, the Lie algebroid $\mathcal{T}^{\tau_{(\al\times TM)^{*}}}(\al\times TM)$ may be identified with the vector bundle $\al^{*}\times(\al\times\al^{*})\times TT^{*}M\to\al^{*}\times T^{*}M$ whose vector bundle projection is $$(\mu,((\xi,\beta),X_{\alpha_q}))\mapsto (\mu,\alpha_q)$$ for $(\mu,((\xi,\beta),X_{\alpha_q}))\in\al^{*}\times(\al\times\al^{*})\times TT^{*}M.$ Therefore, if $(\xi,\beta)\in\al\times\al^{*}$ and $X\in\mathfrak{X}(T^{*}M)$ then one may consider the section
$((\xi,\beta),X)$ given by $$((\xi,\beta),X)(\mu,\alpha_{q})=(\mu,((\xi,\beta),X(\alpha_q)))\hbox{ for }(\mu,\alpha_q)\in\al^{*}\times T_{q}^{*}M.$$ Moreover, $$\lcf((\xi,\beta),X),((\tilde{\xi},\tilde{\xi}),\tilde{X})\rcf_{\tau_{(\al\times TM)^{*}}^{(1)}}=(([\xi,\tilde{\xi}]_{\al},0),[X,\tilde{X}]_{TM}),$$ and the anchor map $\rho_{\tau_{(\al\times TM)^{*}}^{(1)}}:\al^{*}\times(\al\times\al^{*})\times TT^{*}M\to\al^{*}\times\al^{*}\times TT^{*}M$ is defined as $$\rho_{\tau_{(\al\times TM)^{*}}^{(1)}}(\mu,((\xi,\beta),X))=((\mu,\beta),X).$$
\end{example}

\subsection{Symplectic Lie algebroids} In this subsection we
will recall some results given in \cite{LMM} about symplectic Lie
algebroids.

\begin{definition}
A Lie algebroid $(E,\lcf\cdot,\cdot\rcf,\rho)$ over a manifold $M$
is said to be symplectic if it admits a symplectic section $\Omega,$
that is, $\Omega$ is a section of the vector bundle
$\bigwedge^{2}E^{*}\Flder M$ such that:
\begin{enumerate}
\item For all $x\in M,$ the 2-form $\Omega_{x}:E_{x}\times
E_{x}\Flder\R$ in the vector space $E_{x}$ is nondegenerate and
\item $\Omega$ is a 2-cocycle, that is, $d^{E}\Omega=0$.
\end{enumerate}
\end{definition}

%\begin{example}
%\begin{enumerate}
%\item $(\mathcal{T}^{\tau_{{E}^{*}}}E,\lcf\cdot,\cdot\rcf_{\tau_{E^{*}}^{(1)}},\rho_{\tau_{E^{*}}^{(1)}});$ is a symplectic Lie algebroid where
%$\Omega_{E}$ is the canonical symplctic $2$-section of
%$\mathcal{T}^{\tau_{{E}^{*}}}E.$
%\item If $E$ is a symplectic Lie algebroid
%with symplectic section $\Omega$ then the prolongation
%$\mathcal{T}^{\tau_{E}}E$ of $E$ over the vector bundle
%projection $\tau_E:E\to M$ is a symplectic Lie algebroid with
%symplectic section $\Omega^{\bf c}$, the complete lift of
%$\Omega$ (see \cite{LMM}).
%\end{enumerate}
%\end{example}
\subsubsection{The canonical symplectic structure of
$\mathcal{T}^{\tau_{{E}^{*}}}E$}

Let $(E,\lcf\cdot,\cdot\rcf,\rho)$ be a Lie algebroid of rank $n$
over a manifold $M$ of dimension $m$ and
$\mathcal{T}^{\tau_{{E}^{*}}}E$ be the prolongation of $E$ over the
vector bundle projection $\tau_{E^{*}}:E^{*}\Flder M.$ We may
introduce a canonical section $\lambda_{E}$ of
$(\mathcal{T}^{\tau_{{E}^{*}}}E)^{*}$ as follows. If $\mu\in E^{*}$
and $(e,v_{\mu})$ is a point on the fibre of
$\mathcal{T}^{\tau_{{E}^{*}}}E$ over $\mu$ then
\begin{equation}\label{liouvillesection}
\lambda_{E}(\mu)(e,v_{\mu})=\langle\mu,e\rangle.
\end{equation}
$\lambda_{E}$ is called the \textit{Liouville section} of
$\mathcal{T}^{\tau_{{E}^{*}}}E.$ Now, in an analogous way that the
canonical symplectic form is defined from the Liouville $1$-form on
the cotangent bundle, we introduce the $2$-section $\Omega_{E}$ on
$\mathcal{T}^{\tau_{{E}^{*}}}E$ as

\begin{equation}\label{2section}
\Omega_{E}=-d^{\mathcal{T}^{\tau_{E^{*}}}E}\lambda_{E}.
\end{equation}

\begin{proposition}\cite{LMM}
$\Omega_{E}$ is a non-degenerate $2$-section of
$\mathcal{T}^{\tau_{{E}^{*}}}E$ such that
$$d^{\mathcal{T}^{\tau_{{E}^{*}}}E}\Omega_{E}=0.$$
\end{proposition}

Therefore $\Omega_{E}$ is a symplectic $2$-section on
$\mathcal{T}^{\tau_{{E}^{*}}}E$ called \textit{canonical symplectic
section} on $\mathcal{T}^{\tau_{{E}^{*}}}E$.

\begin{example}
If $E$ is the standard Lie algebroid $TM$ then $\lambda_{E}=\lambda$
and $\Omega_{E}=\omega_{M}$ are the usual Liouville $1$-form and
canonical symplectic $2$-form on $T^{*}M$, respectively.
\end{example}

\begin{example}
Let $\al$ be a finite dimensional Lie algebra. Then $\al$ is a Lie
algebroid over a single point $M=\{q\}.$ If $\xi\in\al$ and
$\mu,\alpha\in\al^{*}$ then
$$\lambda_{\al}(\mu)(\xi,\alpha)=\mu(\xi)$$ is the Liouville
1-section on $\al^{*}\times(\al\times\al^{*}).$ Thus, the symplectic
section $\Omega_{\al}$ is
$$\Omega_{\al}(\mu)((\xi,\alpha),(\eta,\beta))=\langle\mu,[\xi,\eta]\rangle-\langle\alpha,\eta\rangle-\langle\beta,\xi\rangle$$ for $\mu\in\al^{*}, (\xi,\alpha),(\eta,\beta)\in\al\times\al^{*}.$
\end{example}

\subsection{Admissible elements on a Lie algebroid}

Let $E$ be a Lie algebroid over $M$ with fiber bundle
projection $\tau_{E}:E\to M$ and anchor map
$\rho:E\to TM.$

\begin{definition}
A tangent vector $v$ at the point $e\in E$ is called
\textit{admissible} if $\rho(e)=T_{e}\tau_{E}(v);$ and a
curve on $E$ is admissible if its tangent vectors are
admissible. The set of admissible elements on $E$ will
be denote $E^{(2)}$.
\end{definition}

Notice that $v$ is admissible if and only if $(e,e,v)$ is in
$\mathcal{T}^{\tau_{E}}E.$ We will consider
$E^{(2)}$ as the subset of the prolongation of $E$ over
$\tau_{E}$, that is, $E^{(2)}\subset
E_{\rho}\times_{T\tau_{E}} TE$
is given by
$$E^{(2)}=\{(e,v_e)\in E\times TE\mid \rho(e)=T\tau_{E}(v_e)\}.$$ Other authors call this
set $\hbox{Adm}(E)$ (see \cite{Pepin2007} and
\cite{Eduardoalg}).

A curve $\sigma:I\subset\mathbb{R}\to E$ is said to be an admissible curve on $E$ if it satisfies $\rho(\sigma(t))=\dot{\gamma}(t)$ where $\gamma=\tau_{E}(\sigma(t))$ is a curve on $M$. Locally, admisible curves on $E$ are characterized by the so-called admissibility condition. A curve $\gamma(t)=(x^i(t),y^A(t))$ on $E$ is admissible if it satisfies the admissibility condition $\displaystyle{\dot{x}^{i}(t)=\rho_{A}^{i}(x^{i}(t))y^{A}(t)}$. Therefore, locally, $E^{(2)}$ is determined by $(\gamma(0),\dot{\gamma}(0))$ where $\gamma$ is an admissible curve on $E$. Admissible curves on $E$ are also called $E$-path \cite{Eduardoho}.

%with initial condition $x(0)=\gamma(0)$

We consider $E^{(2)}$ as the substitute of the second order tangent
bundle in classical mechanics. If $(x^{i})$ are local coordinates on
$M$ and $\{e_{A}\}$ is a basis of sections of $E$ then
we denote $(x^{i},y^{A})$ the corresponding local coordinates on
$E$ and $(x^{i},y^{A};z^{A},v^{A})$ local coordinates on
$\mathcal{T}^{\tau_{E}}E$ induced by the
basis of sections $\{e_{A}^{(1)},e_{A}^{(2)}\}$ of
$\mathcal{T}^{\tau_{E}}E$ (see subsection
\ref{prolongationliealgebroidsubsection}). Therefore, the set
$E^{(2)}$ is locally characterized by the condition
$\{(x^{i},y^{A};z^{A},v^{A})\in
\mathcal{T}^{\tau_{E}}E\mid y^{A}=z^{A}\},$ that is
$(x^{i},y^{A},v^{A}):=(x^{i},y^{A},\dot{y}^{A})$ are local
coordinates on $E^{(2)}.$

We denote the canonical inclusion of $E^{(2)}$ on the prolongation
of $E$ over $\tau_{E}$ as \begin{eqnarray*}
i_{E^{(2)}}:E^{(2)}&\hookrightarrow& \mathcal{T}^{\tau_{E}}E,\\
(x^{i},y^{A},\dot{y}^{A})&\longmapsto& (x^{i}, y^{A}, y^{A},
\dot{y}^{A}).
\end{eqnarray*}

\begin{example}
Let $M$ be a differentiable manifold of dimension $m$, if $(x^{i})$
are local coordinates on $M,$ then $\{\frac{\partial}{\partial
x^{i}}\}$ is a local basis of $\mathfrak{X}(M)$ and then we have
fiber local coordinates $(x^{i},\dot{x}^{i})$ on $TM.$ The
corresponding local structure functions of the Lie algebroid
$\tau_{TM}:TM\to M$ are
$$\mathcal{C}_{ij}^{k}=0\hbox{ and } \rho_{i}^{j}=\delta_{i}^{j},\hbox{ for }
i,j,k\in\{1,\ldots,m\}.$$ In this case, we have seen that the
prolongation Lie algebroid over $\tau_{TM}$ is just the tangent
bundle $T(TM)$ where the Lie algebroid structure of the vector
bundle $T(TM)\to TM$ is as we have described above as the tangent
bundle of a manifold.

The set of admissible elements is given by
$$E^{(2)}=\{(x^{i},v^{i},\dot{x}^{i},w^{i})\in
T(TM)\mid\dot{x}^{i}=v^{i}\}$$ and observe that this subset is just
the second-order tangent bundle of a manifold $M,$ that is,
$E^{(2)}=T^{(2)}M$. Admissible curves on $E^{(2)}=T^{(2)}M$ are given by $$\sigma(t)=(x^{i}(t),\dot{x}^{i}(t),
\ddot{x}^{i}(t)).$$

\end{example}

\begin{example}
Consider a Lie algebra $\al$ as a Lie algebroid over a point $\{ e\}$. Given a basis of section $\{e_{A}\}$ and element $\xi\in\al$ can be written as $\xi=e_A\xi^A$ and given that the anchor map is given by $\rho(\xi)\equiv 0$, every curve $\xi(t)\in\al$ is an admisible curve. The set of admisible elements  is described by the cartesian product of two copies of the Lie algebra, $2\al$. Local coordinates on $2\al$ are determined by the basis of sections of $\al$, $\{e_{A}\}$ and $\{e_A^{(1)},e_{A}^{(2)}\}$, the basis of the prolongation Lie algebroid introduced in Example \ref{ejemploprolongado1}. They are denoted by $(\xi^1,\xi^2)$ and also $(\xi^1,\xi^2):=(\xi(0), \dot{\xi}(0))\in 2\al$ where $\xi(t)$ is admissible. 

\end{example}

\begin{example}

Let $G$ be a Lie group and we assume that $G$ acts free and properly
on $M$. We denote by $\pi:M\Flder \widehat{M}=M/G$ the associated
principal bundle. The tangent lift of the action gives a free and
proper action of $G$ on $TM$ and $\widehat{TM}=TM/G$ is a quotient
manifold. Then we consider the Atiyah algebroid $\widehat{TM}$ over
$\widehat{M}$.

According to example \ref{Atiyah case}, the basis
$\{\hat{e}_{i},\hat{e}_{B}\}$ induce local coordinates
$(x^{i},y^{i},\bar{y}^{B})$. From this basis one can induces a basis
of the prolongation Lie algebroid, namely
$\{\hat{e}_{i}^{(1)},\hat{e}_{B}^{(1)}\}$. This basis induce adapted
coordinates
$(x^{i},y^{i},\bar{y}^{B},\dot{y}^{i},\dot{\bar{y}}^{B})$ on
$\widehat{T^{(2)}M}=(T^{(2)}M)/G$.

\end{example}

\section{Second-order variational problems on Lie algebroids}\label{seccion3}

The geometric description of mechanics in terms of Lie
algebroids gives a general framework to obtain all the relevant
equations in mechanics (Euler-Lagrange, Euler-Poincar\'e,
Lagrange-Poincar\'e,...). In this section we use the
notion of Lie algebroid and prolongation of a Lie algebroid
described in $\S$\ref{LALG} to derive the Euler-Lagrange equations and
Hamilton equations on Lie algebroids. Next, after introduce the constraint algorithm for presymplectic Lie algebroids and study vakonomic mechanics on Lie algebroids, we study the geometric formalism for second-order constrained variational problems using and adaptation of the classical Skinner-Rusk formalism for the second-order constrained systems on Lie algebroids. 

\subsection{Mechanics on Lie algebroids}

In \cite{Eduardoalg} (see also \cite{LMM}) a geometric formalism for
Lagrangian mechanics on Lie algebroids was introduced. It was
developed in the prolongation $\te^{\tau_E}E$ of a Lie algebroid $E$
(see $\S$\ref{LALG}) over the vector bundle projection
$\tau_{E}:E\to M$. The prolongation of the Lie algebroid is playing
the same role as $TTQ$ in the standard mechanics. We first introduce
the canonical geometrical structures defined on $\te^{\tau_E}E$ to
derive the Euler-Lagrange equations on Lie algebroids.

Two canonical objects on ${\mathcal T}^{\tau_{E}}E$ are \emph{the
Euler section} $\Delta$ and \emph{the vertical endomorphism $S$}. Considering the
local basis of sections of $\mathcal{T}^{\tau_{E}}E$, $\{ e_{A}^{(1)}, e_{A}^{(2)}\}$, 
$\Delta$ is the section of ${\mathcal T}^{\tau_{E}}E\to E$ locally
defined by
\begin{equation}
\label{Lioulo} \Delta = y^{A}e_{A}^{(2)}
\end{equation}
and $S$ is the section of the vector bundle $({\mathcal
T}^{\tau_{E}}E)\otimes ({\mathcal T}^{\tau_{E}}E)^*\to E$ locally
characterized by the following conditions:
\begin{equation}\label{endverlo}
Se_{A}^{(1)} = e_{A}^{(2)}, \makebox[.3cm]{} Se_{A}^{(2)} = 0, \makebox[.3cm]{}
\mbox{ for all } A.
\end{equation}
Finally, a section $\xi$ of ${\mathcal T}^{\tau_{E}}E\to E$ is said
to be a \emph{second order differential equation} (SODE) on $E$ if
$S(\xi)=\Delta$ or, alternatively, $pr_1(\xi(e))=e$, for all $e\in
E$ (for more details, see \cite{LMM}).

Given a Lagrangian function $L\in C^{\infty}(E)$ we introduce the
\emph{Cartan 1-section}
$\Theta_L\in\Gamma({(\mathcal{T}^{\tau_E}E)^*})$, the \emph{Cartan
2-section} $\omega_L\in\Gamma({\wedge^2(\mathcal{T}^{\tau_E}E)^*})$
and the \emph{Lagrangian energy} $E_L\in C^{\infty}(E)$ as
%\begin{equation}\label{Cartan-forms}
$$\Theta_L=S^*(d^{{\mathcal T}^{\tau_{E}}E}L) , \qquad \omega_L =
-d^{{\mathcal T}^{\tau_{E}}E}\Theta_L\qquad
E_L=\mathcal{L}^{\mathcal{T}^{\tau_{E}}E}_{\Delta} L-L.$$
%\end{equation}
If $(x^i, y^{A})$ are local fibred coordinates on $E$, $(\rho^i_{A},
{\mathcal C}^{C}_{AB})$ are the corresponding local structure
functions on $E$ and $\{ e_{A}^{(1)}, e_{A}^{(2)}\}$ the
corresponding local basis of sections of $\mathcal{T}^{\tau_{E}}E$
then
\begin{equation}
\label{omegaL} \omega_L
=\frac{\partial^2L}{\partial y^{A}\partial y^{B}}e^{A}_{(1)}\we e^{B}_{(2)}+\frac{1}{2}\lp\frac{\der^2L}{\der x^i\der y^{A}}\rho_{B}^i-\frac{\der^2L}{\der x^i\der y^{B}}\rho_{A}^i+\frac{\der L}{\der y^{A}}\mathcal{C}^{C}_{AB}\rp e^{A}_{(1)}\we e^{B}_{(1)},\\\\
\end{equation}
\begin{equation}
\label{EL} E_L=\frac{\partial L}{\partial y^{A}}y^{A}-L.
\end{equation}

A curve $t\mapsto c(t)$ on $E$ is a solution of the
\emph{Euler-Lagrange equations} for $L$ if
\begin{itemize}
\item[-] $c$ is \emph{admissible} (that is, $\rho(c(t))=\dot{m}(t)$, where
$m=\tau_{E}\circ c$) and
\item[-] $\displaystyle{i_{(c(t), \dot{c}(t))}\omega_L(c(t))-d^{{\mathcal T}^{\tau_{E}}E}E_L
(c(t))=0}$, for all $t$.
\end{itemize}
If $c(t)=(x^i(t), y^{A}(t))$ then $c$ is a solution of the
Euler-Lagrange equations for $L$ if and only if
\begin{equation}\label{ELalgebroid}
\dot x^i=\rho_{A}^iy^{A},\,\,\,\frac{\de}{\de t}\frac{\der L}{\der y^{A}}+\frac{\der L}{\der y^{C}}\mathcal{C}_{AB}^{C}y^{B}-\rho_{A}^i\frac{\der L}{\der x^i}=0.
\end{equation}

Observe that, if $E$ is the standard Lie algebroid $TQ$ then the above
equations are the classical Euler-Lagrange equations for $L: TQ \to
\R$.

On the other hand, the Lagrangian function $L$ is said to be
\textit{regular} if $\omega_L$ is a symplectic section. In such a
case, there exists a unique solution $\xi_L$ verifying
\begin{equation}\label{symp}
i_{\xi_L}\omega_L-d^{{\mathcal T}^{\tau_{E}}E}E_L=0\; .
\end{equation}

In addition, one can check that $i_{S\xi_{L}}\omega_{L} =
i_{\Delta}\omega_{L} $ which implies that $\xi_{L}$ is a SODE
section. Thus, the integral curves of $\xi_L$ (that is, the integral
curves of the vector field $\rho_{1}(\xi_L)$) are solutions of the
Euler-Lagrange equations for $L$. $\xi_L$ is called the
\emph{Euler-Lagrange section} associated with $L$.

From (\ref{omegaL}), we deduce that the Lagrangian $L$ is regular
if and only if the matrix
$$\displaystyle{(W_{AB})=\Big(\frac{\partial^2 L}{\partial
y^{A}\partial y^{B}}\Big)}$$ is regular. Moreover, the local
expression of $\xi_L$ is
\[
\xi_L=y^Ae_{A}^{(1)}+f^Ae_{A}^{(2)} ,
\]
where the functions $f^A$ satisfy the linear equations
%\begin{equation}\label{free-forces1}
\[
\frac{\partial^2L}{\partial y^{B}\partial y^{A}}f^{B}+\frac{\der^2L}{\der x^i\der y^{A}}\rho_{B}^iy^{B}+\frac{\der L}{\der y^{C}}\mathcal{C}_{AB}^{C}y^{B}-\rho_{A}^i\frac{\der L}{\der x^i}=0,\,\,\,\forall\,A.
\]
%\end{equation}

Another possibility is when the matrix
$\displaystyle{(W_{AB})=\Big(\frac{\partial^2 L}{\partial
y^{A}\partial y^{B}}\Big)}$ is singular. This type of Lagrangian
is called \emph{singular} or \emph{degenerate Lagrangian}. In such a
case, $\omega_L$ is not a symplectic section and Equation
(\ref{symp}) has no solution, in general, and even if it exists it
will not be unique. In the next subsection, we will give the extension
of the classical Gotay-Nester-Hinds algorithm \cite{GoNesHinds78}
for presymplectic systems on Lie algebroids given in \cite{IMMS},
which in particular will be applied to optimal control problems.

For an arbitrary Lagrangian function $L:E\to\R$, we introduce the
\emph{Legendre transformation} associated with $L$ as the smooth map
$leg_L:E\to E^*$ defined by
$$leg_L(e)(e')=\frac{d}{dt}\bigg{|}_{t=0}L(e+te'),$$
for $e,e'\in E_x$. Its local expression is
\begin{equation}\label{leg-L}
leg_L(x^i,y^A)=(x^i,\displaystyle\frac{\partial L}{\partial y^{A}}).
\end{equation}

The Legendre transformation induces a Lie algebroid morphism
$${\mathcal T}\,leg_L:{\mathcal T}^{\tau_{E}}E\to{\mathcal T}^{\tau_{E^{*}}}E$$
over $leg_L:E\to E^*$ given by
$$({\mathcal T}\,leg_L)(e,v)=(e,(Tleg_L)(v)),$$
where $Tleg_L:TE\to TE^*$ is the tangent map of $leg_L:E\to E^*$.

We have that (see \cite{LMM} for details)
\begin{equation}\label{pullbacks}
({\mathcal T}\,leg_L,leg_L)^*(\lambda_E)=\Theta_L,\;\;\;({\mathcal
T}\,leg_L,leg_L)^*(\Omega_E)=\omega_L.
\end{equation} where $\lambda_{E}$ is the Liouville section indroduced in \eqref{liouvillesection} and $\Omega_{E}$ is the canonical symplectic section on ${\mathcal T}^{\tau_{E^{*}}}E$.

On the other hand, from (\ref{leg-L}), it follows that the
Lagrangian function $L$ is regular if and only if $leg_L:E\to E^*$
is a local diffeomorphism.

Next, we will assume that $L$ is \emph{hyperregular}, that is,
$leg_L:E\to E^*$ is a global diffeomorphism. Then, the pair
$({\mathcal T}\,leg_L,leg_L)$ is a Lie algebroid isomorphism.
Moreover, we may consider the \emph{Hamiltonian function}
$H:E^*\to\R$ defined by
$$H=E_L\circ leg_L^{-1}$$
and the \emph{Hamiltonian section} $\xi_H\in\Gamma({\mathcal
T}^{\tau_{E^{*}}}E)$ which is characterized by the condition
$$i_{\xi_H}\Omega_E=d^{{\mathcal T}^{\tau_{E^{*}}}E}H.$$

The integral curves of the vector field $\rho_{1}(\xi_H)$ on $E^*$
satisfy the \emph{Hamilton equations for $H$}
\[
\frac{\de x^i}{\de t}=\rho_{A}^i\frac{\der H}{\der p_{A}},\,\,\,\,\,\,\,\,\,\frac{\de p_{A}}{\de t}=-\rho_{A}^i\frac{\der H}{\der x^i}-p_{C}\mathcal{C}_{AB}^{C}\frac{\der H}{\der p_{B}}.
\] for $i\in\{1,\dots,m\}$ and $A\in\{1,\dots,n\}$ (see
\cite{LMM}).

In addition, the Euler-Lagrange section $\xi_L$ associated with
$L$ and the Hamil\-to\-nian section $\xi_H$ are $({\mathcal
T}\,leg_L,leg_L)$-related, that is,
$$\xi_H\circ leg_L={\mathcal T}\,leg_L\circ\xi_L.$$

Thus, if $\gamma:I\to E$ is a solution of the Euler-Lagrange
equations associated with $L$, then $\mu=leg_L\circ\gamma:I\to E^*$
is a solution of the Hamilton equations for $H$ and, conversely, if
$\mu:I\to E^*$ is a solution of the Hamilton equations for $H$ then
$\gamma=leg_L^{-1}\circ\mu$ is a solution of the Euler-Lagrange
equations for $L$ (for more details, see \cite{LMM}).

\begin{example}
Consider the Lie algebroid $E=TQ$, the fiber bundle of a manifold
$Q$ of dimension $m$. If $(x^{i})$ are local coordinates on $Q$,
then $\displaystyle{\Big{\{}\frac{\partial}{\partial x^{i}}\Big{\}}}$ is a local
basis of $\mathfrak{X}(Q)$ and we have fiber local coordinates
$(x^{i},\dot{x}^{i})$ on $TQ$. The corresponding structure functions
are $\mathcal{C}_{ij}^{k}=0$ and $\rho_{i}^{j}=\delta_{i}^{j}$ for
$i,j,k\in\{1,\ldots,m\}.$ Therefore given a Lagrangian function
$L:TQ\to\R$ the Euler-Lagrange equations associated to $L$ are
$$\frac{d}{dt}\left(\frac{\partial
L}{\partial\dot{x}^{i}}\right)=\frac{\partial L}{\partial
x^{i}},\quad i=1,\ldots,m.$$ Moreover, given a Hamiltonian function
$H:T^{*}Q\to\R,$ the Hamilton equations associated to $H$ are
$$\dot{x}^{i}=\frac{\partial H}{\partial
p_i},\quad\dot{p}_{i}=-\frac{\partial H}{\partial x^{i}},\quad
i=1,\ldots,m$$ where $(x^{i},p_{i})$ are local coordinates on
$T^{*}Q$ induced by the local coordinates $(x^{i})$ and the local
basis $\{dx^{i}\}$ of $T^{*}Q$ (see \cite{Blo} for example).
\end{example}

\begin{example}
Consider as a Lie algebroid the finite dimensional Lie algebra
$(\mathfrak{g},[\cdot,\cdot]_{\mathfrak{g}})$ over a point. If
$e_{A}$ is a basis of $\mathfrak{g}$ and
$\widetilde{\mathcal{C}}_{AB}^{C}$ are the structure constants of
the Lie algebra, the structures constant of the Lie algebroid
$\mathfrak{g}$ with respect to the basis $\{e_{A}\}$ are
$\mathcal{C}_{AB}^{C}=\widetilde{\mathcal{C}}_{AB}^{C}$ and
$\rho_{A}^{i}=0.$ Denote by $(y^{A})$ and $(\mu_{A})$ the local
coordinates on $\mathfrak{g}$ and $\mathfrak{g}^{*}$ respectively,
induced by the basis $\{e_{A}\}$ and its dual basis $\{e^{A}\}$
respectively. Given a Lagrangian function $L:\mathfrak{g}\to\R$ then
the Euler-Lagrange equations for $L$ are just the Euler-Poincar\'e
equations
$$\frac{d}{dt}\left(\frac{\partial L}{\partial
y^{A}}\right)=\frac{\partial L}{\partial
y^{C}}\mathcal{C}_{AB}^{C}y^{B}.$$ Given a Hamiltonian function
$H:\mathfrak{g}^{*}\to\R$ the Hamilton equations on
$\mathfrak{g}^{*}$ read as the Lie-Poisson equations for $H$
$$\dot{\mu}=ad^{*}_{\frac{\partial H}{\partial\mu}}\mu$$ (see
\cite{Blo} for example).
\end{example}

\begin{example}
Let $G$ be a Lie group and assume that $G$ acts free and properly on
$M$. We denote by $\pi:M\Flder \widehat{M}=M/G$ the associated
principal bundle. The tangent lift of the action gives a free and
proper action of $G$ on $TM$ and $\widehat{TM}=TM/G$ is a quotient
manifold. Then we consider the Atiyah algebroid $\widehat{TM}$ over
$\widehat{M}$.

According to Example \ref{Atiyah case}, the basis
$\{\hat{e}_{i},\hat{e}_{B}\}$ induce local coordinates
$(x^{i},y^{i},\bar{y}^{B})$ on $\widehat{TM}$. Given a Lagrangian
function $\ell:\widehat{TM}\to\R$ on the Atiyah algebroid
$\widehat{TM}\to\widehat{M},$ the Euler-Lagrange equations for
$\ell$ are

\begin{align}
\frac{\partial\ell}{\partial x^{j}}-\frac{d}{dt}\left(\frac{\partial\ell}{\partial y^{j}}\right)&=\frac{\partial\ell}{\partial\bar{y}^{A}}\left(\mathcal{B}_{ij}^{A}y^{i}+c_{DB}^{A}\mathcal{A}_{j}^{B}\bar{y}^{B}\right)\quad\forall j,\nonumber\\
\frac{d}{dt}\left(\frac{\partial\ell}{\partial\bar{y}^{B}}\right)&=\frac{\partial\ell}{\partial\bar{y}^{A}}\left(C_{DB}^{A}\bar{y}^{D}-c_{DB}^{A}\mathcal{A}_{i}^{D}y^{i}\right)\quad\forall B,\nonumber
\end{align} which are the Lagrange-Poincar\'e equations associated to a $G$-invariant Lagrangian $L:TM\to\R$ (see \cite{CeMaRa} and \cite{LMM} for example)
where $c_{AB}^{C}$ are the structure constants of the Lie algebra
according to Example \ref{Atiyah case}.
\end{example}

\subsection{Constraint algorithm for presymplectic Lie
algebroids}\label{presymplectic}

In this section we introduce the constraint algorithm for
presymplectic Lie algebroids given in \cite{IMMS} which generalizes
the well-known Gotay-Nester-Hinds algorithm \cite{GoNesHinds78}.
First we give a review of the Gotay-Nester-Hinds algorithm and then
we introduce the construction given in \cite{IMMS} to the case of
Lie algebroids.

\subsubsection{The Gotay-Nester-Hinds algorithm of
constraints}
In this subsection we will briefly review the constraint algorithm of constraints for presymplectic systems (see \cite{GoNe} and \cite{GoNesHinds78}).

Take the following triple $(M, \Omega, H)$ consisting of a smooth
manifold $M$, a closed 2-form $\Omega$ and a differentiable function
$H: M\rightarrow \R$. On $M$ we consider the equation
\begin{equation}\label{pres}
i_X\Omega=dH.
\end{equation}
Since we are not assuming that $\Omega$ is nondegenerate (that is,
$\Omega$ is not, in general, symplectic) then Equation (\ref{pres})
has no solution in general, or the solutions are not defined
everywhere. In the most favorable case, Equation (\ref{pres}) admits
a global (but not necessarily unique) solution $X$. In this case, we
say that the system admits global dynamics. Otherwise, we select the
subset of points of $M$, where such a solution exists. We denote by
$M_2$ this subset and we will assume that it is a submanifold of
$M=M_1$. Then the equations (\ref{pres}) admit a solution $X$ defined
at all points of $M_2$, but $X$ need not be tangent to $M_2$, hence,
does not necessarily induce a dynamics on $M_2$. So we impose an
additional tangency condition, and we obtain a new submanifold $M_3$
along which there exists a solution $X$, but, however, such $X$
needs to be tangent to $M_3$. Continuing this process, we obtain a
sequence of submanifolds
\[
\cdots M_s \hookrightarrow \cdots \hookrightarrow M_2 \hookrightarrow M_1=M
\]
where the general description of $M_{l+1}$ is
\[
M_{l+1}=\{p\in M_{l} \textrm{ such that there exists } X_p\in T_pM_l
\textrm{ satisfying }i_{X_p}\Omega(p)=dH(p) \}.
\]
If the algorithm ends at a final constraint submanifold, in the
sense that at some $s\geq 1$ we have  $M_{s+1}=M_s$. We will denote
this final constraint submanifold by $M_f$. It may still happen that
$\dim M_f=0$, that is, $M_f$ is a discrete set of points, and in
this case the system does not admit a proper dynamics. But, in the
case when $\dim M_f>0$, there exists a well-defined solution $X$ of
(\ref{pres}) along $M_f$.

 There is another characterization of the
submanifolds $M_l$ that we will useful in the sequel. If $N$ is a
submanifold of $M$ then we define
\[
TN^{\perp}=\{Z\in T_pM, \ p\in N \textrm{ such that } \Omega(X,Z)=0 \textrm{ for all } X\in T_pN\}.
\]
Then, at any point $p\in M_l$ there exists $X_p\in T_pM_l$ verifying
$i_X\Omega(p)=dH(p)$ if and only if $\langle TM_l^{\perp},dH
\rangle=0$ (see \cite{GoNe,GoNesHinds78}). Hence, we can
define the $l+1$ step of the constraint algorithm as
\[
M_{l+1}:=\{p\in M_l\textrm{ such that }\langle TM_l^{\perp},dH \rangle(p)=0\} \; .
\]

\subsubsection{Constraint algorithm for presymplectic Lie
algebroids}\label{constraintsalgorithmsection} Let $\tau_{E}:E\to M$
be a Lie algebroid and suppose that $\Omega \in \Gamma(\wedge
^2E^*)$. Then, we can define the vector bundle morphism
$\flat_{\Omega}:E\to E^*$ (over the identity of $M$) as follows
$$\flat_{\Omega}(e)=i(e)\Omega(x),\makebox[1cm]{for}e\in E_x.$$

Now, if $x\in M$ and $F_x$ is a subspace of $E_x$, we may introduce
the vector subspace $F_x^\perp$ of $E_x$ given by
$$F_x^\perp=\{e\in E_x \,|\, \Omega(x)(e,f)=0,\forall f\in E_x\}.$$

On the other hand, if $\flat_{\Omega_x}={\flat_\Omega}_{|E_x}$ it is
easy to prove that
\begin{equation}\label{Subset}
\flat_{\Omega_{x}}(F_{x}) \subseteq (F_{x}^\perp)^{0},
\end{equation}
where $(F_{x}^\perp)^{0}$ is the annihilator of the subspace
$F_{x}^\perp$. Moreover, using \begin{equation}\label{Forortho} dim
F_{x}^\perp = dim E_{x} - dim F_{x} + dim (E_{x}^\perp \cap F_{x}).
\end{equation} we obtain that
\[
dim (F_{x}^\perp)^{0} = dim F_{x} - dim (E_{x}^\perp \cap F_{x}) =
dim (\flat_{\Omega_{x}}(E_{x})).
\]
Thus, from (\ref{Subset}), we deduce that
\begin{equation}\label{flat1}
\flat_{\Omega_x}(F_x)=(F_x^\perp)^\circ.
\end{equation}

Next, we will assume that $\Omega$ is a presymplectic 2-section
($d^E\Omega =0$) and that $\alpha \in \Gamma(E^*)$ is a closed
1-section ($d^E\alpha =0$). Furthermore, we will assume that the
kernel of $\Omega$ is a vector subbundle of $E$.

The dynamics of the presymplectic system defined by $(\Omega,
\alpha)$ is given by a section $X\in \Gamma(E)$ satisfying the
dynamical equation
\begin{equation}\label{presym}
i_X\Omega =\alpha\; .
\end{equation}
In general, a section $X$ satisfying (\ref{presym}) cannot be found
in all points of $E$. First, we look for the points where
(\ref{presym}) has sense. We define
\[
M_1  =\{ x\in M \,|\, \exists e\in E_x:\; i(e)\Omega(x)=\alpha
(x)\}
\]
From (\ref{flat1}), it follows that
\begin{equation}\label{q1}
M_1  =\{ x\in M \,|\, \alpha (x)(e)=0,\makebox[1.5cm]{for all}e\in
\hbox{ker}\Omega(x)=E_x^\perp \}.
\end{equation}

If $M_1$ is an embedded submanifold of $M$, then we deduce that
there exists $X:M_1\to E$ a section of $\tau_{E}:E\to M$ along $M_1$
such that (\ref{presym}) holds. But $\rho (X)$ is not, in general,
tangent to $M_1$. Thus, we have to restrict to $E_1=\rho ^{-1}
(TM_1)$. We remark that, provided that $E_1$ is a manifold and
$\tau_1=\tau_{E}\mid_{E_1}:E_1\to M_1$ is a vector bundle,
$\tau_1:E_1\to M_1$ is a Lie subalgebroid of $E\to M$.

Now, we must consider the subset $M_2$ of $M_1$ defined by
$$\begin{array}{cl}
M_2&=\{ x\in M_1 \,|\, \alpha
(x)\in\flat_{\Omega_x}((E_1)_x)=\flat_{\Omega_x}(\rho^{-1}(T_xM_1))\}\\[5pt]
&=\{ x\in M_1 \,|\, \alpha (x)(e)=0,\, \mbox{ for all } e\in
(E_1)_x^\perp=(\rho ^{-1} (T_xM_1))^\perp \}. \end{array}$$

If $M_2$ is an embedded submanifold of $M_1$, then we deduce that
there exists $X:M_2\to E_1$ a section of $\tau_1:E_1\to M_1$ along
$M_2$ such that (\ref{presym}) holds. However, $\rho(X)$ is not, in
general, tangent to $M_2$. Therefore, we have that to restrict to
$E_2=\rho^{-1}(TM_2)$. As above, if
$\tau_2=\tau_{E}\mid_{E_2}:E_2\to M_2$ is a vector bundle, it
follows that $\tau_2:E_2\to M_2$ is a Lie subalgebroid of
$\tau_1:E_1\to M_1$.

Consequently, if we repeat the process, we obtain a sequence of
Lie subalgebroids (by assumption)
$$
\begin{array}{cccccccccccc}
\ldots&\hookrightarrow&M_{k+1}&\hookrightarrow&M_{k}&\hookrightarrow\ldots \hookrightarrow& M_{2} & \hookrightarrow & M_{1}&\hookrightarrow &M_0=M \\
&& \quad\uparrow \tau_{k+1} && \quad\uparrow\tau_{k} && \quad\uparrow\tau_2 && \quad\uparrow\tau_{1}&&\quad\uparrow\tau_{E} \\
\ldots & \hookrightarrow & E_{k+1} & \hookrightarrow & E_k & \hookrightarrow\ldots\hookrightarrow & E_{2} & \hookrightarrow & E_1& \hookrightarrow & E_0=E \end{array}
$$
where
\begin{equation}\label{qkmas1}
M_{k+1}=\{ x\in M_k \,|\, \alpha(x)(e)=0,\mbox{ for all }
e\in(\rho^{-1}(T_xM_k))^\perp\} \end{equation} and
$$E_{k+1}=\rho^{-1}(TM_{k+1}).$$

If there exists $k\in\mathbb{N}$ such that $M_k=M_{k+1}$, then we say that
the sequence stabilizes. In such a case, there exists a
well-defined (but non necessarily unique) dynamics on the final
constraint submanifold $M_f=M_k$. We write \[
{M}_f=M_{k+1}=M_k,\qquad E_f=E_{k+1}=E_k=\rho ^{-1} (TM_k).
\]

\noindent Then, $\tau_f=\tau_k:E_f=E_k\to M_f=M_k$ is a Lie
subalgebroid of $\tau_E:E\longrightarrow M$ (the Lie algebroid
restriction of $E$ to $E_f$). From the construction of the
constraint algorithm, we deduce that there exists a section $X\in
\Gamma(E_f)$, verifying (\ref{presym}). Moreover, if $X\in
\Gamma(E_f)$ is a solution of the equation (\ref{presym}), then
every arbitrary solution is of the form $X'=X+Y$, where
$Y\in\Gamma(E_f)$ and $Y(x)\in\ker \Omega(x)$, for all $x\in M_f$.
In addition, if we denote by $\Omega_f$ and $\alpha_f$ the
restriction of $\Omega$ and $\alpha$, respectively, to the Lie
algebroid $E_f\longrightarrow M_f$, we have that $\Omega_f$ is a
presymplectic 2-section and then any $X\in \Gamma(E_f)$ verifying
Equation (\ref{presym}) also satisfies
\begin{equation}\label{presym2}
i_X\Omega_f=\alpha_f
\end{equation}
but, in principle, there are solutions of (\ref{presym2}) which are
not solutions of (\ref{presym}) since $\ker \Omega\cap E_f \subset
\ker\Omega_f$.

\begin{remark}
Note that one can generalize the previous procedure to the general
setting of \textit{implicit differential equations on a Lie
algebroid}. More precisely, let $\tau_E:E\to M$ be a Lie algebroid
and $S\subset E$ be a submanifold of $E$ (not necessarily a vector
subbundle). Then, the corresponding sequence of submanifolds of $E$
is
\[
\begin{array}{cl}
S_0&=S\\[5pt]
S_1&=S_0\cap \rho ^{-1} \big (T \tau_E(S_0)\big )\\
\vdots & \\
\\
S_{k+1}&=S_k\cap \rho ^{-1}\big (T \tau_E(S_k)\big )
\\
\vdots &
\end{array}
\]
In our case, $S_k=\rho ^{-1} (TM_k)$ (equivalently,
$M_k=\tau_E(S_k)$).\hfill$\diamond$
\end{remark}
\subsection{Vakonomic mechanics on Lie algebroids}\label{vakoliealgebroids}

In this section we will develop a geometrical description for
second-order mechanics on Lie algebroids in the Skinner and Rusk
formalism, given a general geometric framework for the previous
results in this chapter and using strongly the results given in
\cite{IMMS}.

First, we will review the description of vakonomics mechanics on Lie
algebroids given by Iglesias, Marrero, Mart\'in de Diego and Sosa in
\cite{IMMS}. After it we will introduce the notion of admissible
elements on a Lie algebroid  and we will particularize the previous
construction to the case when the Lie algebroid is the prolongation
of a Lie algebroid and the constraint
submanifold is the set of admissible elements. Then we will obtain
the second-order Skinner and Rusk formulation on Lie algebroids.

 Let $\tau_{\widetilde{E}}
:\widetilde{E}\to Q$ be a Lie algebroid of rank $n$ over a manifold
$Q$ of dimension $m$ with anchor map $\rho:\widetilde{E}\to TQ$ and
$L:\widetilde{E}\to \R$ be a Lagrangian function on $\widetilde{E}$.
Moreover, let $\mathcal{M}\subset \widetilde{E}$ be an embedded
submanifold of dimension $n+m-\bar{m}$ such that $\tau
_\mathcal{M}=\tau_{\widetilde{E}}\big{|}_{\mathcal{M}}:\mathcal{M}\to
Q$ is a surjective submersion.

Suppose that $e$ is a point of $\mathcal{M}$ with
$\tau_\mathcal{M}(e)=x\in Q$, $(x^i)$ are local coordinates on
an open subset $U$ of $Q$, $x\in U$, and $\{e_A\}$ is a local
basis of $\Gamma(\widetilde{E})$ on $U$. Denote by $(x^i,y^A)$ the
corresponding local coordinates for $\widetilde{E}$ on the open
subset $\tau^{-1}_{\widetilde{E}}(U)$. Assume that
$$\mathcal{M}\cap\tau^{-1}_{\widetilde{E}}(U)\equiv\{(x^i,y^A)\in\tau^{-1}_{\widetilde{E}}(U) \,|\,
\Phi^\alpha(x^i,y^A)=0,\;\alpha=1,\dots,\bar{m}\}$$ where
$\Phi^\alpha$ are the local independent constraint functions for the
submanifold $\mathcal{M}$.

We will suppose, without loss of generality, that the
$(\bar{m}\times n)$-matrix
$$\Big(\displaystyle{\frac{\partial\Phi^\alpha}{\partial
y^B}}\bigg{|}_{e}\Big)_{\alpha=1,\dots,\bar{m};B=1,\dots,n}$$
is of maximal rank.

Now, using the implicit function theorem, we obtain that there exists
an open subset $\widetilde{V}$ of $(\tau_{\widetilde{E}})^{-1}(U)$,
an open subset $W\subseteq\R^{m+n-\bar{m}}$ and smooth real
functions $\Psi^\alpha:W\to\R,\;\;\alpha=1,\dots,\bar{m},$ such that
$$\mathcal{M}\cap\widetilde{V} \equiv \{(x^i,y^A)\in\widetilde{V} \,|\,
y^\alpha=\Psi^\alpha(x^i,y^a),\hbox{ with }\;\alpha=1,\dots,\bar{m} \hbox{ and }\bar{m}+1\leq a\leq n\}.$$
Consequently, $(x^i,y^a)$ are local coordinates on $\mathcal{M}$ and we
will denote by $\tilde{L}$ the restriction of $L$ to $\mathcal{M}$.

Consider the Whitney sum of ${\widetilde{E}}^*$ and $\widetilde{E}$,
that is, $W=\widetilde{E}\oplus \widetilde{E}^{*}$, and the
canonical projections $pr_1: \widetilde{E}\oplus \widetilde{E}^{*}
\longrightarrow \widetilde{E}$ and $pr_2: \widetilde{E}\oplus
\widetilde{E}^{*} \longrightarrow \widetilde{E}^{*}$. Now, let $W_0$
be the submanifold $W_0=pr_1^{-1} (\mathcal{M})=\mathcal{M}\times _Q
\widetilde{E}^{*}$ and the restrictions $\pi_1={pr_1}|_{W_0}$ and
$\pi_2={pr_2}|_{W_0}$. Also denote by $\nu: W_0\longrightarrow Q$
the canonical projection of $W_0$ over the base manifold.

Next, we consider the prolongation of the Lie algebroid
$\widetilde{E}$ over $\tau_{\widetilde{E}^{*}}:\widetilde{E}^{*}\to
Q$ (res\-pectively, $\nu :W_0\to Q$). We will denote this Lie
algebroid by ${\mathcal T}^{\tau_{\widetilde{E}^*}}\widetilde{E}$
(respectively, ${\mathcal T}^{\nu}\widetilde{E}$). Moreover, we can
prolong $\pi _2:W_0\to \widetilde{E}^*$ to a morphism of Lie
algebroids ${\mathcal T}\pi_2 :{\mathcal T}^{\nu}\widetilde{E}\to
{\mathcal T}^{\tau_{\widetilde{E}
^{*}}}\widetilde{E}$ defined by
$\mathcal{T}\pi_2=(Id,T\pi_2)$.

If $(x^i,p_A)$ are the local coordinates on $\widetilde{E}^*$
associated with the local basis $\{e^A\}$ of
$\Gamma(\widetilde{E}^{*})$, then $(x^i,p_A,y^a)$ are
local coordinates on $W_0$ and we may consider the local basis
$\{\widetilde{e}_{A}^{(1)} ,(\widetilde{e}^{A})^{(2)} ,e_a^{(2)}\}$
of $\Gamma({\mathcal T}^{\nu}\widetilde{E})$ defined by
%\begin{equation}\label{tilbar3}
\begin{align} \widetilde{e}_{A}^{(1)}(\check{e},e^*)&=\left(e_A(x),
\rho_A^i\displaystyle\frac{\partial
}{\partial x^i}\bigg{|}_{(\check{e},e^*)}\right),\quad
(\widetilde{e}^{A})^{(2)}\left(\check{e},e^*\right)=\left(0,\displaystyle\frac{\partial
}{\partial
p_{A}}\bigg{|}_{(\check{e},e^*)}\right),\nonumber\\
e_a^{(2)}\left(\check{e},e^*\right)&=\left(0,\displaystyle\frac{\partial
}{\partial y^{a}}\bigg{|}_{(\check{e},e^{*})}\right),\nonumber\end{align} where $(\check{e},e^*)\in W_0$ and $\nu(\check{e},e^*)=x$. If
$(\lcf \cdot , \cdot \rcf ^{\nu}, \rho^{\nu})$ is the Lie algebroid
structure on ${\mathcal T}^{\nu}\widetilde{E}$, we have that
\[
\lcf \widetilde{e}_{A}^{(1)}, \widetilde{e}_{B}^{(1)} \rcf^{\nu} = {\mathcal
C}_{AB}^{C} \widetilde{e}_{C}^{(1)},
\]
and the rest of the fundamental Lie brackets are zero. Moreover,
\[
\rho^{\nu}(\widetilde{e}_{A}^{(1)}) = \rho_{A}^{i} \displaystyle
\frac{\partial}{\partial x^{i}}, \; \; \rho^{\nu}((\widetilde{e}^{A})^{(2)}) = \displaystyle \frac{\partial}{\partial p_{A}}, \; \;
\rho^{\nu}(e_{a}^{(2)}) = \displaystyle
\frac{\partial}{\partial y^{a}}.
\]

The Pontryagin Hamiltonian $H_{W_0}$ is a function defined on $W_0=
\mathcal{M}\times_Q\widetilde{E}^*$ given by
\[
H_{W_0} (\check{e},e^*)= \langle e^* , \check{e}\rangle
-\tilde{L}(\check{e}),
\]
or, in local coordinates,
\begin{equation}\label{H0}
H_{W_0}(x^i, p_A, y^a)=p_ay^a+p_{\alpha}\Psi^{\alpha}(x^i,
y^a)-\tilde{L}(x^i, y^a)\, .
\end{equation}

Moreover, one can consider the presymplectic 2-section $\Omega
_0=({\mathcal T}\pi_2, \pi_2)^*\Omega_{\widetilde{E}}$, where
$\Omega_{\widetilde{E}}$ is the canonical symplectic section on
${\mathcal T}^{\tau_{\widetilde{E}^{*}}}\widetilde{E}$ defined in
Equation (\ref{2section}). In local coordinates,
\begin{equation}\label{Omega0}
\Omega_0=\widetilde{e}^{A}_{(1)}\wedge \widetilde{e}_A^{(2)} + \frac{1}{2} {\mathcal
C}_{AB}^C p_C \widetilde{e}^A_{(1)}\wedge \widetilde{e}^B_{(1)},
\end{equation} where $\{\widetilde{e}^{A}_{(1)},\widetilde{e}_{A}^{(2)},e^{a}_{(2)}\}$ denotes the dual basis of $\{\widetilde{e}_{A}^{(1)},(\widetilde{e}^{A})^{(2)},e_{a}^{(2)}\}$ .

Therefore, we have the triple $({\mathcal
T}^{\nu}\widetilde{E},\Omega _0,d^{{\mathcal
T}^{\nu}\widetilde{E}}H_{W_0})$ as a presymplectic hamiltonian
system.

\begin{definition}
The vakonomic problem on Lie algebroids consists on finding the
solutions for the equation
\begin{equation}\label{Hamilt}
i_{X}\Omega _0 = d^{{\mathcal T}^{\nu}\widetilde{E}} H_{W_0};
\end{equation}
that is, to solve the constraint algorithm for $({\mathcal
T}^{\nu}\widetilde{E},\Omega _0,d^{{\mathcal T}^{\nu}\widetilde{E}}
H_{W_0})$.
\end{definition}

In local coordinates, we have that
\[
d^{{\mathcal T}^{\nu}\widetilde{E}}H_{W_0} = \left(p_\alpha\frac{\partial \Psi
^\alpha}{\partial x^i} -\frac{\partial \tilde{L}}{\partial
x^i}\right)\rho ^i_A \widetilde{e}^A_{(1)} + \Psi ^\alpha \widetilde{e}_\alpha^{(2)} +y^a \widetilde{e}_a^{(2)} +
\left(p_a+p_\alpha \frac{\partial \Psi ^\alpha}{\partial
y^a}-\frac{\partial \tilde{L}}{\partial y^a}\right)e^a_{(2)}.
\]

If we apply the constraint algorithm,
\[
W_1=\{ w\in \mathcal{M}\times _Q \widetilde{E}^* \, | \, d^{{\mathcal T}^{\nu}\widetilde{E}}
H_{W_0}(w)(Y)=0,\quad \forall Y\in \hbox{ker }\,\Omega _0(w)\}.
\]
Since $\hbox{ker}\,\Omega _0= \hbox{span }\{ e_a^{(2)} \}$, we get
that $W_1$ is locally characterized by the equations
\[
\varphi _a = d^{{\mathcal T}^{\nu}\widetilde{E}} H_{W_0} (e_a^{(2)})=p_a+p_\alpha \frac{\partial \Psi ^\alpha}{\partial
y^a}-\frac{\partial \tilde{L}}{\partial y^a}=0,
\]
or
\[
p_a=\frac{\partial \tilde{L}}{\partial y^a}-p_\alpha
\frac{\partial \Psi ^\alpha}{\partial y^a}, \quad \bar{m}+1\leq a
\leq n .
\]
Let us also look for the expression of $X$ satisfying Eq.
(\ref{Hamilt}). A direct computation shows that
\[
X= y^a\widetilde{e}^{(1)}_a + \Psi ^\alpha \widetilde{e}^{(1)}_\alpha + \Big [
\Big ( \frac{\partial \tilde{L}}{\partial x^i}-p_\alpha
\frac{\partial \Psi ^\alpha}{\partial x^i} \Big ) \rho ^i_A
-y^a{\mathcal C}^B_{Aa}p_B-\Psi ^\alpha {\mathcal
C}^B_{A\alpha}p_B \Big ](\widetilde{e}^A)^{(2)} + \Upsilon ^ae^{(2)}_a.
\]
Therefore, the vakonomic equations are
%\begin{equation}\label{vak-eq}
$$\left \{
\begin{array}{l}
\displaystyle \dot{x}^i=y^a \rho ^i_a+\Psi ^\alpha \rho ^i_\alpha ,\\[10pt]
\displaystyle \dot{p}_{\alpha}=\Big ( \frac{\partial
\tilde{L}}{\partial x^i}-p_\beta \frac{\partial \Psi
^\beta}{\partial x^i} \Big ) \rho ^i_{\alpha} -y^a {\mathcal
C}^B_{\alpha a}p_B-\Psi ^\beta {\mathcal
C}^B_{\alpha\beta}p_B ,\\[10pt]
\displaystyle\frac{d}{dt}\left(\frac{\partial\widetilde{L}}{\partial y^{a}}-\rho_{\alpha}\frac{\partial\Psi^{\alpha}}{\partial y^{a}}\right)=
\Big ( \frac{\partial \tilde{L}}{\partial x^i}-p_\alpha
\frac{\partial \Psi ^\alpha}{\partial x^i} \Big ) \rho ^i_{a} -y^b
{\mathcal
C}^B_{a b}p_B-\Psi ^\alpha {\mathcal C}^B_{a\alpha}p_B.\\[10pt]
\end{array}
\right .$$
%\end{equation}
%\begin{remark}
%Note that the vakonomic equations can be obtained following a
%constrained variational principle (see \cite{IMMS}).
%\end{remark}

Of course, we know that there exist sections $X$ of ${\mathcal
T}^{\nu}\widetilde{E}$ along $W_1$ satisfying (\ref{Hamilt}), but
they may not be sections of $(\rho^\nu)^{-1}(TW_1)={\mathcal
T}^{\nu_1}\widetilde{E}$, in general (here $\nu_{1}:W_{1}\to Q$).
Then, following the procedure detailed in Section
\ref{constraintsalgorithmsection}, we obtain a sequence of embedded
submanifolds \[
\ldots \hookrightarrow W_{k+1}\hookrightarrow W_{k}\hookrightarrow
\ldots \hookrightarrow W_{2}\hookrightarrow W_{1}\hookrightarrow
W_{0}=\mathcal{M}\times_Q\widetilde{E}^*.
\] If the algorithm stabilizes, then we find a final constraint
submanifold $W_f$ on which at least a section $X\in\Gamma( {\mathcal
T}^{\nu_{f}}E)$ verifies
$$(i_{X}\Omega _0 = d^{{\mathcal T}^{\nu}\widetilde{E}}
H_{W_0})\big{|}_{W_f}$$ where $\nu_{f}:W_{f}\to Q$.

One of the most important cases is when $W_f=W_1$. The authors of
\cite{IMMS} have analyzed this case with the following result:
Consider the restriction $\Omega_{1}$ of $\Omega_{0}$ to ${\mathcal
T}^{\nu_{1}}\widetilde{E}$;
\begin{proposition}\label{prop:charact-sympl}
$\Omega_{1}$ is a symplectic section of the Lie algebroid ${\mathcal
T}^{\nu_1}\widetilde{E}$ if and only if for any system of
coordinates $(x^i, p_A, y^a)$ on $W_0$ we have that
\[
\det\left( \frac{\partial^2 \tilde{L}}{\partial y^a \partial
y^b}-p_{\alpha}\frac{\partial^2 {\Psi^{\alpha}}}{\partial y^a
\partial y^b}\right)\not=0,\mbox{ for all point in }W_1.
\]
\end{proposition}
\subsection{Second-order variational problems on Lie algebroids}

In this section we will study second-order variational problems on Lie algebroid. First we introduce the geometric object for the formalism and then we study second-order unconstrained variation problems. After that, we will analyze the constrained case.
\subsubsection{Prolongation of a Lie algebroid over a smooth map
(cont'd)} This subsection is devoted to study some additional properties and
characterizations about the prolongation of a Lie algebroid over a
smooth map (see subsection
\ref{prolongationliealgebroidsubsection}).

%\subsubsection{Admissible elements}

Let $\widetilde{E}$ be a Lie algebroid over $Q$ with fiber bundle
projection $\tau_{\widetilde{E}}:\widetilde{E}\to Q$ and anchor map
$\rho:\widetilde{E}\to TQ.$ Also, let $\tau_E:E\to M$ be a Lie algebroid with anchor map $\rho:E\to TM$
and let $\mathcal{T}^{\tau_{E}}E$ be the $E-$tangent bundle to $E$. Now we
will define the bundle
$\mathcal{T}^{\tau_{E}^{(1)}}(\mathcal{T}^{\tau_{E}}E)$ over
$\mathcal{T}^{\tau_{E}}E$. This bundle plays the
role of $\tau_{T(TM)}:T(TTM)\to T(TM)$ in ordinary Lagrangian
Mechanics.

%\begin{lemma}
%Let us consider the Lie algebroid $E$ over a manifold $Q$ with fiber
%bundle projection $\tau_{E}^{(0)}:E\to Q$. We denote by
%$\mathcal{T}^{\tau_{E}^{(0)}}E$ the prolongation Lie algebroid over
%$E$. If we denote by $\tau_{E}^{(1,0)}$ the fiber bundle projection
%of $\mathcal{T}^{\tau_{E}^{(1)}}E$ over $Q$, then
%$$\mathcal{T}^{\tau_{E}^{(1,0)}}E=\mathcal{T}^{\tau_{E}^{(1)}}\left(\mathcal{T}^{\tau_{E}^{(0)}}E\right).$$
%\end{lemma}

In what follows we will describe the Lie algebroid structure of the
$E$-tangent bundle to the prolongation Lie algebroid over
$\tau_{E}:E\to Q$.

As we know from subsection
\eqref{prolongationliealgebroidsubsection}, the basis of sections
$\{e_{A}\}$ of $E$ induces a local basis of the sections of
$\mathcal{T}^{\tau_{E}}E$ given by
$$e_{A}^{(1)}(e)=\left(e,e_A(\tau_{E}(e)),\rho_{A}^{i}\frac{\partial}{\partial x^{i}}\Big{|}_{e}\right),\quad e^{(2)}_A(e)=\left(e,0,\frac{\partial}{\partial
y^{A}}\Big{|}_{e}\right),$$ for $e\in E$. From this basis we can
induce local coordinates $(x^{i},y^{A};z^{A},v^{A})$ on
$\mathcal{T}^{\tau_{E}}E$. Now, from this basis, we can induce a
local basis of sections of
$\mathcal{T}^{\tau_{E}^{(1)}}(\mathcal{T}^{\tau_{E}}E)$ in the
following way: consider an element
$(e,v_b)\in\mathcal{T}^{\tau_E}E$, then define the components of the
basis $\{e_{A}^{(1,1)},e_{A}^{(2,1)},e_{A}^{(1,2)},e_{A}^{(2,2)}\}$
as \begin{align}
e_{A}^{(1,1)}(e,v_b)&=\left((e,v_b),e_A^{(1)}(e),\rho^{i}_{A}\frac{\partial}{\partial x^{i}}\Big{|}_{(e,v_b)}\right),\quad
e_{A}^{(2,1)}(e,v_b)=\left((e,v_{b}),e_{A}^{(2)}(e),\frac{\partial}{\partial y^{A}}\Big{|}_{(e,v_b)}\right),\nonumber\\
e_{A}^{(1,2)}(e,v_b)&=\left((e,v_b),0,\frac{\partial}{\partial z^{A}}\Big{|}_{(e,v_b)}\right),\qquad e_{A}^{(2,2)}(e,v_b)=\left((e,v_b),0,\frac{\partial}{\partial v^{A}}\Big{|}_{(e,v_b)}\right).\nonumber
\end{align}
The basis
$\{e_{A}^{(1,1)},e_{A}^{(2,1)},e_{A}^{(1,2)},e_{A}^{(2,2)}\}$ induces
local coordinates $(x^{i},y^{A},z^{A},v^{A},b^{A},c^{A},d^{A},w^{A})$ on
$\mathcal{T}^{\tau_{E}^{(1)}}\left(\mathcal{T}^{\tau_{E}}E\right)$.
If we denote by
$(\mathcal{T}^{\tau_{E}^{(1)}}(\mathcal{T}^{\tau_{E}}E),\lcf\cdot,\cdot\rcf_{\tau_{{E}}^{(2)}},\rho_2)$
the Lie algebroid structure of the fiber bundle 
$\mathcal{T}^{\tau_{E}^{(1)}}(\mathcal{T}^{\tau_{E}}E)$, it is
characterized by \begin{align}
\rho_{2}(e_A^{(1,1)})(e,v_{b})&=\left((e,v_b),\rho_A^i\displaystyle\frac{\partial
}{\partial x^i}\bigg{|}_{(e,v_b)}\right),\quad
\rho_2(e_{A}^{(2,1)})(e,v_b)=\left((e,v_b),\displaystyle\frac{\partial}{\partial y^{A}}\bigg{|}_{(e,v_b)}\right),\nonumber\\
\rho_2(e_A^{(1,2)})(e,v_b)&=\left((e,v_b),\displaystyle\frac{\partial
}{\partial z^{A}}\bigg{|}_{(e,v_b)}\right),\quad
\rho_2(e_{A}^{(2,2)})(e,v_b)=\left((e,v_b),\displaystyle\frac{\partial}{\partial v^{A}}\bigg{|}_{(e,v_b)}\right),\nonumber\\
\lcf e_{A}^{(1,1)},e_{B}^{(1,1)}\rcf_{\tau_{E}^{(2)}}&=\mathcal{C}_{AB}^{C}{e}_{C}^{(1,1)},\nonumber\\
\lcf e_{A}^{(1,1)},e_{B}^{(1,2)}\rcf_{\tau_{E}^{(2)}}&=\lcf e_A^{(1,2)},e_{B}^{(1,2)}\rcf_{\tau_{E}^{(2)}}=0,\nonumber\\
\lcf e_{A}^{(1,1)},e_{A}^{(2,2)}\rcf_{\tau_{E}^{(2)}}&= \lcf e_{A}^{(2,1)},e_{A}^{(2,2)}\rcf_{\tau_{E}^{(2)}}=\lcf e_A^{(1,2)},
e_{A}^{(2,1)}\rcf_{\tau_{E}^{(2,1)}}=\lcf e_{A}^{(1,1)},e_{B}^{(2,1)}\rcf_{\tau_{E}^{(2)}}=0.\nonumber
\end{align} for all $A,B$ and $C$ where
$\mathcal{C}_{AB}^{C}$ are the structure constants of $E$.

In the same way, from the basis
$\{\widetilde{e}_{A}^{(1)},(\widetilde{e}^{A})^{(2)}\}$ of sections
of $\mathcal{T}^{\tau_{E^{*}}}E$ given by
$$\widetilde{e}_{A}^{(1)}(e^{*})=\left(e^{*},e_{A}(\tau_{E^{*}}(e^{*})),\rho_{A}^{i}\frac{\partial}{\partial x^{i}}\bigg{|}_{e^{*}}\right),\quad
(\widetilde{e}^{A})^{(2)}(e^{*})=\left(e^{*},0,\frac{\partial}{\partial p_{A}}\bigg{|}_{e^{*}}\right),$$ where $e^{*}\in E$, we construct the set $\{\widetilde{e}_{A}^{(1,1)},(\widetilde{e}^{A})^{(2,1)},\widetilde{e}_{A}^{(1,2)},(\widetilde{e}^{A})^{(2,2)}\}$, the basis of sections of
$\mathcal{T}^{\tau_{(\mathcal{T}^{\tau_{E}}E)^{*}}}\mathcal{T}^{\tau_{E}}E$.
In what follows $(x^{i},y^A,p_A,\bar{p}_A)$ denotes local
coordinates on $\mathcal{T}^{\tau_{E^{*}}}E$ induced by the basis
$\{\widetilde{e}_{A}^{(1)},(\widetilde{e}^{A})^{(2)}\}$.
This basis is given by \begin{align}
\widetilde{e}_{A}^{(1,1)}(\alpha^{*})&=\left(\alpha^{*},e_{A}^{(1)}(\tau_{(\mathcal{T}^{\tau_{E}}E)^{*}}(\alpha^{*})),\rho_{A}^{i}\frac{\partial}{\partial x^{i}}\Big{|}_{\alpha^{*}}\right),\quad (\widetilde{e}^A)^{(1,2)}(\alpha^{*})=\left(\alpha^{*},0,\frac{\partial}{\partial p_{A}}\Big{|}_{\alpha^{*}}\right)\nonumber\\
(\widetilde{e}_A)^{(2,1)}(\alpha^{*})&=\left(\alpha^{*},e_{A}^{(2)}(\tau_{(\mathcal{T}^{\tau_{E}}E)^{*}}(\alpha^{*})),\frac{\partial}{\partial y^{A}}\Big{|}_{\alpha^{*}}\right),\quad
(\widetilde{e}^A)^{(2,2)}(\alpha^{*})=\left(\alpha^{*},0,\frac{\partial}{\partial \bar{p}_{A}}\Big{|}_{\alpha^{*}}\right).\nonumber
\end{align} where $\alpha^{*}\in (\mathcal{T}^{\tau_{E}}E)^{*}$ and $\tau_{(\mathcal{T}^{\tau_{E}}E)^{*}}:(\mathcal{T}^{\tau_{E}}E)^{*}\to E$ is the vector bundle projection.

The Lie algebroid structure
$(\mathcal{T}^{\tau_{(\mathcal{T}^{\tau_{E}}E)^{*}}}(\mathcal{T}^{\tau_{E}}E);
\lcf\cdot,\cdot\rcf_2, {\rho}_2)$ is given by
\begin{align}
{\rho}_2(\widetilde{e}_{A}^{(1,1)}(\alpha^{*}))&=\left(\alpha^{*},\rho_{A}^{i}\frac{\partial}{\partial x^{i}}\bigg{|}_{\alpha^{*}}\right),\;\;\; {\rho}_2((\widetilde{e}^{A})^{(2,1)}(\alpha^{*}))=\left(\alpha^{*},\frac{\partial}{\partial y^{A}}\bigg{|}_{\alpha^{*}}\right),\nonumber\\
{\rho}_2(\widetilde{e}_{A}^{(1,2)}(\alpha^{*}))&=\left(\alpha^{*},\frac{\partial}{\partial p_{A}}\bigg{|}_{\alpha^{*}}\right),\;\;\; {\rho}_2((\widetilde{e}^{A})^{(2,2)}(\alpha^{*}))=\left(\alpha^{*},\frac{\partial}{\partial \bar{p}_{A}}\bigg{|}_{\alpha^{*}}\right),\nonumber
\end{align}
where the unique non-zero Lie bracket is
$\lcf\widetilde{e}_{A}^{(1,1)},\widetilde{e}_{B}^{(1,1)}\rcf_{2}=\mathcal{C}_{AB}^{C}\widetilde{e}_{C}^{(1,1)}$.
This basis induces local coordinates
$(x^{i},y^{A},p_{A},\bar{p}_{A}, q^{A},\bar{q}^{A};
l_{A},\bar{l}_{A})$ on $\mathcal{T}^{\tau_{(\mathcal{T}^{\tau_{E}}E)^{*}}}\mathcal{T}^{\tau_{E}}E$.

\subsubsection{Second-order unconstrained problem on Lie
algebroids}\label{seconorderunconstrained} 
Next, we will study second-order problem on Lie
algebroids. Consider the Whitney sum
of $\left(\mathcal{T}^{\tau_{E}}E\right)^{*}$ and
$\mathcal{T}^{\tau_{E}}E$,
$W=\mathcal{T}^{\tau_{E}}E\times_{E}\left(\mathcal{T}^{\tau_{E}}E\right)^{*}
$ and its canonical projections $pr_1:W\to \mathcal{T}^{\tau_{E}}E$
and $pr_2:W\to\left(\mathcal{T}^{\tau_{E}}E\right)^{*}$. Now, let
$W_0$ be the submanifold
$W_0=pr_{1}^{-1}(E^{(2)})=E^{(2)}\times_{E}\left(\mathcal{T}^{\tau_{E}}E\right)^{*}$
and the restrictions $\pi_{1}=pr_1\mid_{W_0}$ and
$\pi_2=pr_2\mid_{W_0}$. Also we denote by $\nu:W_0\to E$ the
canonical projection. The diagram in Figure
\ref{figuraSKRKLiealgebroid} illustrates the situation.
\begin{figure}[h]
$$\xymatrix{
  &&W_0=  E^{(2)}\times_{E}\left(\mathcal{T}^{\tau_{E^{*}}}E\right)^{*}
 \ar[lld]_{\pi_1} \qquad\qquad\ar[dd]^{\nu} \ar[rrd]^{\pi_2}&&\\
  E^{(2)}\ar[rrd]^{\tau_{E}^{(2,1)}} && && \left(\mathcal{T}^{\tau_{E}}E\right)^{*}  \ar[lld]^{\tau_{(\mathcal{T}^{\tau_{E}}E)^{*}}}\\
  && E &&
   }$$
   \caption{Second order Skinner and Rusk formalism on Lie algebroids}\label{figuraSKRKLiealgebroid}
\end{figure}

Consider the prolongations of $\mathcal{T}^{\tau_{E}}E$ by
$\tau_{(\mathcal{T}^{\tau_{E}}E)^{*}}$ and by $\nu$, respectively.
We will denote these Lie algebroids by
$\mathcal{T}^{\tau_{(\mathcal{T}^{\tau_{E}}E)^{*}}}(\mathcal{T}^{\tau_{E}}E)$
and $\mathcal{T}^{\nu}\mathcal{T}^{\tau_{E}}E$ respectively.
Moreover, we can prolong
$\pi_{2}:W_0\to(\mathcal{T}^{\tau_{E}}E)^{*}$ to a morphism of Lie
algebroids
$\mathcal{T}\pi_{2}:\mathcal{T}^{\nu}\mathcal{T}^{\tau_{E}}E\to\mathcal{T}^{\tau_{(\mathcal{T}^{\tau_{E}}E)^{*}}}(\mathcal{T}^{\tau_{E^{*}}}E)$
defined by $\mathcal{T}\pi_2=(Id,T\pi_2)$.

We denote by $(x^{i},y^{A},p_A,\bar{p}_A)$ local coordinates on
$(\mathcal{T}^{\tau_{E}}E)^{*}$ induced by
$\{e^{A}_{(1)},e^{A}_{(2)}\}$, the dual basis of the basis
$\{e_{A}^{(1)},e_{A}^{(2)}\}$, a basis of $\mathcal{T}^{\tau_{E}}E$.
Then, $(x^{i},y^{A},p_{A},\bar{p}_{A},z^{A})$ are local coordinates
in $W_0$ and we may consider $\{\widetilde{e}^{(1,1)}_{A},\widetilde{e}^{(2,1)}_{A},(\widetilde{e}^{A})^{(1,2)},(\widetilde{e}^{A})^{(2,2)},\check{e}^{(1,2)}_{A}\}$, the local basis of $\Gamma(\mathcal{T}^{\nu}\mathcal{T}^{\tau_{E}}E)$ defined as \begin{align}
&\widetilde{e}_{A}^{(1,1)}(\check{\alpha},\alpha^{*})=\left((\check{\alpha},\alpha^{*}),e_{A}^{(1)}(\tau_{(\mathcal{T}^{\tau_{E}}E)^{*}}(\alpha^{*})),\rho_{A}^{i}\frac{\partial}{\partial x^{i}}\Big{|}_{(\check{\alpha},\alpha^{*})}\right),\nonumber\\
&(\widetilde{e}^A)^{(1,2)}(\check{\alpha},\alpha^{*})=\left((\check{\alpha},\alpha^{*}),0,\frac{\partial}{\partial p_{A}}\Big{|}_{(\check{\alpha},\alpha^{*})}\right),\nonumber\\
&\widetilde{e}_A^{(2,1)}(\check{\alpha},\alpha^{*})=\left((\check{\alpha},\alpha^{*}),e_{A}^{(2)}(\tau_{(\mathcal{T}^{\tau_{E}}E)^{*}}(\alpha^{*})),\frac{\partial}{\partial y^{A}}\Big{|}_{(\check{\alpha},\alpha^{*})}\right),\nonumber\\
&(\widetilde{e}^A)^{(2,2)}(\check{\alpha},\alpha^{*})=\left((\check{\alpha},\alpha^{*}),0,\frac{\partial}{\partial \bar{p}_{A}}\Big{|}_{(\check{\alpha},\alpha^{*})}\right),\quad
\check{e}_{A}^{(1,2)}(\check{\alpha},\alpha^{*})=\left((\check{\alpha},\alpha^{*}),0,\frac{\partial}{\partial z^{A}}\Big{|}_{(\check{\alpha},\alpha^{*})}\right)\nonumber
\end{align} for $\alpha^{*}\in (\mathcal{T}^{\tau_{E}}E)^{*}$, $\check{\alpha}\in E^{(2)}$, $(\check{\alpha},\alpha^{*})\in W_0$, and $\tau_{(\mathcal{T}^{\tau_{E}}E)^{*}}:(\mathcal{T}^{\tau_{E}}E)^{*}\to E$ is the canonical projection.

If $(\lcf\cdot,\cdot\rcf^{\nu},\rho^{\nu})$ is the Lie algebroid
structure on $\mathcal{T}^{\nu}\mathcal{T}^{\tau_{E}}E$, we have that
$\displaystyle{\lcf\widetilde{e}_A^{(1,1)},\widetilde{e}_B^{(1,1)}\rcf ^{\nu}=\mathcal{C}_{AB}^{C}\widetilde{e}_{C}^{(1,1)},}$
and the rest of the fundamental Lie brackets are zero. Moreover,
\begin{align}
\rho^{\nu}(\widetilde{e}_{A}^{(1,1)}(\check{\alpha},\alpha^{*}))&=\left((\check{\alpha},\alpha^{*}),\rho_A^{i}\frac{\partial}{\partial
x^{i}}\bigg{|}_{(\check{\alpha},\alpha^{*})}\right),\nonumber\\
\rho^{\nu}((\widetilde{e}^{A})^{(1,2)}(\check{\alpha},\alpha^{*}))&=\left((\check{\alpha},\alpha^{*}),\frac{\partial}{\partial p_{A}}\bigg{|}_{(\check{\alpha},\alpha^{*})}\right),\quad
\rho^{\nu}(\check{e}_{A}^{(1,2)}(\check{\alpha},\alpha^{*}))=\left((\check{\alpha},\alpha^{*}),\frac{\partial}{\partial z^{A}}\bigg{|}_{(\check{\alpha},\alpha^{*})}\right),\nonumber\\
\rho^{\nu}(\widetilde{e}_{A}^{(2,1)}(\check{\alpha},\alpha^{*}))&=\left((\check{\alpha},\alpha^{*}),\frac{\partial}{\partial y^{A}}\bigg{|}_{(\check{\alpha},\alpha^{*})}\right),\quad
\rho^{\nu}((\widetilde{e}^{A})^{(2,2)}(\check{\alpha},\alpha^{*}))=\left((\check{\alpha},\alpha^{*}),\frac{\partial}{\partial \bar{p}_{A}}\bigg{|}_{(\check{\alpha},\alpha^{*})}\right).\nonumber
\end{align}

The Pontryagin Hamiltonian $H_{W_0}$ is a function in $W_0$ given by
$$H_{W_0}(\check{\alpha},\alpha^{*})=\langle\alpha^{*},\check{\alpha}\rangle-L(\check{\alpha}),$$
or in local coordinates
$$H_{W_0}(x^{i},y^{A},p_{A},\overline{p}_{A},z^{A})=\overline{p}_{A}z^{A}+p_{A}y^{A}-L(x^{i},y^{A},z^{A}).$$

Moreover, one can consider the presymplectic $2$-section
$\Omega_{0}=(\mathcal{T}\pi_{2},\pi_{2})^{*}\Omega_{E}$, where
$\Omega_{E}$ is the canonical symplectic section on
$\mathcal{T}^{\tau_{E^{*}}}E$. In local coordinates,
$$\Omega_0=\widetilde{e}_{(1,1)}^{A}\wedge(\widetilde{e}_{A})^{(1,2)}+\widetilde{e}^{A}_{(2,1)}\wedge(\widetilde{e}_{A})^{(2,2)}+\frac{1}{2}\widetilde{\mathcal{C}}_{AB}^{C}p_{C}\widetilde{e}_{(1,1)}^{A}\wedge\widetilde{e}_{(1,1)}^{B}.$$ Here, the basis $\{\widetilde{e}_{(1,1)}^{A},
\widetilde{e}^{A}_{(2,1)},(\widetilde{e}_{A})^{(1,2)},(\widetilde{e}_{A})^{(2,2)},\check{e}^{A}_{(1,2)}\}$
denotes the dual basis of the basis of sections for
$\mathcal{T}^{\tau_{(\mathcal{T}^{\tau_{E}}E)^{*}}}\mathcal{T}^{\tau_{E}}E$, denoted by  
$\{\widetilde{e}^{(1,1)}_{A},\widetilde{e}^{(2,1)}_{A},(\widetilde{e}^{A})^{(1,2)},(\widetilde{e}^{A})^{(2,2)},\check{e}^{(1,2)}_{A}\}$.

Therefore, the triple
$\left(\mathcal{T}^{\nu}\mathcal{T}^{\tau_{E}}E,\Omega_0,d^{\mathcal{T}^{\nu}\mathcal{T}^{\tau_{E}}E}H_{W_0}\right)$
is a presymplectic Hamiltonian system.

The second-order problem on the Lie algebroid $\tau_{E}:E\to M$
consists on finding the solutions of the equation
$$i_{X}\Omega_0=d^{\mathcal{T}^{\nu}\mathcal{T}^{\tau_{E}}E}H_{W_0},$$ that is, to solve
the constraint algorithm for
$\left(\mathcal{T}^{\nu}\mathcal{T}^{\tau_{E}}E,\Omega_0,d^{\mathcal{T}^{\nu}\mathcal{T}^{\tau_{E}}E}H_{W_0}\right)$.

In adapted coordinates, \begin{align}d^{\mathcal{T}^{\nu}\mathcal{T}^{\tau_{E}}E}H_{W_0}=&-\rho_{A}^{i}\frac{\partial L}{\partial x^{i}}\widetilde{e}_{(1,1)}^{A}+\left(p_{A}-\frac{\partial L}{\partial y^{A}}\right)\widetilde{e}_{(2,1)}^{A}+\left(\bar{p}_{A}-\frac{\partial L}{\partial z^{A}}\right)\check{e}^{A}_{(2,1)}\nonumber\\
&+z^{A}(\widetilde{e}_A)^{(2,2)}+y^{A}(\widetilde{e}_A)^{(1,2)}.\nonumber\end{align}

If we apply the constraint algorithm, since $\hbox{ker
}\Omega_0=\hbox{span }\{\check{e}_{A}^{(2,1)}\}$ the first constraint
submanifold $W_1$ is locally characterized by the equation

$$\varphi_{A}=d^{\mathcal{T}^{\nu}\mathcal{T}^{\tau_{E}}E}H_{W_0}(\check{e}_{A}^{(2,1)})=\bar{p}_{A}-\frac{\partial L}{\partial
z^{A}}=0,$$ or $$\bar{p}_{A}=\frac{\partial L}{\partial z^{A}}.$$ Looking for the expression of $X$ satisfying the equation for the second-order problem we have that the second-order equations are \begin{align}
\dot{x}^{i}&=\rho_{A}^{i}y^{A},\nonumber\\
\dot{p}_{A}&=\rho_{A}^{i}\frac{\partial L}{\partial x^{i}}+\mathcal{C}_{AB}^{C}p_{C}y^{B},\nonumber\\
\dot{\bar{p}}_{A}&=-p_A+\frac{\partial L}{\partial y^{A}},\nonumber\\
\bar{p}_{A}&=\frac{\partial L}{\partial z^{A}}.\nonumber
\end{align}

After some straightforward computations the last equations are
equivalent to the following equations:
\begin{equation}\label{ecuacionesorden2algebroides}
0=\frac{d^2}{dt^2}\frac{\partial L}{\partial
    z^{A}}+\mathcal{C}_{AB}^{C}y^{B}\frac{d}{dt}\left(\frac{\partial
    L}{\partial z^{A}}\right)-\frac{d}{dt}\frac{\partial
    L}{\partial
    y^{A}}-\mathcal{C}_{AB}^{C}y^{B}\left(\frac{\partial
    L}{\partial y^{A}}\right)+\rho_{A}^{i}\frac{\partial
    L}{\partial x^{i}}.
\end{equation}

As in the previous section, it is possible to apply the constraint
algorithm \eqref{constraintsalgorithmsection} to obtain a final
constraint submanifold where we have at least a solution which is
dynamically compatible. The algorithm is exactly the same but
applied to the equation
$i_{X}\Omega_0=d^{\mathcal{T}^{\nu}\mathcal{T}^{\tau_{E}}E}H_{W_0}$. Observe that the first constraint submanifold $W_1$  is determined
by the conditions $$\varphi_{A}=\bar{p}_{A}-\frac{\partial L}{\partial z^{A}}=0.$$

If we denote by $\Omega_{W_1}$ the pullback of the presymplectic
2-section $\Omega_{W_0}$ to $W_1$, then we deduce the following:

\begin{proposition}\label{theorem-2algebroides}
  $\Omega_{W_1}$ is a symplectic section of the Lie algebroid $\mathcal{T}^{\nu_{1}}\mathcal{T}^{\tau_{E}}E$ if and only if
$$\left(
\displaystyle{\frac{\partial^2 L}{\partial z^{A}\partial z^{B}}}
\right)$$ is nondegenerate along $W_1$, where $\nu_{1}=\nu\mid_{W_{1}}:W_{1}\to E$.
\end{proposition}
 \begin{remark} Proposition \ref{theorem-2algebroides} is the same result than the theorem given in \cite{IMMS} explained in section \ref{vakoliealgebroids} to the particular case when the $M=E^{(2)}$.\hfill$\diamond$\end{remark}

\begin{example}
Observe that we can particularize the equations
\eqref{ecuacionesorden2algebroides} to the case of Atiyah algebroids
to obtain the second-order Lagrange-Poincar\'e equations.

Let $G$ be a Lie group and we assume that $G$ acts free and properly
on $M$. We denote by $\pi:M\Flder \widehat{M}=M/G$ the associated
principal bundle. The tangent lift of the action gives a free and
proper action of $G$ on $TM$ and $\widehat{TM}=TM/G$ is a quotient
manifold. Then we consider the Atiyah algebroid $\widehat{TM}$ over
$\widehat{M}$.

According to example \ref{Atiyah case}, the basis
$\{\hat{e}_{i},\hat{e}_{B}\}$ induce local coordinates
$(x^{i},y^{i},\bar{y}^{B})$. From this basis one can induces a basis
of the prolongation Lie algebroid, namely
$\{\hat{e}_{i}^{(1)},\hat{e}_{B}^{(1)}\}$. This basis induce adapted
coordinates
$(x^{i},y^{i},\bar{y}^{B},\dot{y}^{i},\dot{\bar{y}}^{B})$ on
$\widehat{T^{(2)}M}=(T^{(2)}M)/G$.

Given a Lagrangian function $\ell:\widehat{T^{(2)}M}\to\R$ over the
set of admissible elements of the Atiyah algebroid
$\widehat{TTM}\to\widehat{TM}$, where $\widehat{TTM}=(TTM)/G$, the
Euler-Lagrange equations for $\ell$ are \begin{align}
\frac{\partial\ell}{\partial x^{j}}-\frac{d}{dt}\left(\frac{\partial\ell}{\partial y^{j}}\right)+\frac{d^{2}}{dt^{2}}\left(\frac{\partial\ell}{\partial\dot{y}^{j}}\right)&=\left(\frac{d}{dt}\left(\frac{\partial\ell}{\partial\dot{\bar{y}}^{A}}\right)-\frac{\partial\ell}{\partial\bar{y}^{A}}\right)\left(\mathcal{B}_{ij}^{A}y^{i}+c_{DB}^{A}\mathcal{A}_{j}^{B}\bar{y}^{B}\right)\quad\forall j,\nonumber\\
\frac{d^{2}}{dt^{2}}\left(\frac{\partial\ell}{\partial\dot{\overline{y}}^{B}}\right)-\frac{d}{dt}\left(\frac{\partial\ell}{\partial\bar{y}^{B}}\right)&=\left(\frac{d}{dt}\left(\frac{\partial\ell}{\partial\dot{\bar{y}}^{A}}\right)-\frac{\partial\ell}{\partial\bar{y}^{A}}\right)\left(c_{DB}^{A}\bar{y}^{D}-c_{DB}^{A}\mathcal{A}_{i}^{D}y^{i}\right), \forall B\nonumber
\end{align} which are the second-order Lagrange-Poincar\'e equations associated to a $G$-invariant Lagrangian $L:T^{(2)}M\to\R$ (see \cite{GHMRV12} and \cite{GHR12}) where $c_{AB}^{C}$ are the structure constants of the Lie algebra according to Example \ref{Atiyah case}.

Observe that If $G=\{e\}$, the identity of $G$, $\widehat{T^{(2)}M}=T^{(2)}M$ and the second-order Lagrange-Poincar\'e equations become into the second-order Euler-Lagrange equations \cite{ldm}, \cite{LR1}
$$0=\frac{d^2}{dt^2}\left(\frac{\partial L}{\partial
    \dot{y}^{A}}\right)-\frac{d}{dt}\left(\frac{\partial
    L}{\partial
    y^{A}}\right)+\frac{\partial
    L}{\partial x^{i}}.$$

If $G=M$, $\widehat{T^{(2)}M}=2\al$ after a left-trivialization, and the second-order Lagrange-Poincar\'e equations become into the second-order Euler-Poincar\'e equations \cite{CoMdD}, \cite{GHMRV10}, \cite{GHMRV12}
$$0=\frac{d^2}{dt^2}\left(\frac{\partial L}{\partial
    \dot{y}^{A}}\right)+c_{AB}^{C}y^{B}\frac{d}{dt}\left(\frac{\partial
    L}{\partial \dot{y}^{A}}\right)-\frac{d}{dt}\frac{\partial
    L}{\partial
    y^{A}}-c_{AB}^{C}y^{B}\left(\frac{\partial
    L}{\partial y^{A}}\right).$$
\end{example}

\subsubsection{Second-order constrained problem on Lie
algebroids}\label{scocproblemalg}

 Now, we will consider second-order
mechanical systems subject to second-order constraints. Let
$\mathcal{M}\subset E^{(2)}$ be an embedded submanifold of dimension
$n+m-\bar{m}$ (locally determined by the vanishing of the constraint
functions $\Phi^{\alpha}:\mathcal{M}\to\R$, $\alpha=1,\ldots,m$)
such that the bundle projection
$\tau_{E}^{(2,1)}\mid_{\mathcal{M}}:\mathcal{M}\to E$ is a
surjective submersion.

%Locally, this condition means that the matrix
%$$\left(\frac{\partial\Phi^{\alpha}}{\partial x^{i}},\frac{\partial\Phi^{\alpha}}{\partial y^{A}},\frac{\partial\Phi^{\alpha}}{\partial
%z^{A}}\right)$$ has maximum rank at all points of $\mathcal{M}$.

We will suppose that the $(\bar{m}\times n)-$matrix
$\left(\frac{\partial\Phi^{\alpha}}{\partial z^{B}}\right)$ with
$\alpha=1,\ldots,\bar{m}$ and $B=1,\ldots,n$ is of maximal rank. Then,
we will use the following notation $z^{A}=(z^{\alpha},z^{a})$ for
$1\leq A\leq n,$ $1\leq\alpha\leq\bar{m}$ and $\bar{m}+1\leq a\leq
n$. Therefore, using the implicit function theorem we can write
$$z^{\alpha}=\Psi^{\alpha}(x^{i},y^{A},z^{a}).$$ Consequently we can
consider local coordinates on $\mathcal{M}$ by $(x^{i},y^{A},z^{a})$
and we will denote by $\widetilde{L}$ the restriction of $L$ to
$\mathcal{M}$.

\begin{proposition}[\cite{Mack}]\label{proposition mackenzie}
Let $(E, \ab, \rho)$ be a Lie algebroid over a manifold $M$ with
projection $\tau_E:E\rightarrow M$ and anchor map with constant
rank. Consider a submanifold $N$ of $M$. If $\tau_E
\big{|}_{\rho^{-1}(TN)}:\rho^{-1}(TN)\rightarrow M$ is a vector
subbundle, then $\rho^{-1}(TN)$ is a Lie algebroid over $N.$
\end{proposition}

Let us take the submanifold
$\overline{W}_0=pr_{1}^{-1}(\mathcal{M})=\mathcal{M}\times_{E}\left(\mathcal{T}^{\tau_{E}}E\right)^{*}$
and the restrictions of $\overline{W}_0$ of the canonical projections
$\pi_1$ and $\pi_2$ given by $\pi_1=pr_1\mid_{\overline{W}_0}$ and
$\pi_2=pr_1\mid_{\overline{W}_0}$. We will denote local coordinates
on $\overline{W}_{0}$ by $(x^{i},y^{A},p_{A},\bar{p}_{A},z^{a})$.

Therefore, proceeding as in the unconstrained case one can construct
the presymplectic Hamiltonian system
$(\overline{W}_0,\Omega_{\overline{W}_0},H_{\overline{W}_0})$, where
$\Omega_{\overline{W}_0}$ is the presymplectic $2$-section on
$\overline{W}_0$ and the Hamiltonian function
$H:\overline{W}_0\to\R$ is locally given by
$$H_{\overline{W}_0}(x^{i},y^{A},p_{A},\overline{p}_{A},z^{a})=p_Ay^{A}+\overline{p}_az^{a}+\overline{p}_{\alpha}\Psi^{\alpha}(x^{i},y^{A},z^{a})-\widetilde{L}(x^{i},y^{A},z^{a}).$$

With these two elements it is possible to write the following
presymplectic system
\begin{equation}\label{dynamiceqSKconstrainedLiealgebroids}
i_{X}\Omega_{\overline{W}_0}=d^{(\rho^{\nu})^{-1}(T\overline{W}_0)}H_{\overline{W}_0},
\end{equation} where $(\rho^{\nu})^{-1}(TW_0)$ denotes the Lie subalgebroid of $\mathcal{T}^{\nu}\mathcal{T}^{\tau_{E}}E$ over $\overline{W}_0\subset W_0$.

 To characterize the equations we will adopt an ``extrinsic point
of view", that is, we will work on the full space $W_0$ instead of
in the restricted space $\overline{W_0}$. Consider an arbitrary
extension  $L: E^{(2)}\to \R$ of $L_{\mathcal M}: {\mathcal M}\to
\R$. The main idea is to take into account that Equation
(\ref{dynamiceqSKconstrainedLiealgebroids}) is equivalent to
\[
\left\{
\begin{array}{rcl}
i_X\Omega_{{W}_0}-d^{\mathcal{T}^{\nu}\mathcal{T}^{\tau_{E}}E} {H}&\in& \hbox{ann }(\rho^{\nu})^{-1}(T_{x}\overline{W}_0)\; ,\\
X&\in&(\rho^{\nu})^{-1}(T_{x}\overline{W}_0)\;\hbox{ and } x\in\overline{W}_{0},
\end{array}
\right.
\]
where $H:W_0\to\R$ is the function defined in the last section and
$\hbox{ann}$ denotes the set of sections
$\widetilde{X}\in\Gamma((\mathcal{T}^{\nu}\mathcal{T}^{\tau_{E}}E)^{*})$ such that
$\langle \widetilde{X},Y\rangle=0$ for all
$Y\in(\rho^{\nu})^{-1}(TW_0)$.

Assuming that ${\mathcal M}$ is determined by the vanishing of
$\overline{m}$-independent constraints
\[
\Phi^{\alpha}(x^{i},y^{A},z^{a})=0, \ 1\leq \alpha\leq \overline{m}\; ,
\]
then, locally, $ \hbox{ann
}(\rho^{\nu})^{-1}(T\overline{W}_0)=\hbox{span }\{
d^{\mathcal{T}^{\nu}\mathcal{T}^{\tau_{E}}E}\Phi^{\alpha}\}\, , $ and therefore the
previous equations are rewritten as
 \[
\left\{
\begin{array}{rcl}
i_X\Omega_{{W}_0}-d^{\mathcal{T}^{\nu}\mathcal{T}^{\tau_{E}}E}{H}&=& \lambda_{\alpha}d^{\mathcal{T}^{\nu}\mathcal{T}^{\tau_{E}}E}\Phi^{\alpha}\, ,\\
X(x)\in (\rho^{\nu})^{-1}(T_{x}\overline{W}_0)&&\hbox{for all }x\in\overline{W}_0\; ,
\end{array}
\right.
\]
where $\lambda_\alpha$ are Lagrange multipliers to be determined.

Proceeding as in the previous section, one can obtain the following
system of equations for
$\widetilde{L}=L+\lambda_{\alpha}\Phi^{\alpha}$ \begin{align}\label{ecuacionesvakonomoalgebroides2}
0&=\frac{d^2}{dt^2}\frac{\partial \widetilde{L}}{\partial
    z^{A}}+\mathcal{C}_{AB}^{C}y^{B}\frac{d}{dt}\left(\frac{\partial
    \widetilde{L}}{\partial z^{A}}\right)-\frac{d}{dt}\frac{\partial
    \widetilde{L}}{\partial
    y^{A}}-\mathcal{C}_{AB}^{C}y^{B}\left(\frac{\partial
    \widetilde{L}}{\partial y^{A}}\right)+\rho_{A}^{i}\frac{\partial
    \widetilde{L}}{\partial x^{i}}\\
    0&=\Phi^{\alpha}(x^{i},y^{A},z^{A}).\nonumber
\end{align}
Here the first constraint submanifold $\overline{W}_{1}$ is
determined by the condition
\begin{eqnarray*}
0&=&\bar{p}_{A}-\frac{\partial L}{\partial z^{A}}+\lambda_{\alpha}\frac{\partial\Phi^{\alpha}}{\partial z^{A}}\\
0&=&\Phi^{\alpha}(x^{i},y^{A},z^{A}).
\end{eqnarray*}
%\todo{ La condicion de regularidad es en terminos de l tilde, sin el multiplicador}
If we denote by $\Omega_{\overline{W}_{1}}$ the pullback of the
presymplectic section $\Omega_{\overline{W}_0}$ to
$\overline{W}_{1}$, then we can deduce that
$\Omega_{\overline{W}_{1}}$ is a symplectic section if and only if

\begin{equation}\label{cdalgebroids}
\left(
\begin{array}{ccc}
\displaystyle{\frac{\partial^2 L}{\partial z^{A}\partial z^{B}}+\lambda_{\alpha}\frac{\partial^2 \Phi^{\alpha}}{\partial z^{A}\partial z^{B}}}& \ &\displaystyle{\frac{\partial \Phi^{\alpha}}{\partial z^{A}}}\\
\displaystyle{\frac{\partial \Phi^{\alpha}}{\partial z^{B}}}&\ &\textbf{0}
\end{array}
\right)
\end{equation} is nondegenerate.

\section{Application to optimal control of mechanical systems}\label{section4}

In this section we study optimal control problems of mechanical systems defined on Lie algebroids. First we treat with fully actuated system and next with underactuated systems. Optimality conditions for the optimal control of the controlled Elroy's Beany system are derived.
 
 Optimal control problems can be seen as higher-order variational problems (see \cite{Blo} and \cite{BlochCrouch2}). Higher-order variational problems are given by
 $$\min_{q(\cdot)}\int_{0}^{T}L(q^i, \dot{q}^{i},\ldots,q^{(k)i})dt,$$
subject to boundary conditions. The relationship between higher-order  variational problems and optimal control problems of mechanical systems comes from the fact that Euler-Lagrange equations are represented by a second-order Newtonian system and  mechanical control systems have the form $F(q^i,\dot{q}^{i},\ddot{q}^{i})=u$, where $u$ are the control inputs. Then, if $C$ is a given cost function, 
$$\min_{(q(\cdot),u(\cdot))}\int_{0}^{T}C(q^i,\dot{q}^{i},u)dt,$$ is equivalent to a higher-order variational problem with $k=2$.

\subsection{Optimal control problems of fully-actuated mechanical systems on Lie algebroids}

Let  $(E,\llbracket\cdot,\cdot\rrbracket,\rho)$ be a Lie algebroid over
$Q$ with bundle projection $\tau_{E}:E\to Q$. The dynamics is specified fixing a Lagrangian $L:E\to\R$. External forces are modeled,
in this case, by curves $u_F:\R\to E^{*}$ where $E^{*}$ is the dual bundle $\tau_{E^{*}}:E^{*}\to\ Q$.  

Given local coordinates $(q^i)$ on $Q$, and fixing a basis of sections $\{e_{A}\}$ of $\tau_{E}:E\to Q$ we can
induce local coordinates $(q^i,y^{A})$ on $E$; that is, every element $b\in E_q=\tau_E^{-1}(q)$ is expressed univocally as $b=y^{A}e_{A}(q)$.  

It is possible to adapt the derivation of the Lagrange-d'Alembert
principle to study fully-actuated mechanical controlled systems
on Lie algebroids (see \cite{CoMa} and \cite{Ma2}). Let $q_0$ and $q_T$ fixed in $Q$, consider an admissible curve $\xi:I\subset\mathbb{R}\to E$
which satisfies the principle
$$0=\delta\int_{0}^{T}L(\xi(t))dt+\int_{0}^{T}\langle u_{F}(t),\eta(t)\rangle dt,$$
where $\eta(t)\in {E}_{\tau_{E}(\xi(t))}$ and $u_{F}(t)\in E^{*}_{\tau_{E}(\xi(t))}$ defines
the control force (where we are assuming they are arbitrary). The infinitesimal variations in the variational principle are given by
$\delta\xi=\eta^{\mathbf{C}}$, for all time-dependent sections
$\eta\in\Gamma(\tau_{E})$, with $\eta(0)=0$ and $\eta(T)=0$, where
$\eta^{\mathbf{C}}$ is a time-dependent vector field on $E$, the
\textit{complete lift}, locally defined by
$$\eta^{\mathbf{C}}=\rho_{A}^i\eta^{A}\frac{\partial}{\partial q^i}+
(\dot{\eta}+\mathcal{C}^{A}_{BC}\eta^{B}y^{C})\frac{\partial}{\partial
y^{A}}$$ (see \cite{CoMa,Eduardo,Eduardo1,Eduardoalg}). Here the structure functions $\mathcal{C}_{BC}^{A}$ are determined by $\lcf e_{B},e_{C}\rcf=\mathcal{C}_{BC}^{A}e_{A}$.

From the Lagrange-d'Alembert
principle one easily derives the controlled
Euler-Lagrange equations by using standard variational calculus  \begin{align*}
\frac{d}{dt}\left(\frac{\partial L}{\partial
y^{A}}\right)-\rho_{A}^{i}\frac{\partial L}{\partial
q^{i}}+\mathcal{C}_{AB}^{C}(q)y^{B}\frac{\partial
L}{\partial y^{C}}=&(u_{F})_{A},\\
\frac{dq^i}{dt}=&\rho_{A}^{i}y^{A}.
\end{align*}
where $(u_{F})_{A}(t)=\langle u_F(t), e_{A}(q(t))\rangle$ are the local components of $u_F$ fixed the system of coordinates $(q^i)$ on $Q$ and the basis of section $\{e_{A}\}$.

The control force $u_{F}$ is chosen such that it minimizes the
cost functional
$$\int_0^{T}C(q^i,y^{A},(u_{F})_{A})dt,$$ where
$C:E\oplus E^{*}\to\R$ is the cost function associated with the optimal control problem.

Therefore, the optimal control problem consists on finding an admissible curve $\xi(t)=(q^{i}(t),y^{A}(t))$ solution of the controlled Euler-Lagrange equations,  the  boundary conditions and minimizing the cost functional for $C: E\oplus E^{*}\to\R$. This optimal control problem can be equivalently  solved as a second-order variational problem by defining the second-order Lagrangian $\widetilde{L}:E^{(2)}\to\R$ as
\begin{equation}\label{costLtilde}\widetilde{L}(q^i,y^{A},z^{A})=C\left(q^i,y^{A},\frac{d}{dt}\left(\frac{\partial
L}{\partial y^{A}}\right)-\rho_{A}^{i}\frac{\partial
L}{\partial
q^{i}}+\mathcal{C}_{AB}^{C}(q)y^{B}\frac{\partial
L}{\partial y^{C}}\right)\end{equation} where we are considering local coordinates $(q^{i},y^{A},z^{A})$ on $E^{(2)}$.

Consider $W_0=E^{(2)}\times\left(\mathcal{T}^{\tau_{E^{*}}}E\right)^{*}$ with local coordinates $(q^i,y^A,p_a,\bar{p}_A,z^A)$. The optimality conditions are determined by 
\begin{align}
\dot{q}^i&=\rho_{A}^{i}y^A,\nonumber\\
\dot{p}_{A}&=\rho_{A}^{i}\frac{\partial C}{\partial q^{i}}+\mathcal{C}_{AB}^{C}p_Cy^B,\nonumber\\
\dot{\bar{p}}_A&=-p_A+\frac{\partial C}{\partial y^{A}},\nonumber\\
\bar{p}_A&=\frac{\partial C}{\partial z^{A}}.\nonumber
\end{align} The constraint submanifold $W_1$ is determined by $\displaystyle{\bar{p}_{A}-\frac{\partial C}{\partial z^A}=0.}$ If the matrix $\displaystyle{\left(\frac{\partial^{2}C}{\partial z^{A}\partial z^{B}}\right)}$ is non-singular then we can write the previous equations as an explicit system of ordinary differential equations. This regularity assumption is equivalent to the condition that the constraint algorithm stops at the first constraint submanifold $W_1$.
Proceeding as in the previous section, after some computations,
the dynamics associated with the second-order Lagrangian $\widetilde{L}:E^{(2)}\to\R$ (and therefore the optimality conditions for the optimal control problem) is given by the second-order Euler-Lagrange equations on Lie algebroids  
\begin{equation}\label{ecuacionesvakonomoalgebroides2}
\frac{d^2}{dt^2}\left(\frac{\partial \widetilde{L}}{\partial
    z^{A}}\right)+\mathcal{C}_{AB}^{C}(q)y^{B}\frac{d}{dt}\left(\frac{\partial
    \widetilde{L}}{\partial z^{C}}\right)-\frac{d}{dt}\frac{\partial
    \widetilde{L}}{\partial
    y^{A}}-\mathcal{C}_{AB}^{C}(q)y^{B}\frac{\partial
    \widetilde{L}}{\partial y^{C}}+\rho_{A}^{i}\frac{\partial
    \widetilde{L}}{\partial q^{i}}=0, 
\end{equation} together with the admissibility condition $\displaystyle{\frac{dq^i}{dt}=\rho_{A}^{i}y^{A}}.$

\begin{remark}
Alternatively, one can define the Lagrangian
$\widetilde{L}:E^{(2)}\to\R$ in terms of the Euler-Lagrange operator as
$$\widetilde{L}=C\circ(\tau^{\tiny{E^{(2)}}}_{E}\oplus\mathcal{EL}(L)):E^{(2)}
\to\R,$$ where $\mathcal{EL}(L):E^{(2)}\to E^{*}$ is the \textit{Euler-Lagrange operator} which locally reads as
$$\mathcal{EL}(L)=\left(\frac{d}{dt}\frac{\partial L}{\partial y^{A}}-\rho_{A}^{i}\frac{\partial
L}{\partial
q^{i}}+\mathcal{C}_{AB}^{D}(q)y^{B}\frac{\partial
L}{\partial y^{D}}\right)e^{A}.$$ Here $\{e^{A}\}$ is the dual basis of $\{e_{A}\},$ the basis of sections of $E$ and $\tau^{\tiny{E^{(2)}}}_{E}:E^{(2)}\to E$ is the canonical projection between $E^{(2)}$ and $E$ given by the map $E^{(2)}\ni (q^{i},y^{A},z^{A})\mapsto(q^{i},y^{A})\in E.$\hfill$\diamond$
\end{remark}

\begin{example}\textbf{An illustrative example: optimal control of a fully actuated rigid body and cubic splines on Lie groups}

%\todo{cambiar por http://authors.library.caltech.edu/28008/1/97-03.pdf}

We consider the motion of a rigid body where the configuration space is the Lie group $G=SO(3)$ and $\mathfrak{so}(3)\equiv \R^3$ its Lie algebra. The motion of the rigid body is invariant under $SO(3)$. The reduced Lagrangian function for this system defined on the Lie algebroid $E=\mathfrak{so}(3)$, $\ell:\mathfrak{so}(3)\rightarrow\R$ is given by  $$\ell(\Omega_1,\Omega_2,\Omega_3)=\frac{1}{2}\left(I_1\Omega_1^2+I_2\Omega_2^2+I_3\Omega_3^2\right).$$ 

Denote by $t\to R(t)\in SO(3)$ a curve. The columns of the
matrix $R(t)$ represent the directions of the principal axis of the
body at time $t$ with respect to some reference system. 
Consider the following left invariant control problem. First, we have the
reconstruction equation:
$$\dot{R}(t)=R(t)
\left(
  \begin{array}{ccc}
    0& -\Omega_3(t) & \Omega_2(t) \\
    \Omega_3(t) & 0 & -\Omega_1(t) \\
    -\Omega_2(t) & \Omega_1(t) & 0 \\
  \end{array}
\right)=R(t)\left(\Omega_1(t)E_1+\Omega_2(t)E_2+\Omega_3(t)E_3\right)$$
where \[E_1:=\left(
               \begin{array}{ccc}
                 0 & 0 & 0 \\
                 0 & 0 & -1 \\
                 0 & 1 & 0 \\
               \end{array}
             \right), \qquad E_2:=\left(
                                   \begin{array}{ccc}
                                      0 & 0 & 1 \\
                                      0 & 0 & 0 \\
                                      -1 & 0 & 0 \\
                                    \end{array}
                                  \right), \qquad E_3:=\left(
                                                         \begin{array}{ccc}
                                                           0 & -1 & 0 \\
                                                           1 & 0 & 0 \\
                                                           0 & 0 & 0 \\
                                                         \end{array}
                                                       \right)\]
and the equations for the angular velocities $\Omega_i$ with $i=1,2,3$:
\begin{align*}
I_1\dot{\Omega}_1(t)&=(I_2-I_3)\Omega_2(t)\Omega_3(t)+u_1(t)\\
I_2\dot{\Omega}_2(t)&=(I_3-I_1)\Omega_3(t)\Omega_1(t)+u_2(t)\\
I_3\dot{\Omega}_3(t)&=(I_1-I_2)\Omega_1(t)\Omega_2(t)+u_3(t)
\end{align*}
where $I_1,I_2,I_3$ are the moments of inertia and $u_1, u_2, u_3$ denote
the applied torques playing the role of controls of the system.

The optimal control problem for the rigid body consists on finding  the trajectories $(R(t), \Omega(t), u(t))$  with fixed initial and final conditions
$(R(0), \Omega(0)),$ $(R(T), \Omega(T))$ respectively and minimizing the cost functional
$$\mathcal{A} =\int_{0}^{T} \mathcal{C}(\Omega, u_1, u_2,u_3) dt=\frac{1}{2} \int_{0}^{T} (u_1^2+u_2^2+u_3^2)\, dt.$$

This optimal control problem is equivalent to solve the following second-order (unconstrained) variational problem
$$\min \widetilde{\mathcal{J}} = \int_{0}^{T} \widetilde{L}(\Omega, \dot\Omega)dt$$  
where
\[
\widetilde{L}(\Omega, \dot\Omega) = \mathcal{C}\left(\Omega, I_1\dot\Omega_1 - (I_2-I_3)\Omega_2\Omega_3 ,I_2\dot\Omega_2 - (I_3-I_1)\Omega_3\Omega_1, I_3\dot\Omega_3 - (I_1-I_2)\Omega_1\Omega_2 \right)\; ,
\] that is,  $$\widetilde{L}(\Omega, \dot\Omega)=\frac{1}{2}\left[\left(I_1\dot\Omega_1-(I_2-I_3)
\Omega_2\Omega_3\right)^2+\left(I_2\dot\Omega_2-(I_3-I_1)\Omega_3\Omega_1\right)^2+\left(I_3\dot\Omega_3-(I_1-I_2)
\Omega_1\Omega_2\right)^2\right].
$$
%$$+c_2\left(\Omega_1^2+\Omega_2^2+\Omega_3^2\right)\; .$$

Next, for simplicity, we consider the particular case $I_1=I_2=I_3=1$. The second order Lagrangian is given by 

$$\widetilde{L}(\Omega, \dot\Omega)=\frac{1}{2}\left(\dot\Omega_1^2+\dot\Omega_2^2+\dot\Omega_3^2\right).$$ The Pontryagin bundle is $W_0=2\mathfrak{so}(3)\times 2\mathfrak{so}(3)^{*}$ with induced coordinates $$(\Omega_1,\Omega_2,\Omega_3,\dot{\Omega}_1,\dot{\Omega}_2,\dot{\Omega}_3,p_1,p_2,p_3,\bar{p}_1,\bar{p}_2,\bar{p}_3).$$ The first constraint submanifold is given by $$W_{1}=\{\bar{p}_1-\dot{\Omega}_1=0,\quad\bar{p}_2-\dot{\Omega}_2=0,\quad\bar{p}_3-\dot{\Omega}_3=0\}.$$ Observe that $$\left(\frac{\partial^2\widetilde{L}}{\partial\dot{\Omega}_A\dot{\Omega}_{B}}\right)=\mathbf{I}_{3\times 3}$$ where $\mathbf{I}_{3\times 3}$ denotes the $3\times 3$ identity matrix. Thus, the constraint algorithm stops at the first constraint submanifold $W_1$. 

We can write the equations of motion for the optimal control system as: 

\begin{align}
\dot{p}_1&=p_3\Omega_2-p_2\Omega_3,\quad \frac{d}{dt}\Omega_1=\dot{\Omega}_1,\nonumber\\
\dot{p}_2&=p_1\Omega_3-p_3\Omega_1,\quad \frac{d}{dt}\Omega_2=\dot{\Omega}_2,\nonumber\\
\dot{p}_3&=p_2\Omega_1-p_1\Omega_2,\quad \frac{d}{dt}\Omega_3=\dot{\Omega}_3,\nonumber\\
\dot{\bar{p_1}}&=-p_1,\quad \dot{\Omega}_1=\bar{p}_1,\nonumber\\
\dot{\bar{p_2}}&=-p_2,\quad \dot{\Omega}_2=\bar{p}_2,\nonumber\\
\dot{\bar{p_3}}&=-p_3,\quad \dot{\Omega}_3=\bar{p}_3.\nonumber
\end{align} After some strighforward computations, previous equations can be reduced to  $$
\dddot{\Omega}_1=\Omega_3\ddot{\Omega}_2-\Omega_2\ddot{\Omega}_3,\quad
\dddot{\Omega}_2=\Omega_1\ddot{\Omega}_3-\Omega_3\ddot{\Omega}_1,\quad
\dddot{\Omega}_3=\Omega_2\ddot{\Omega}_1-\Omega_1\ddot{\Omega}_2.
$$ or in short notation, $$\dddot{\Omega}=-\Omega\times\ddot{\Omega}.$$ The previous equations are the equations given by L. Noakes, G. Heinzinger and B. Paden, \cite{splines3} for cubic splines on $SO(3)$.

Finally, we would like to comment that the regularity condition provides the existence of a unique solution of the dynamics along the submanifold $W_1$. Therefore, there exists a unique vector field  $X\in {\mathfrak X}(W_1)$ which
satisfies $i_{X}\Omega_{W_{1}} = d {H}_{W_{1}}$.
 In consequence, we have a unique control input which extremizes (minimizes) the objective function ${\mathcal A}$.
If we take the flow $F_{t}: W_{1}\rightarrow W_{1}$ of the vector
field $X$ then we have that $F_{t}^{*}\Omega_{W_1} = \Omega_{W_1}$.
Obviously, the Hamiltonian function

$$H_{W_0}(\Omega,\dot{\Omega} ,p,\overline{p})=\overline{p}_{A}\dot{\Omega}_{A}+p_{A}\dot{\Omega}_{A}-\frac{1}{2}\left(\dot\Omega_1^2+\dot\Omega_2^2+\dot\Omega_3^2\right)$$ is preserved by the solution of the optimal control problem, that is
$\widetilde{H}\big{|}_{W_1}\circ F_t=\widetilde{H}\big{|}_{W_1}$.

\end{example}

\subsection{Optimal control problems of underactuated mechanical systems on Lie algebroids}
Now, suppose that our mechanical control system is underactuated, that is, the number of control inputs is less than the dimension of the configuration space. The class of underactuated mechanical systems is abundant in real life for different reasons; for instance, as a result of design choices motivated by the search of less cost engineering devices or as a result of a failure regime in fully actuated mechanical systems. Underactuated systems include spacecrafts, underwater vehicles, mobile robots, helicopters, wheeled vehicles and underactuated manipulators.
In the general situation, the dynamics is specified fixed a
Lagrangian $L:E\to\R$ where $(E,\lcf\cdot,\cdot\rcf,\rho)$ is a Lie
algebroid over a manifold $Q$ with fiber bundle projection
$\tau_{E}:E\to Q$.

If we take local coordinates $(q^{i})$ on $Q$ and a local basis
$\{e_A\}$ of sections of $E$, then we have the corresponding local
coordinates $(q^{i},y^A)$ on $E$. Such coordinates determine the
local structure functions $\rho_{A}^{i}$ and
$\mathcal{C}_{AB}^{C}$ and then the Euler-Lagrange equations on Lie algebroids can be written as $$\frac{d}{dt}\left(\frac{\partial L}{\partial
y^{A}}\right)-\rho_{A}^{i}\frac{\partial L}{\partial
q^{i}}+\mathcal{C}_{AB}^{C}y^{B}\frac{\partial
L}{\partial y^{A}}=0.$$ These equations are precisely the components of the \textit{Euler-Lagrange operator} $\mathcal{E}L:E^{(2)}\to E^{*}$, which locally reads $$\mathcal{E}L=\left(\frac{d}{dt}\left(\frac{\partial L}{\partial
y^{A}}\right)-\rho_{A}^{i}\frac{\partial L}{\partial
q^{i}}+\mathcal{C}_{AB}^{C}y^{B}\frac{\partial
L}{\partial y^{A}}\right)e^{A},$$ where $\{e^{A}\}$ is the dual basis of $\{e_{A}\}$ (see \cite{CoMa}). In terms of the Euler-Lagrange operator, the equations of motion just read $\mathcal{E}L=0$.

In the underactuated case, we model the set of control forces by the vector subbundle $\hbox{span}\{e^{a}\}\subset E^{*}$ and the forces are given by $u_F=(u_{F})_{a} e^a$.

Now, we add controls in our picture. Assume that the controlled Euler-Lagrange equations are \begin{equation}\label{controlledalgebroids}
\left(\frac{d}{dt}\left(\frac{\partial L}{\partial
y^{A}}\right)-\rho_{A}^{i}\frac{\partial L}{\partial
q^{i}}+\mathcal{C}_{AB}^{C}y^{B}\frac{\partial
L}{\partial y^{A}}\right)e^{A}=u_{a}e^{a},
\end{equation} where we are denoting as $\{e^{A}\}=\{e^{a},e^{\alpha}\}$ the dual basis of $\{e_{A}\}$ and $u_a$ are admissible control parameters. Using the basis of sections of $E,$ equations \eqref{controlledalgebroids} can be rewritten as
\begin{eqnarray}\label{controlledalgebroids2}
\frac{d}{dt}\left(\frac{\partial L}{\partial
y^{a}}\right)-\rho_{a}^{i}\frac{\partial L}{\partial
q^{i}}+\mathcal{C}_{aB}^{C}y^{B}\frac{\partial
L}{\partial y^{C}}&=&u_{a},\\
\frac{d}{dt}\left(\frac{\partial L}{\partial
y^{\alpha}}\right)-\rho_{\alpha}^{i}\frac{\partial L}{\partial
q^{i}}+\mathcal{C}_{\alpha B}^{C}y^{B}\frac{\partial
L}{\partial y^{C}}&=&0.\nonumber
\end{eqnarray}

The optimal control problem consists on finding an admissible
curve $\gamma(t)=(q^{i}(t),y^{A}(t), u(t))$ of the state
variables and control inputs given initial and final boundary
conditions $(q^{i}(0),y^{A}(0))$ and $(q^{i}(T),y^{A}(T))$,
respectively, solving the controlled Euler-Lagrange equations
\eqref{controlledalgebroids2} and minimizing $$
\mathcal{A}(q^{i},y^{A}, u_{a}) = \int_{0}^{T} C(q^{i},y^{A}, u_a)dt
\, ,
$$ where $C \colon E \times U \to \R$ denotes the cost function.

To solve this optimal control problem is equivalent to solve the
following second-order problem:
\begin{align*}
&\textnormal{min } \widetilde{L}(q^{i}(t),y^{A}(t),z^{A}(t)) \\
&\textnormal{subject to } \Phi^{\alpha}(q^{i}(t),y^{A}(t),z^{A}(t)) \, , \, \alpha= 1,\ldots,m
\end{align*}
where $\widetilde{L}, \Phi^{\alpha} \in \mathcal{C}^{\infty}(E^{(2)})$. Here
$$
\widetilde{L}(q^{i}(t),y^{A}(t),z^{A}(t))
=C\left(q^{i}(t),y^{A}(t),F_{a}(x^{i}(t),y^{A}(t),z^{A}(t))\right) \, ,
$$
where 
$$
F_{a}(q^{i}(t),y^{A}(t),z^{A}(t)) =
\frac{d}{dt}\left(\frac{\partial L}{\partial
y^{a}}\right)-\rho_{a}^{i}\frac{\partial L}{\partial
q^{i}}+\mathcal{C}_{aB}^{C}y^{B}\frac{\partial
L}{\partial y^{C}} \, .
$$
The Lagrangian $\widetilde{L}$ is subjected to the second-order
constraints:
$$
\Phi^{\alpha}(q^{i}(t),y^{A}(t),z^{A}(t))=
\frac{d}{dt}\left(\frac{\partial L}{\partial
y^{\alpha}}\right)-\rho_{\alpha}^{i}\frac{\partial L}{\partial
q^{i}}+\mathcal{C}_{\alpha B}^{C}y^{B}\frac{\partial
L}{\partial y^{C}} \, ,
$$ which determines a submanifold $\mathcal{M}$ of $E^{(2)}.$

\begin{remark}
Observe that the cost function is not completely defined in $E\oplus E^{*}$, it is only defined in a smaller subset of this space because $\displaystyle{
\frac{d}{dt}\left(\frac{\partial L}{\partial
y^{a}}\right)-\rho_{a}^{i}\frac{\partial L}{\partial
q^{i}}+\mathcal{C}_{aB}^{C}y^{B}\frac{\partial
L}{\partial y^{C}} }$ only belongs to the vector subbundle $\hbox{span}\{e^{a}\}\subset E^{*}$. That is, in the case of fully actuated system the cost function would be defined in the full space $E^*$, and when we are dealing with an underactuated systems, the cost function is defined in a proper subset of $E^*$. Next, for simplicity, we assume that $C: E\oplus E^*\rightarrow {\mathbb R}.$\hfill$\diamond$

\end{remark}
Observe that from the constraint equations we have that
$$\frac{\partial^{2}L}{\partial y^{\alpha}\partial y^{\beta}}z^{\beta}+\frac{\partial ^{2}L}{\partial y^{\alpha}\partial y^{a}}z^{a}-\rho_{\alpha}^{i}\frac{\partial L}{\partial q^{i}}+\mathcal{C}_{\alpha B}^{C}y^{B}\frac{\partial L}{\partial y^{C}}=0.$$
Therefore, assuming that the matrix
$\displaystyle{W_{\alpha\beta}=\left(\frac{\partial^{2}L}{\partial y^{\alpha}\partial
y^{\beta}}\right)}$ is regular, we can write the equations as
$$
z^{\alpha}=-W^{\alpha\beta}\left(\frac{\partial^{2}L}{\partial y^{\beta}\partial y^{a}}z^{a}-\rho_{\beta}^{i}\frac{\partial L}{\partial q^{i}}+\mathcal{C}_{\beta B}^{C}y^{B}\frac{\partial L}{\partial y^{C}}\right)
=G^{\alpha}(q^{i},y^{A},z^{a})$$ where $W^{\alpha\beta}=(W_{\alpha\beta})^{-1}.$

Therefore, we can choose coordinates $(q^{i},y^{A},z^{a})$ on
$\mathcal{M}$. This choose allows us to consider an intrinsic point
of view, that is, to work directly on
$\overline{W}=\mathcal{M}\times\left(\mathcal{T}^{\tau_{E}}E\right)^{*}$
avoiding the use of the Lagrange multipliers.

Define the restricted Lagrangian $\widetilde{L}_{\mathcal{M}}$ by
$\widetilde{L}\big{|}_{\mathcal{M}}:\mathcal{M}\to\R$ and take
induced coordinates on
$\overline{W}$, $(q^{i},y^{A},z^{a},p_{A},\overline{p}_{A})$. Applying the same procedure than in section \ref{scocproblemalg} we
derive the following system of equations
%
%We consider the presymplectic 2-section on $\overline{W}_{0}$.
% \[
%  (\Omega_{\overline{W}_0})=\check{\widetilde{e}}^{A}_{(1,1)}\wedge\check{\widetilde{e}}_{(1,2)}^{A}+\check{\widetilde{e}}^{A}_{(2,1)}\wedge\check{\widetilde{e}}^{A}_{(2,2)}+\frac{1}{2}\widetilde{\mathcal{C}}_{AB}^{C}p_{C}\check{\widetilde{e}}^{A}_{(1,1)}\wedge\check{\widetilde{e}}_{(1,1)}^{B},
%  \] and define the Hamiltonian $\bar{H}:\overline{W}_{0}\to\R$ by
%  \[
%  \bar{H}(x^{i},y^{A},p_{A},\bar{p}_A,z^{a})=y^Ap_A+ z^a\bar{p}_a+ G^{\alpha}(x^{i}, y^A, z^a)\bar{p}_{\alpha}-
%\widetilde{L}_{\mathcal M }(x^{i}, y^A, z^a).
%\]
%In consequence,
%\begin{eqnarray*}
%d\bar{H}&=&
%-\rho_{A}^{i}\left(\frac{\widetilde{L}_{\mathcal{M}}}{\partial x^{i}}-\bar{p}_{\beta}\frac{\partial G^{\beta}}{\partial x^{i}}\right) \check{\widetilde{e}}^A_{(1,1)}+\left(p_A-\frac{\partial \widetilde{L}_{\mathcal M }}{\partial y^{A}}+\bar{p}_{\beta}\frac{\partial G^\beta}{\partial y^{A}}\right)\check{\widetilde{e}}^A_{(2,1)}\\
%&&+\left(\bar{p}_a-\frac{\partial \widetilde{L}_{\mathcal M }}{\partial z^{a}}+\bar{p}_{\beta}\frac{\partial G^{\beta}}{\partial z^{a}}\right)\check{e}^a_{(1,2)}+y^A\check{\widetilde{e}}_{(1,2)}^{A}+z^a\check{\widetilde{e}}^a_{(2,2)}+G^{\alpha}\check{\widetilde{e}}^{\alpha}_{(2,2)}\; .
%\end{eqnarray*}
%The conditions for the section $X$ satisfying equations $i_X
%\Omega_{\overline{W}_0}=d\bar{H}$ are
\begin{align}
\dot{q}^{i}&=\rho_{A}^{i}y^{A},\nonumber\\
\frac{dy^a}{dt}&=z^a,\nonumber\\
\frac{d y^{\alpha}}{dt} &= G^{\alpha}(q^{i}, y^A, z^a),\nonumber\\
\frac{d p_A}{dt} &= \rho_{A}^{i}\left(\frac{\partial \widetilde{L}_{\mathcal M }}{\partial q^{i}}-\bar{p}_{\beta}\frac{\partial G^{\beta}}{\partial q^{i}}\right)+{\mathcal C}_{AB}^Cp_Cy^B,\nonumber\\
\frac{d \bar{p}_{A}}{dt} &=-p_A+\frac{\partial \widetilde{L}_{\mathcal M }}{\partial y^{A}}-\bar{p}_{\beta}\frac{\partial G^{\beta}}{\partial y^{A}},\nonumber\\
\bar{p}_{a} &= \frac{\partial \widetilde{L}_{\mathcal M }}{\partial z^{a}}-\bar{p}_{\beta}\frac{\partial G^{\beta}}{\partial z^{a}}.\nonumber
\end{align}

To shorten the number of unknown variables involved in the previous
set of equations, we can write them using as variables $(q^{i}, y^A,
z^a, \overline{p}_{\alpha})$
\begin{align}
\dot{q}^{i}&=\rho_{A}^{i}y^{A},\nonumber\\
\frac{d y^{\alpha}}{dt} &= G^{\alpha}(q^{i}, y^A, z^a),\nonumber\\
0&=\frac{d^2}{dt^2}\left(\frac{\partial \widetilde{L}_{\mathcal M }}{\partial z^{a}}-\overline{p}_{\beta}\frac{\partial G^{\beta}}{\partial z^{a}}\right)-{\mathcal C}_{Aa}^by^A\left(\frac{d}{dt}\left[\frac{\partial \widetilde{L}_{\mathcal M }}{\partial z^{b}}-\overline{p}_{\beta}\frac{\partial G^{\beta}}{\partial z^{b}}\right]\right)-\frac{d}{dt}\left(\frac{\partial \widetilde{L}_{\mathcal M }}{\partial y^{a}}-\overline{p}_{\beta}\frac{\partial G^{\beta}}{\partial y^{a}}\right)\nonumber\\
& +{\mathcal C}_{Aa}^{C}y^A\left(\frac{\partial \widetilde{L}_{\mathcal M }}{\partial {y}^{C}}-\overline{p}_{\beta}\frac{\partial G^{\beta}}{\partial {y}^{C}}\right)+\rho_{a}^{i}\left(\frac{\partial \widetilde{L}_{\mathcal M }}{\partial q^{i}}-\overline{p}_{\beta}\frac{\partial G^{\beta}}{\partial q^{i}}\right)
-{\mathcal C}_{Aa}^{\gamma}y^A\frac{d\overline{p}_{\gamma}}{dt}.\nonumber\\
0&=\frac{d^2\overline{p}_{\alpha}}{dt^2}+{\mathcal C}_{A\alpha}^{\beta}y^A\frac{d\overline{p}_{\beta}}{dt}
-{\mathcal C}_{A\alpha}^Cy^A\left[\frac{\partial \widetilde{L}_{\mathcal M }}{\partial {y}^{C}}-\overline{p}_{\beta}\frac{\partial G^{\beta}}{\partial {y}^{C}}\right]-\frac{d}{dt}\left[\frac{\partial \widetilde{L}_{\mathcal M }}{\partial {y}^{\alpha}}-\overline{p}_{\beta}\frac{\partial G^{\beta}}{\partial {y}^{\alpha}}\right]
\nonumber\\
&
+\rho_{\alpha}^{i}\left(\frac{\partial \widetilde{L}_{\mathcal M }}{\partial q^{i}}-\overline{p}_{\beta}\frac{\partial G^{\beta}}{\partial q^{i}}\right)+{\mathcal C}_{A\alpha}^by^A\left(\frac{d}{dt}\left[\frac{\partial \widetilde{L}_{\mathcal M }}{\partial z^{b}}-\overline{p}_{\beta}\frac{\partial G^{\beta}}{\partial z^{b}}\right]\right)
-{\mathcal C}_{A\alpha}^by^A\left[\frac{\partial \widetilde{L}_{\mathcal M }}{\partial {y}^{b}}-\overline{p}_{\beta}\frac{\partial G^{\beta}}{\partial {y}^{b}}\right]\nonumber
\end{align}

If the matrix \[ \left(\frac{\partial^2 \widetilde{L}_{\mathcal M
}}{\partial z^{a}\partial z^b}\right)
\]
is regular then we can write the previous equations as an explicit
system of third-order differential equations. This regularity
assumption is equivalent to the condition that the constrain
algorithm  stops at the first constraint submanifold. In this submanifold there exists a unique solution for the boundary value problem determined by the optimal control problem. 
\begin{example}\textbf{Optimal control of an underactuated Elroy's beanie:}
This mechanical system is probably the simplest example of a dynamical
system with a non-Abelian Lie group symmetry. It consists of two
planar rigid bodies connected through their centers of mass (by a rotor let's say) moving freely
in the plane (see \cite{Blo} and \cite{Ostrowski}). The main (i.e. more massive) rigid body has the capacity to apply a torque to the connected rigid body. 

The configuration space is
$Q=SE(2)\times S^1$ with coordinates $(x, y, \theta,
\psi)$, where the first three coordinates  describe the position and
orientation of the center of mass of the first body and the last one describe
the relative orientation between both bodies.

\begin{center}
\begin{figure}[h!]
\includegraphics[scale=.35]{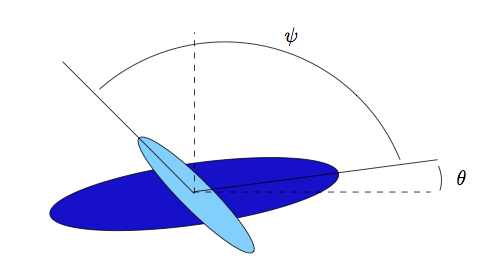} 
\caption{Top View of Elroy's beanie.}\label{Elroybeanie}
\end{figure}
\end{center}

The
Lagrangian $L: TQ\to \R$ is
\[
L= \frac{1}{2}m
(\dot{x}^2+\dot{y}^2)+\frac{1}{2}
I_1\dot{\theta}^2+\frac{1}{2}I_2
(\dot{\theta}+\dot{\psi})^2-V(\psi)
\]
where $m$ denotes the mass of the system and
$I_1$ and $I_2$ are the inertias of the first and
the second body, respectively; additionally, we
also consider a potential function of the form
$V(\psi)$.
The kinetic energy is associated with the Riemannian metric ${\mathcal G}$ on $Q$ given by
\[
{\mathcal G}=m(dx^2+dy^2)+(I_1+I_2)d\theta^2+I_2d\theta\otimes
d\psi +I_2d\psi\otimes d\theta+I_2 d\psi^2.
\]

The system is $SE(2)$-invariant for the action  \[
\Phi_g(q)=\left( z_1+x\cos\alpha-y\sin \alpha,
z_2+x\sin\alpha+y\cos \alpha, \alpha+\theta, \psi \right)
\]
where $g=(z_1, z_2, \alpha)$.

Let $\{\xi_1, \xi_2, \xi_3\}$ be the standard basis of
$\frak{se}(2)$,
\[
{}[\xi_1,\xi_2]=0,\qquad [\xi_1,\xi_3]=-\xi_2\qquad, [\xi_2,
\xi_3]=\xi_1\; .
\]
The quotient space $\widehat{Q}=Q/SE(2)=(SE(2)\times S^1)/SE(2)\simeq
S^1$ is naturally parameterized by the coordinate $\psi$. The
Atiyah algebroid $TQ/SE(2)\to \widehat{Q}$ is identified with the
vector bundle: $ \tau_{\bar{A}}: \bar{A}=TS^1\times {\mathfrak
se}(2)\to S^1. $ The canonical basis of sections of
$\tau_{\bar{A}}$ is: ${\displaystyle \left\{
\frac{\partial}{\partial \psi}, \xi_1, \xi_2, \xi_3\right\}}. $
Since the metric ${\mathcal G}$ is also $SE(2)$-invariant we
obtain a bundle metric $\hat{\mathcal G}$ and a $\hat{\mathcal
G}$-orthonormal basis of sections:
\[
\left\{
\displaystyle X_1=\sqrt{\frac{I_1+I_2}{I_1I_2}}\left(
\frac{\partial}{\partial \psi}-\frac{I_2}{I_1+I_2} \xi_3\right),
X_2=\frac{1}{\sqrt{m}}\xi_1, X_3=\frac{1}{\sqrt{m}}\xi_2, X_4=\frac{1}{\sqrt{I_1+I_2}}\xi_3\right\}
\]
In the coordinates $(\psi, v^1, v^2, v^3, v^4)$ induced by the orthonormal basis of sections, the reduced Lagrangian is $$\bar{L}=\frac{1}{2}\left( (v^1)^2+(v^2)^2 +(v^3)^2+(v^4)^2\right)-V(\psi)\; .$$

Additionally, we deduce that
\[
\begin{array}{ll}
\displaystyle \lcf X_1, X_2 \rcf_{\bar{A}}= -\sqrt{\frac{I_2}{I_1(I_1+I_2)}}X_3, & \displaystyle\lcf X_1, X_3 \rcf_{\bar{A}}=\sqrt{\frac{I_2}{I_1(I_1+I_2)}}X_2,\\[10pt]
\displaystyle \lcf X_1, X_4 \rcf_{\bar{A}}=0, &\displaystyle \lcf X_2, X_3 \rcf_{\bar{A}}=0,\\[10pt]
\displaystyle \lcf X_2, X_4 \rcf_{\bar{A}}=-\frac{1}{\sqrt{I_1+I_2}}X_3,& \displaystyle\lcf X_3, X_4 \rcf_{\bar{A}}=\frac{1}{\sqrt{I_1+I_2}}X_2.
\end{array}
\]
Therefore, the non-vanishing structure functions are
\[
\displaystyle C_{12}^3=-\sqrt{\frac{I_2}{I_1(I_1+I_2)}}, \quad
C_{13}^2=\sqrt{\frac{I_2}{I_1(I_1+I_2)}},\quad \displaystyle
C_{24}^3=-\frac{1}{\sqrt{I_1+I_2}}, \quad
C_{34}^2=\frac{1}{\sqrt{I_1+I_2}}.
\]
Moreover,
\[
\rho_{\bar{A}}(X_1)=\sqrt{\frac{I_1+I_2}{I_1I_2}}
\frac{\partial}{\partial \psi}, \quad \rho_{\bar{A}}(X_2)=0, \quad\rho_{\bar{A}}(X_3)=0,\quad\rho_{\bar{A}}(X_4)=0.
\]
The local expression of the Euler-Lagrange equations for the reduced Lagrangian system $\bar{L}:\bar{A}\to \R$ is:
\begin{eqnarray*}
\dot{\psi}&=&\sqrt{\frac{I_1+I_2}{I_1I_2}}v^1,\\
\dot{v}^1&=&-\sqrt{\frac{I_1+I_2}{I_1I_2}}
\frac{\partial V}{\partial \psi},\\
\dot{v}^2&=&-\sqrt{\frac{I_2}{I_1(I_1+I_2)}}v^1v^3+\frac{1}{\sqrt{I_1+I_2}}v^3v^4,\\
\dot{v}^3&=&\sqrt{\frac{I_2}{I_1(I_1+I_2)}}v^1v^2-\frac{1}{\sqrt{I_1+I_2}}v^2v^4,\\
\dot{v}^4&=&0.
\end{eqnarray*}

Next we introduce controls in our picture. Let $u(t)\in\mathbb{R}$ be a control input that permits to steer the system from an initial position to a desired position  by controlling only the variable $\psi$. Therefore the controlled Euler-Lagrange equations are now

\begin{eqnarray*}
\dot{\psi}&=&\sqrt{\frac{I_1+I_2}{I_1I_2}}v^1,\\
\dot{v}^1&=&-\sqrt{\frac{I_1+I_2}{I_1I_2}}
\frac{\partial V}{\partial \psi}+u,\\
\dot{v}^2&=&-\sqrt{\frac{I_2}{I_1(I_1+I_2)}}v^1v^3+\frac{1}{\sqrt{I_1+I_2}}v^3v^4,\\
\dot{v}^3&=&\sqrt{\frac{I_2}{I_1(I_1+I_2)}}v^1v^2-\frac{1}{\sqrt{I_1+I_2}}v^2v^4,\\
\dot{v}^4&=&0.
\end{eqnarray*}

{}From the second equation we obtain the feedback control law:

$$u=\dot{v}^1+\sqrt{\frac{I_1+I_2}{I_1I_2}}
\frac{\partial V}{\partial \psi}.$$

 %\frac{I_1I_2}{2(I_1+I_2)}\left(\ddot{\psi}+\frac{\partial V}{\partial\psi}\right)^2
 
The optimal control problem consists
of finding trajectories of the states variables and controls inputs,
satisfying the previous equations subject to given initial and final conditions
and minimizing the cost functional,
$$\min_{(v,\psi, \dot{\psi},u)}\int_{0}^{T}C(v,\psi, \dot{\psi},u)dt=\min_{(\psi, \dot{\psi}, \Omega,u)}\int_{0}^{T}\frac{1}{2}u^{2}dt$$ where $v=(v^1,v^2,v^3,v^4)$.

Our optimal control problem is equivalent to solving the following
second-order variational problem with second-order constraints given
by  
$$\min_{(v,\dot{v},\psi,\dot{\psi},\ddot{\psi})}\widetilde{L}(v,\dot{v},\psi,\dot{\psi},\ddot{\psi})=C\left(v,\psi,\dot{\psi},\dot{v}^1+\sqrt{\frac{I_1+I_2}{I_1I_2}}
\frac{\partial V}{\partial \psi}\right),$$ where $\widetilde{L}:T^{(2)}\mathbb{S}^{1}\times
2\widetilde{SE(2)}\to\R$, subject the second-order constraints
$\Phi^{\alpha}:T^{(2)}\mathbb{S}^{1}\times 2\widetilde{SE(2)}\to\R,$
$\alpha=1,\ldots,4,$

\begin{align}
\Phi^{1}(v,\dot{v},\psi,\dot{\psi},\ddot{\psi})&=\dot{\psi}-\sqrt{\frac{I_2+I_1}{I_2I_1}}v^1,\nonumber\\
\Phi^{2}(v,\dot{v},\psi,\dot{\psi},\ddot{\psi})&=\dot{v}^2-\frac{1}{\sqrt{I_1+I_2}}v^3v^4+\sqrt{\frac{I_2}{I_1(I_1+I_2)}}v^1v^3,\nonumber\\
\Phi^{3}(v,\dot{v},\psi,\dot{\psi},\ddot{\psi})&=\dot{v}^3+\frac{1}{\sqrt{I_1+I_2}}v^2v^4-\sqrt{\frac{I_2}{I_1(I_1+I_2)}}v^1v^2,\nonumber\\
\Phi^{4}(v,\dot{v},\psi,\dot{\psi},\ddot{\psi})&=\dot{v}_{4}.\nonumber
\end{align} Therefore, the constraint submanifold $\mathcal{M}$ of $T^{(2)}\mathbb{S}^{1}\times 2\widetilde{SE(2)}$ is given by \begin{align*}\mathcal{M}=\Big{\{}(v,\dot{v},\psi,\dot{\psi})\mid& \dot{\psi}=\sqrt{\frac{I_2+I_1}{I_2I_1}}v^1, \dot{v}^2=\frac{1}{\sqrt{I_1+I_2}}v^3v^4-\sqrt{\frac{I_2}{I_1(I_1+I_2)}}v^1v^3, \\ &\qquad\qquad\qquad\qquad\dot{v}^3=-\frac{1}{\sqrt{I_1+I_2}}v^2v^4+\sqrt{\frac{I_2}{I_1(I_1+I_2)}}v^1v^2, \dot{v}^4=0\Big{\}}.\end{align*}
We consider the submanifold $W_0=\mathcal{M}\times2\widetilde{SE(2)}^{*}$ with induced coordinates $$(v^1,v^2,v^3,v^4,\psi,\dot{v}^{1},p_1,p_2,p_3,p_4,\bar{p}_1,\bar{p}_2,\bar{p}_3,\bar{p}_4).$$ Now, we consider the restriction $\widetilde{L}_{\mathcal{M}}$ given by $$\widetilde{L}_{\mathcal{M}}=\frac{1}{2}\left(\dot{v}^1+\sqrt{\frac{I_1+I_2}{I_1I_2}}
\frac{\partial V}{\partial \psi}\right)^2.$$

Moreover, the first constraint submanifold $W_{1}$ is determined by $$W_1=\Big{\{} z\in W_0\mid\bar{p}_1-\dot{v}^1-\sqrt{\frac{I_1+I_2}{I_1I_2}}\left(\frac{\partial V}{\partial\psi}+\bar{p}_1\right)-\sqrt{\frac{I_2}{I_1(I_1+I_2)}}\left(\bar{p}_{3}v^{2}-\bar{p}_{2}v^{3}\right)=0\Big{\}}.$$ Observe that $$\det\left(\frac{\partial^2\widetilde{L}_{\mathcal{M}}}{\partial\dot{v}^1\partial\dot{v}^{1}}\right)\neq 0.$$ Thus, the constraint algorithm stops at the first constraint submanifold $W_1$. 

Finally, in a similar fashion as the unconstrained situation, we would like to point out that the regularity condition provides the existence of a unique solution of the dynamics along the submanifold $W_1$.

Then,  we  can  write  the  equations  determining necessary conditions for  the  optimal  control problem:

\begin{align*}
\dot{p}_{1}&=\frac{I_1+I_2}{I_1I_2}\left(\dot{v}_{1}+\sqrt{\frac{I_1+I_2}{I_1I_2}}\frac{\partial V}{\partial\psi}\right)\frac{\partial^{2}V}{\partial\psi\partial\psi}-\sqrt{\frac{I_2}{I_1(I_1+I_2)}}p_3v^2,\\
\dot{p}_2&=-\frac{p_3v^4}{\sqrt{I_1+I_2}}+\sqrt{\frac{I_2}{I_1(I_1+I_2)}}p_3v^1,\\
\dot{p}_3&=\frac{p_2v^4}{\sqrt{I_1+I_2}}-\sqrt{\frac{I_2}{I_1(I_1+I_2)}}p_2v^1,\\
\dot{p}_4&=\frac{1}{\sqrt{I_1+I_2}}(p_3v^2-p_2v^3),\\
\dot{\bar{p}}_{1}&=-p_1+\bar{p}_1\sqrt{\frac{I_1+I_2}{I_1I_2}}+\sqrt{\frac{I_2}{I_1(I_1+I_2)}}(\bar{p}_3v^2-\bar{p}_2v^3),\\
\dot{\bar{p}}_{2}&=-p_2+\bar{p}_3\left(\frac{v_4}{\sqrt{I_1+I_2}}+v^1\sqrt{\frac{I_2}{I_1(I_1+I_2)}}\right),\\
\dot{\bar{p}}_{3}&=-p_3+\bar{p}_2\left(-\frac{v_4}{\sqrt{I_1+I_2}}+v^1\sqrt{\frac{I_2}{I_1(I_1+I_2)}}\right),\\
\dot{\bar{p}}_4&=-p_4+\frac{1}{\sqrt{I_1+I_2}}(\bar{p}_2v^3-\bar{p}_3v^2),\\
\bar{p}_1&=\dot{v}^1+\sqrt{\frac{I_1+I_2}{I_1I_2}}\left(\frac{\partial V}{\partial\psi}+\bar{p}_1\right)+\sqrt{\frac{I_2}{I_1(I_1+I_2)}}\left(\bar{p}_{3}v^{2}-\bar{p}_{2}v^{3}\right),\\
\dot{\psi}&=\sqrt{\frac{I_2+I_1}{I_2I_1}}v^1,\quad \dot{v}^2=\frac{1}{\sqrt{I_1+I_2}}v^3v^4-\sqrt{\frac{I_2}{I_1(I_1+I_2)}}v^1v^3, \\
\dot{v}^3&=-\frac{1}{\sqrt{I_1+I_2}}v^2v^4+\sqrt{\frac{I_2}{I_1(I_1+I_2)}}v^1v^2,\quad \dot{v}^4=0.
\end{align*}
\end{example}

\section*{Acknowledgments} 
This work has been partially supported by MICINN (Spain) Grant MTM 2013-42870-P and NSF grant INSPIRE-1343720. L. Colombo would like to thank David Mart\'in de Diego for fruitful comments and discussions and to propose the initial ideas of this work.
% You may incorporate your references as follows in your main tex file.
% Using BibTex is not recommended but can be handled.

\medskip
% The data information below will be filled by AIMS editorial staff
%Received xxxx 20xx; revised xxxx 20xx.
\medskip

\end{document}